\documentclass[a4paper,fleqn,usenatbib]{mnras}

\usepackage{newtxtext,newtxmath}
\usepackage[T1]{fontenc}
\usepackage{graphicx}	% Including figure files
\usepackage{amsmath}	% Advanced maths commands
\usepackage{amssymb}	% Extra maths symbols
\usepackage{bm}

\usepackage{hyperref}
\usepackage{natbib}
\usepackage{amsmath}
\usepackage[super]{nth} % for 1st, 2nd, etc

\usepackage{booktabs} % to input pandas tables

\usepackage{tikzexternal}
\tikzsetexternalprefix{fig/}

\newcommand{\press}{p}
\newcommand{\ptot}{\press_T}
\newcommand{\xlab}{\bm x}
\newcommand{\xcell}{\bm \xi}
\newcommand{\diffxlab}{\mathrm{d} \xlab}
\newcommand{\diffxcell}{\mathrm{d} \xcell}
\newcommand{\diffS}{\mathrm{d} S}
\newcommand{\massm}{M}
\newcommand{\wgt}{w}
\newcommand{\degree}{d}
\newcommand{\cns}{\bm u}
\newcommand{\chrv}{\bm a}
\newcommand{\Lchrm}{\bm{\mathsf{L}}}
\newcommand{\Rchrm}{\bm{\mathsf{R}}}
\newcommand{\vel}{\bm v}
\newcommand{\flux}{\bm f}
\newcommand{\bfunc}{\bm \phi}
\newcommand{\idm}{\mathcal{I}}
\newcommand{\dotp}[2]{{#1}{\bm \cdot}{#2}}
\newcommand{\innerp}[2]{\left<{#1},{#2}\right>}
\newcommand{\powellq}{\bm q}
\newcommand{\Bfield}{\bm B}
\newcommand{\avg}[1]{\bar {#1}}
\newcommand{\slope}[2]{{#1}_{,#2}}
\newcommand{\Bslope}[2]{\Bfield_{{#1},{#2}}}
\newcommand{\slopeshort}{\sigma}
\newcommand{\facenormal}{\bm n}

\newcommand{\Bnormal}{B^n}
\newcommand{\divB}{\dotp{\bm \nabla}{\Bfield}}

\newcommand{\athena}{Athena}
\newcommand{\Arepo}{AREPO}
\newcommand{\ArepoDG}{AREPO-DG}
\newcommand{\SymPy}{SymPy}

\newcommand{\Emag}{E_\mathrm{mag}}
\newcommand{\Etherm}{E_\mathrm{th}}
\newcommand{\Etot}{E_\mathrm{tot}}
\newcommand{\Ekin}{E_\mathrm{kin}}

\newcommand{\AffilHITS}{1}
\newcommand{\AffilZAH}{2}
\newcommand{\AffilExe}{3}
\newcommand{\AffilMPA}{4}
\newcommand{\AffilTIFR}{5}
\newcommand{\AffilWrzb}{6}

\title[Discontinuous Galerkin Magnetohydrodynamics]{High-order Magnetohydrodynamics for Astrophysics with an Adaptive Mesh Refinement Discontinuous Galerkin Scheme}

\author[T. Guillet et al.]{%
Thomas Guillet$^{\AffilHITS,\AffilZAH,\AffilExe}$\thanks{E-mail: \href{mailto:T.A.Guillet@exeter.ac.uk}{T.A.Guillet@exeter.ac.uk}},
R\"udiger Pakmor$^{\AffilHITS,\AffilMPA}$,
Volker Springel$^{\AffilHITS,\AffilZAH,\AffilMPA}$,
\newauthor
Praveen Chandrashekar$^{\AffilTIFR}$,
Christian Klingenberg$^{\AffilWrzb}$
\\
$^{\AffilHITS}$Heidelberg Institute for Theoretical Studies, Schloss-Wolfsbrunnenweg 35, 69118 Heidelberg, Germany\\
$^{\AffilZAH}$Zentrum f\"ur Astronomie der Universit\"at Heidelberg, Astronomisches Rechen-Institut, M\"onchhofstr. 12--14, 69120 Heidelberg, Germany \\
$^{\AffilExe}$Department of Physics and Astronomy, Stocker Road, University of Exeter, EX4~4QL, United Kingdom\\
$^{\AffilMPA}$Max-Planck-Institut f\"ur Astrophysik, Karl-Schwarzschild-Str. 1, 85741 Garching, Germany\\
$^{\AffilTIFR}$TIFR Center for Applicable Mathematics, Bangalore, India\\
$^{\AffilWrzb}$Department of Mathematics, Universit\"at W\"urzburg, Emil Fischer Str. 40, 97074 W\"urzburg, Germany\\
}

\date{Accepted XXX. Received YYY; in original form ZZZ}

\pubyear{2018}

\begin{document}
\label{firstpage}
\pagerange{\pageref{firstpage}--\pageref{lastpage}}
\maketitle

% Abstract of the paper
\begin{abstract}
%This is a simple template for authors to write new MNRAS papers.
%The abstract should briefly describe the aims, methods, and main results of the paper.
%It should be a single paragraph not more than 250 words (200 words for Letters).
%No references should appear in the abstract.
%%
Modern astrophysical simulations
aim to accurately model an ever-growing array of physical processes,
including the interaction of fluids with magnetic fields,
under increasingly stringent performance and scalability requirements
driven by present-day trends in computing architectures.
Discontinuous Galerkin methods have recently gained some traction in astrophysics,
because of their arbitrarily high order and controllable numerical diffusion,
combined with attractive characteristics for high performance computing.
In this paper, we describe and test our implementation of a discontinuous Galerkin (DG) scheme for ideal magnetohydrodynamics
in the {\ArepoDG} code.
Our DG-MHD scheme relies on a modal expansion of the solution on Legendre polynomials
inside the cells of an Eulerian octree-based AMR grid.
The divergence-free constraint of the magnetic field is enforced using one out of two distinct cell-centred schemes:
either a Powell-type scheme based on nonconservative source terms,
or a hyperbolic divergence cleaning method.
The Powell scheme relies on a basis of locally divergence-free vector polynomials inside each cell to represent the magnetic field.
Limiting prescriptions are implemented to ensure non-oscillatory and positive solutions.
We show that the resulting scheme is accurate and robust:
it can achieve high-order and low numerical diffusion, as well as accurately capture strong MHD shocks.
In addition, we show that our scheme exhibits a number of attractive properties for astrophysical simulations,
such as lower advection errors and better Galilean invariance at reduced resolution,
together with more accurate capturing of barely resolved flow features.
We discuss the prospects of our implementation, and DG methods in general,
for scalable astrophysical simulations.
\end{abstract}

% Select between one and six entries from the list of approved keywords.
% Don't make up new ones.

%%%%%%%%%%%%%%%%%%%%%%%%%%%%%%%%%%%%%%%%%%%%%%%%%%

%%%%%%%%%%%%%%%%% BODY OF PAPER %%%%%%%%%%%%%%%%%%

\section{Introduction}

Numerical simulations in astrophysics have established themselves as a critical tool
to study the complex interactions of physical processes in the Universe.
Galaxy formation simulations, in particular,
are faced with the challenge of accounting for ever expanding models of the physics,
while requiring increasingly accurate numerical methods and scalable high-performance implementations.

Among these physical processes,
cosmic magnetic fields are now recognized as playing a key role in star formation in the interstellar medium,
and possibly impacting the dynamics of gas in galaxies at larger scales
as they get amplified by multiple dynamo processes which are still only partially understood.
For this reason, many astrophysical codes have been implementing magnetohydrodynamics (MHD) solvers,
to compute the joint evolution of gas and magnetic fields
\citep[e.g.][]{stone_zeus-2d:_1992,fryxell_flashadaptivemesh_2000,fromang_high_2006,mignone_plutonumericalcode_2007,stone_athena:_2008,collins_cosmological_2010,pakmor_magnetohydrodynamics_2011}.
These methods have been applied to setups of growing physical complexity,
from the study of magnetized turbulence in the interstellar medium
\citep{stone_dissipationcompressiblemagnetohydrodynamic_1998,maclow_energydissipationrate_1999,schekochihin_simulationssmallscaleturbulent_2004,federrath_comparingstatisticsinterstellar_2010},
to magnetic field evolution in isolated galaxy setups
\citep[e.g.][]{wang_magnetohydrodynamicsimulationsdisk_2009,dubois_magnetisedwindsdwarf_2010,pakmor_simulations_2013,rieder_smallscaledynamofeedbackdominated_2016},
and more recently, the amplification of magnetic fields in galaxies in full cosmological context
\citep{pakmor_magneticfieldscosmological_2014,pakmor_magneticfieldformation_2017,rieder_smallscaledynamofeedbackdominated_2017b}.

The increasing CPU and memory requirements of these simulations are confronted today by modern trends in computing architectures.
It is now well recognized that advances in computing power are more due to the power-efficient integration of multiple processor cores,
rather than single-core performance growth.
In addition, memory technology has not been keeping up with the steady gains in floating-point computing power,
lagging behind both in terms of performance and capacity.
As a result, codes have to turn towards increasingly parallel and compute-efficient numerical methods
to face the expanding computing and memory requirements of the physical models they attempt to simulate.

In recent years, discontinuous Galerkin (DG) methods have gained traction in the computational fluid dynamics community in general, and in astrophysics in particular,
as a promising framework for scalable and accurate high-order methods for hyperbolic problems.
DG schemes lie at the crossroads of spectral element and finite volume methods: based on the expansion of the solution inside simulation volumes on a set of chosen basis functions (typically polynomials),
DG schemes evolve the expansion coefficients (the so-called weights) in time
based on a weak formulation of the governing equations.
The popularity of DG stems from the fact that it provides a clean framework for discretizing hyperbolic problems at any order of spatial accuracy,
together with attractive data locality:
regardless of spatial order, DG schemes in theory require communication only between directly neighbouring cells,
unlike high-order reconstruction-based finite volume methods.
As a result, DG methods have been the focus of intense research over the last three decades
\cite[see e.g.][]{
cockburn_tvb_1989-1,
cockburn_localdiscontinuousgalerkin_1998,
cockburn_locally_2004,
li_locally_2005,
hesthaven_nodaldiscontinuousgalerkin_2008,
li_central_2011,
shu_brief_2013,
balsara_higherorderaccuratespacetime_2017}.

Because DG methods can easily be setup to operate at any spatial order,
their numerical dissipation is controllable and can be reduced as required,
which is of great interest to simulate the high Reynolds number flows of astrophysical plasmas.
%%%
From a pure hydrodynamics point of view, \cite{bauer_subsonic_2012,nelson_moving_2013} show that numerical dissipation can result in spurious heating of the intergalactic medium from turbulent motions, preventing cooling and accretion of the hot ambient gas.
This result is corroborated by \cite{zhu_turbulence_2013} in the context of higher-order reconstruction-based methods.
%%%
In the context of decaying isothermal magnetized supersonic turbulence,
the code comparison study of \citet{kritsuk_comparing_2011} has shown that higher-order codes
provide a larger turbulence spectral bandwidth and increased effective Reynolds number
compared to second-order schemes.
%%%
Even at second order,
recent studies by \cite{mocz_discontinuous_2014} and \cite{schaal_astrophysical_2015}
show that DG methods demonstrate overall superior accuracy compared to finite volume methods,
in particular due to lower advection errors and reduced angular momentum diffusion.
Beyond second order, \cite{schaal_astrophysical_2015} find that, on smooth problems at least,
increasing the DG order reduces advection errors,
and decreases the total time-to-solution for a given target accuracy.

These properties make DG schemes attractive for fixed grid Eulerian astrophysical codes,
where large structures such as galaxies may travel across the grid with large bulk advection velocities.
As a result, there has been significant recent interest in DG methods in astrophysics,
for hydrodynamics \citep{schaal_astrophysical_2015,velascoromero_planetdiscinteractions_2018},
ideal magnetohydrodynamics \citep{mocz_discontinuous_2014,dumbser_simple_2016,karami_halashi_reconstructed_2016},
non-ideal MHD \citep{boscheri_arbitrary-lagrangianeulerian_2017},
as well as special- and general-relativistic MHD \citep{zanotti_solving_2015,kidder_spectre:_2017,anninos_cosmosdghpadaptivediscontinuous_2017,zhao_rungekutta_2017,fambri_aderdiscontinuousgalerkin_2018}.

In this paper, we present {\ArepoDG}, our extension of the hydrodynamical DG code TENET of \citet{schaal_astrophysical_2015}, to ideal MHD.
Like its predecessor, this work leverages the infrastructure of the moving mesh code {\Arepo} \citep{springel_pursimuove_2010},
but instead of a moving Voronoi mesh,
relies on {\Arepo}'s optional support for fixed Eulerian adaptive grids.
The goal is to develop new parallel Eulerian solvers within this existing framework
which explores different accuracy and scalability compromises than the moving mesh solution.

Our code adopts a modal Legendre polynomial basis for the expansion of hydrodynamical variables.
Two different methods are implemented to control the divergence of the magnetic field:
in a first scheme, which we refer to as the Powell method,
the magnetic field components are expanded within each cell onto a special basis of locally divergence-free (LDF) vector polynomials
\citep{cockburn_locally_2004,li_locally_2005,yakovlev_locally_2013,zhao_positivity-preserving_2014,zhao_rungekutta_2017},
and additional source terms are introduced following the formulation of \citet{powell_solution-adaptive_1999}.
For this Powell method, the LDF basis only ensures that the magnetic field is \emph{locally} divergence-free inside a cell,
but not \emph{globally} divergence-free,
since the normal component of the magnetic field is not guaranteed to be continuous across cell interfaces.
The other implemented divergence control method relies on hyperbolic divergence cleaning as formulated by \citet{dedner_hyperbolic_2002},
for which we typically use the same Legendre basis
both for hydrodynamical variables and for the magnetic field components.
With this scheme and choice of basis, the magnetic field is formally neither locally nor globally divergence-free,
but its divergence is instead dynamically damped and advected away using an additional scalar field.

The DG weights are evolved in time through explicit time integration of the semi-discrete scheme,
in the so-called Runge-Kutta DG (RKDG) framework introduced by \cite{cockburn_tvb_1989}.
In this paper, we discuss our implementation,
detailing the numerical ingredients required for code stability and accuracy,
with a special focus on the treatment of the divergence of the magnetic field.
Our scheme introduces two new elements:
different discretizations of the Powell term in the DG framework in \ref{sec:powell-discretization},
and a non-linear divergence-free slope limiting procedure for the magnetic field in \ref{sec:limiting-div0}.
We also cover some important aspects of the method related to performance and computing efficiency.

The present paper is organised as follows.
In Section~\ref{sec:gov-eqs}, we review the governing equations for ideal MHD in their conserved formulation.
In Section~\ref{sec:dg-scheme}, we describe our DG discretization and time integration scheme,
with a focus on aspects specific to MHD, in particular the divergence-free basis for the Powell scheme.
Section~\ref{sec:num-ingredients} details the numerical ingredients required to make the scheme accurate and stable with MHD;
we discuss in particular the issues related to oscillation limiting,
control of the divergence of the magnetic field,
and enforcement of solution positivity.
In Section~\ref{sec:results}, we present results on a number of test problems,
showing that the method achieves high order in smooth regions, while capturing shocks and discontinuities.
We show that higher orders indeed reduce advection errors and dissipation even in case of strong MHD shocks,
and help to efficiently capture barely resolved flow features.
We also present test problems specifically aimed at testing the control of the magnetic field divergence.
In Section~\ref{sec:discussion}, we discuss prospects for this method and astrophysical simulations,
also covering some aspects related to implementation efficiency and performance.
We also consider some challenges and future prospects for our scheme---and DG methods in general---%
towards large-scale production astrophysical simulations.

\section{Governing equations}
\label{sec:gov-eqs}

The ideal MHD equations may be written as a system of conservation laws of the form:
\begin{equation}
  \frac{\partial \cns}{\partial t} + \sum_{\alpha=1}^{3} \frac{\partial \flux_\alpha(\cns)}{\partial x_\alpha} = 0,
  \label{eq:cns-eq}
\end{equation}
where the index $\alpha$ runs over all space dimensions.
The vector $\cns$ holds the conserved variables of the system:
\begin{equation}
  \cns = \begin{bmatrix}
       \rho \\ \rho \vel \\ E \\ \Bfield
     \end{bmatrix},
     \quad E = \rho u + \frac{1}{2}\rho \vel^2 + \frac{1}{2}\Bfield^2,
  \label{eq:cns-vec}
\end{equation}
where
$\rho$ is the density,
$\vel$ the fluid velocity,
$E$ the total energy density,
$\Bfield$ the magnetic field vector,
$\press$ the hydrodynamic (thermal) pressure,
$u$ is the specific internal energy,
which we assume to be related to the pressure $p$ by the ideal gas equation of state with adiabatic index $\gamma$:
\begin{equation}
  p = \rho u (\gamma - 1),
\end{equation}
from which we will write the total energy as
\begin{equation}
  E = \frac{\press}{\gamma-1} + \frac{1}{2}\rho \vel^2 + \frac{1}{2}\Bfield^2.
\end{equation}
The fluxes $\flux_\alpha(\cns)$ for ideal MHD are given by
\begin{equation}
  \flux = \begin{bmatrix}
    \rho \vel \\
    \rho \vel \otimes \vel + \ptot \idm_3 - \Bfield \otimes \Bfield \\
    (E + \ptot) \vel - (\dotp{\vel}{\Bfield}) \Bfield \\
    \Bfield \otimes \vel - \vel \otimes \Bfield
  \end{bmatrix},
  \label{eq:mhd-flux}
\end{equation}
where $\idm_3$ is the $3 \times 3$ identity matrix, and $\ptot$ is the total pressure defined by
\begin{equation}
  \ptot = p + \frac{1}{2}\Bfield^2.
\end{equation}

For physical solutions, the Maxwell equations impose the additional constraint
\begin{equation}
  \divB = 0
  \label{eq:divB}
\end{equation}
everywhere and at all times.
The form of the flux \eqref{eq:mhd-flux} guarantees that initial divergence-free solutions will remain divergence-free under the dynamical evolution dictated by \eqref{eq:cns-eq}.
However, this is not necessarily true for approximate numerical solutions, which require special divergence control measures.

\section{Discontinuous Galerkin Scheme}
\label{sec:dg-scheme}

We now describe our discontinuous Galerkin scheme,
which follows the overall RKDG framework of
\cite{cockburn_tvb_1989,cockburn_tvb_1989-1}.
The implementation adopts the general setup of \cite{schaal_astrophysical_2015},
but is adapted to MHD.
In the following, we present the more general case of the Powell divergence control scheme,
which relies on a basis of locally divergence-free polynomials for the magnetic field components
\citep{cockburn_locally_2004,li_locally_2005,yakovlev_locally_2013}.
The other divergence control scheme, hyperbolic cleaning, requires only the use of the Legendre basis.
We discuss both implementations of divergence control in Section~\ref{sec:divB}.

\subsection{Representation of the conserved state}

The computational domain is partitioned into non-overlapping cubic cells, structured as an AMR octree.
Within each cell $K$, the local conserved state vector $\cns^K(\xlab,t)$ is expanded as a linear combination of $N$ time-independent basis functions $\bfunc_l^K(\xlab)$, defining time-dependent weights $\wgt_l^K(t)$:
\begin{equation}
  \cns^K(\xlab,t) = \sum_{l=1}^N \wgt^K_l(t) \bfunc_l^K(\xlab).
  \label{eq:u-exp}
\end{equation}
For each basis index $l$, $\bfunc_l^K$ is a vector of 8 time-independent scalar fields over $K$, one for each conserved component, and $\wgt_l^K(t)$ is a time-dependent scalar function.

The weights $\wgt^K_l$ are defined independently in each cell, resulting in fields which are discontinuous at interfaces.
Since all cells are cubic, we can define reference basis functions $\bfunc_l$ independent of cell $K$ by simple translation and rescaling of world coordinates $\xlab$ to reference cell coordinates $\xcell \in [-1, 1]^3$:
\begin{equation}
  \bfunc_l^K(\xlab) = \bfunc_l(\xcell), \qquad
  \xcell = \frac{2}{\Delta x^K}(\xlab - \xlab_c^K)
  \label{eq:x-xi}
\end{equation}
where $\Delta x^K, \xlab_c^K$ are the cell size and cell centre in world coordinates.

We further define the $L^2$ inner product on conserved vector fields in the reference cell:
\begin{equation}
  \innerp{\cns_1}{\cns_2} \coloneqq \int_{[-1,1]^3} \dotp{\cns_1(\xcell)}{\cns_2(\xcell)} \; \diffxcell,
  \label{eq:innerp}
\end{equation}
where the dot product on the right hand side is the usual dot product of $\mathbb{R}^8$, i.e. across conserved components.
By choosing a set of basis functions orthogonal with respect to this inner product, we can readily invert \eqref{eq:u-exp} by taking the inner product against each basis function:
\begin{equation}
  \wgt^K_l(t) = \massm_l^{-1} \innerp{\cns^K}{\bfunc_l},
\end{equation}
where
\begin{equation}
  \massm_l \coloneqq \innerp{\bfunc_l}{\bfunc_l}
\end{equation}
is the diagonal of the mass matrix $\innerp{\bfunc_l}{\bfunc_{l'}}$, all non-diagonal entries being zero.
We now discuss the two types of basis functions making up our global DG basis $\{\bfunc_l\}, l \in \{1,\ldots,N\}$.

\subsection{Legendre basis for hydrodynamics}

Following \cite{schaal_astrophysical_2015}, we expand the hydrodynamical variables $\rho$, $\rho \vel$ and $E$ independently onto 5 scalar Legendre polynomials of degree $\leq \degree$.
In 3 dimensions, scalar trivariate Legendre polynomials are indexed by 3 integers $k_1, k_2, k_3$ with $k_1+k_2+k_3 \leq d$, and take the form:
\begin{equation}
  P_{k_1,k_2,k_3}(\xcell) = \ell_{k_1}(\xi_1) \ell_{k_2}(\xi_2) \ell_{k_3}(\xi_3), \quad k_1+k_2+k_3 \leq d,
  \label{eq:legendre-3d}
\end{equation}
where $\ell_k$ is the univariate Legendre polynomial of degree $k$.
The $P_{k_1,k_2,k_3}$ form a basis of the space of scalar trivariate polynomials of degree $\leq d$, which has dimension
\begin{align}
  n(d) = \frac{1}{6} (d+1)(d+2)(d+3).
\end{align}
For brevity, we collectively label $(k_1, k_2, k_3)$ with a single integer in $\{1, \ldots, n(d)\}$.
Note that the univariate Legendre polynomials $\ell_k$ are orthogonal for the $L^2$ inner product on $[-1,1]$:
\begin{equation}
  \int_{[-1,1]} \ell_{k}(x) \ell_{k'}(x) \,\mathrm{d}x = 0 \Leftrightarrow k \neq k',
\end{equation}
and as a result, the polynomials of \eqref{eq:legendre-3d} are orthogonal for the same inner product on the reference cell $[-1,1]^3$.

Since we are expanding 5 conserved components independently, the Legendre subset of our basis functions will have dimension $N_\text{L} = 5 n(d)$.
We can choose the basis indexing $l$ so that the first $n(d)$ components $l \in \{1,\ldots,n(d)\}$ correspond to the Legendre expansion of $\rho$, then $l \in \{n(d)+1,\ldots,2n(d)\}$ correspond to $\rho v_1$, etc.
To write the Legendre $\bfunc_l$ explicitly in terms of vectors of conserved components:
\begin{equation}
  \bfunc_{l}(\xcell) = \begin{bmatrix}
    P_{l}(\xcell) \\
    \bm 0 \\
    0 \\
    \bm 0
  \end{bmatrix}, \ l \in \{1, \ldots, n(d)\}, \quad \text{for }\rho
\end{equation}
up to
\begin{equation}
  \bfunc_{4n(d)+l}(\xcell) = \begin{bmatrix}
    0 \\
    \bm 0 \\
    P_{l}(\xcell) \\
    \bm 0
  \end{bmatrix}, \ l \in \{1, \ldots, n(d)\}, \quad \text{for } E.
\end{equation}
Since the multivariate Legendre polynomials $P$ are $L^2$-orthogonal, the $\bfunc$ we just constructed are orthogonal for the inner product \eqref{eq:innerp}.

\subsection{Divergence-free basis for the magnetic field}
\label{sec:div0}

Locally divergence-free (LDF) basis functions for discontinuous Galerkin techniques were first introduced by \cite{cockburn_locally_2004} for the linear Maxwell equations, and extended by \cite{li_locally_2005} to non-linear MHD in the context of RKDG methods.
Divergence-free bases have also been used for DG in the context of relativistic MHD by \cite{zhao_rungekutta_2017}.
The idea is to enforce the divergence-free condition \eqref{eq:divB} inside the cells by expanding the magnetic field $\Bfield$ in \eqref{eq:u-exp} onto a basis of divergence-free vector polynomials, which in 3D has dimension:
\begin{equation}
  N_\text{div} = 3 n(d) - n(d-1).
  \label{eq:ndiv}
\end{equation}
We will discuss the properties of this basis for divergence control in Section~\ref{sec:divB},
and we refer the reader to Appendix~\ref{sec:apdx-div0} for more details related to properties and constructions of such bases in 2D and 3D.
In two dimensions, suitable basis functions may be found for example in \cite{cockburn_locally_2004,li_locally_2005} up to $d=2$,
or \cite{zhao_rungekutta_2017} up to $d=3$.

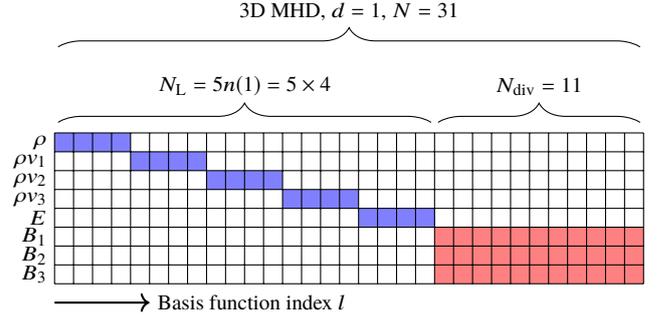
\begin{figure}
  \centering
  \tikzsetnextfilename{tikz-mhd-basis-3d}
  \begin{tikzpicture}[scale=0.25, yscale=-1]
    \coordinate (pos) at (0, 0);

    \foreach \alphaL in {0, ..., 4}
      \fill[LegendreCells] (pos) rectangle ++(4, 1) coordinate (pos);

    \fill[DivFreeCells] (pos) rectangle ++(11, 3);
    \draw (0,0) grid (5*4+11, 5+3);

    \node[left] at (0, 0.5) {$\rho$};
    \node[left] at (0, 1.5) {$\rho v_1$};
    \node[left] at (0, 2.5) {$\rho v_2$};
    \node[left] at (0, 3.5) {$\rho v_3$};
    \node[left] at (0, 4.5) {$E$};
    \node[left] at (0, 5.5) {$B_1$};
    \node[left] at (0, 6.5) {$B_2$};
    \node[left] at (0, 7.5) {$B_3$};

    \draw [brace,xshift=0,yshift=-5pt] (0, 0) -- (20, 0) node[above,midway,yshift=10pt] {$N_\text{L} = 5n(1) = 5 \times 4$};
    \draw [brace,xshift=0,yshift=-5pt] (20, 0) -- +(11, 0) node[above,midway,yshift=10pt] {$N_\text{div} = 11$};
    \draw [brace,xshift=0,yshift=-5pt] (0, -4) -- (31, -4) node[above,midway,yshift=10pt] {3D MHD, $d=1$, $N = 31$};

    \draw[->,thick] (0, 9) -- +(5, 0) node[right] {Basis function index $l$};
  \end{tikzpicture}
  \caption{\emph{Structure of the DG basis functions for 3D MHD with degree $d=1$ polynomials}, showing non-zero contributions (coloured cells) of each basis function $\bfunc_l$ ($l$-th column) to components of the conserved state vector $\cns$ (rows).
    The Legendre subspace for hydrodynamical variables has total dimension $N_\text{L}=20$,
    whereas the divergence-free subspace for $\Bfield$ is of dimension $N_\text{div} = 11$.
    Cells in white correspond to zero.
    While the Legendre subspace has a tensor product structure,
    all 3 components of $\Bfield$ in the divergence-free basis are coupled.
  }
  \label{fig:basis}
\end{figure}

The resulting total dimension of the DG basis with its Legendre and divergence-free subspaces is
\begin{equation}
  N = N_\text{L} + N_\text{div} = 8 n(d) - n(d-1).
  \label{eq:dimN3d}
\end{equation}
Its general structure is represented in Fig.~\ref{fig:basis}.
As detailed in Appendix~\ref{sec:apdx-div0}, the divergence-free basis is constructed so that its vectors are orthogonal with respect to the inner product \eqref{eq:innerp}.

The resulting combined Legendre and divergence-free basis is of dimension $N = N_\text{L} + N_\text{div}$, and is orthogonal:
\begin{equation}
  \forall (l, l') \in \{1,\ldots,N\}^2, \quad \innerp{\bfunc_l}{\bfunc_{l'}} = \delta_{ll'} M_l, \quad M_l > 0,
  \label{eq:basis-ortho}
\end{equation}
which results from the block structure shown in Fig.~\ref{fig:basis}, together with the orthogonality of the Legendre polynomials and divergence-free basis vectors.
The constants $M_l$ are fixed by the choice of basis polynomials.

LDF bases require slightly less storage per cell than expanding the components of $\Bfield$ as independent Legendre fields,
due to the fewer degrees of freedom arising from the divergence constraint.
However, we note that LDF bases prevent some optimizations that are possible with Legendre bases,
thus requiring more floating point operations.
They also require storing more precomputed data (such as the basis function values and gradients)
compared to Legendre bases.
We come back to these points in the discussion in Section~\ref{sec:perf-div0}.

\subsection{The special case of 2D MHD}

Two-dimensional MHD problems can be seen as being governed by the 3D equations,
but with imposed translation invariance along the third dimension of space, i.e.
$\partial / \partial x_3 = \partial / \partial \xi_3 \equiv 0$.
In this case, the divergence-free condition on $\Bfield$ becomes $\divB = \partial B_1 / \partial x_1 + \partial B_2 / \partial x_2 = 0$.
The component $B_3$ is still present in the 2D equations, but may vary independently from $B_1$, $B_2$.

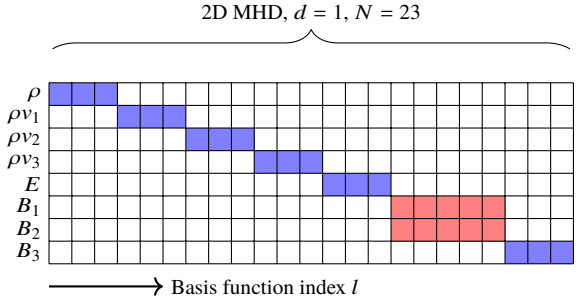
\begin{figure}
  \centering
  \tikzsetnextfilename{tikz-mhd-basis-2d}
  \begin{tikzpicture}[scale=0.30, yscale=-1]
    \coordinate (pos) at (0, 0);

    \foreach \alphaL in {0, ..., 4}
      \fill[LegendreCells] (pos) rectangle ++(3, 1) coordinate (pos);

    \fill[DivFreeCells] (pos) rectangle ++(5, 2) coordinate (pos);

    \fill[LegendreCells] (pos) rectangle ++(3, 1) coordinate (pos);

    \draw (0,0) grid (6*3+5, 5+3);

    \node[left] at (0, 0.5) {$\rho$};
    \node[left] at (0, 1.5) {$\rho v_1$};
    \node[left] at (0, 2.5) {$\rho v_2$};
    \node[left] at (0, 3.5) {$\rho v_3$};
    \node[left] at (0, 4.5) {$E$};
    \node[left] at (0, 5.5) {$B_1$};
    \node[left] at (0, 6.5) {$B_2$};
    \node[left] at (0, 7.5) {$B_3$};

    \draw [brace,xshift=0,yshift=-5pt] (0, -1) -- (23, -1) node[above,midway,yshift=10pt] {2D MHD, $d=1$, $N = 23$};

    \draw[->,thick] (0, 9) -- +(5, 0) node[right] {Basis function index $l$};
  \end{tikzpicture}
  \caption{\emph{Structure of the DG basis functions for 2D MHD with degree $d=1$ polynomials.}
    See Fig.~\ref{fig:basis} for a detailed explanation of the 3D case.
    In the two-dimensional setup, $\partial/\partial x_3 \equiv 0$ and only $B_1, B_2$ are coupled by the divergence-free constraint.
    $B_1, B_2$ are therefore expanded on a 2D divergence-free basis,
    whereas $B_3$ is represented by an additional Legendre scalar component.
  }
  \label{fig:basis2d}
\end{figure}

In this case, we therefore follow \cite{li_locally_2005} and expand $(B_1, B_2)$ on a 2D divergence-free basis, and treat $B_3$ as an independent Legendre scalar component. The resulting basis function structure is shown in Fig.~\ref{fig:basis2d}.

In 2D, the number of Legendre basis functions for a single scalar field is
\begin{align}
  n_{2D}(d) = \frac{1}{2} (d+1)(d+2),
\end{align}
and the dimension of the divergence-free basis for a 2D vector field follows from \eqref{eq:dimVdiv}:
\begin{equation}
  N_{\text{div},2D} = 2 n_{2D}(d) - n_{2D}(d-1).
  \label{eq:ndiv2d}
\end{equation}
Since 2D MHD has one divergence-free vector field and six Legendre scalar fields
(five for hydrodynamics, and one for the out-of-plane component of the magnetic field),
the resulting total dimension of the DG basis for 2D MHD is therefore
\begin{equation}
  N_{2D} = 8 n_{2D}(d) - n_{2D}(d-1),
\end{equation}
which is analogous to the 3D case \eqref{eq:dimN3d}.

\subsection{Semi-discrete DG scheme}

Under the decomposition \eqref{eq:u-exp}, the time dependence of the solution is completely captured by the weights $\wgt^K_l(t)$ for each cell $K$.
To obtain the dynamics of the weights, we write the evolution equation in weak form, using our basis functions as test functions.

Within a given cell $K$, we can write the evolution equation \eqref{eq:cns-eq} transformed to coordinates of the reference cell $[-1,1]^3$:
\begin{equation}
  \frac{\partial \cns}{\partial t} + \frac{\Delta x^K}{2} \sum_{\alpha} \frac{\partial \flux_\alpha}{\partial \xi_\alpha} = \bm s,
  \label{eq:cns-eq-ref}
\end{equation}
where we have added a generic source term $\bm s$ for generality.
The ${\Delta x^K}/{2}$ factor arises from the mapping \eqref{eq:x-xi} from world coordinates $\xlab$ to cell coordinates $\xcell$.
For brevity, we will omit the superscripts indicating the local cell in the rest of the text, with the understanding that all quantities involved are always local to a given cell $K$ unless specified otherwise.
We then project this equation onto our space of DG functions by taking the inner product \eqref{eq:innerp} with the basis functions $\bfunc_l$:
\begin{equation}
  \label{eq:wdot-integrals}
  \frac{\partial}{\partial t} \innerp{\cns}{\bfunc_l} + \frac{\Delta x}{2} \sum_\alpha \innerp{\frac{\partial \flux_\alpha}{\partial \xi_\alpha}}{\bfunc_l} = \innerp{\bm s}{\bfunc_l}
\end{equation}
which holds for all $l \in \{1, \ldots, N\}$.

Using the expansion \eqref{eq:u-exp} of the solution and orthogonality \eqref{eq:basis-ortho} of the $\bfunc_l$, the first term of the sum reduces to $M_l \dot \wgt_l(t) = M_l \mathrm{d} \wgt_l / \mathrm{d} t$.
The second term involves a volume integral over the reference cell which can be integrated by parts, yielding
\begin{align}
  \label{eq:wdot}
  M_l \dot \wgt_l(t) &+ \frac{\Delta x^K}{2} \left[
  \int_{\partial ([-1,1]^3)} \sum_{\alpha=1}^3 (\dotp{\flux_\alpha}{\bfunc_l}) {\bm n}_\alpha \,\diffS \right. \\
  &- \left. \int_{[-1,1]^3} \sum_{\alpha=1}^3 \left(\dotp{\flux_\alpha}{\frac{\partial \bfunc_l}{\partial \xi_\alpha}}\right) \,\diffxcell \right]
   = \int_{[-1,1]^3}  \left(\dotp{\bm s}{\bfunc_l}\right) \,\diffxcell, \nonumber
\end{align}
where $\bm n_\alpha$ is the outwards pointing normal 3-vector to the face, and the dot product $\dotp{}{}$ is the dot product of \eqref{eq:innerp}, i.e. the dot product of $\mathbb{R}^8$ across conserved components. Note that for the Legendre $\bfunc_l$, this dot product effectively only selects one component at a time from $\flux_\alpha$ and $\bm s$ due to the structure of the $\bfunc_l$ represented in Fig.~\ref{fig:basis}.

Note that in \eqref{eq:wdot}, the fluxes $\flux_\alpha$ appearing in the volume integral are evaluated using the analytical fluxes \eqref{eq:mhd-flux}.
The $\flux_\alpha$ at the faces, however, are computed using a numerical flux function which solves a 1D interface Riemann problem at the face, following the traditional finite volume technique.
We discuss our choice of numerical flux in Section~\ref{sec:riemann-solver}.

This equation directly yields the time evolution $\dot \wgt_l(t)$ of the weights for cell $K$, provided that we can compute the face and volume integrals.
We use Gauss-Legendre quadrature with $d+1$ nodes per dimension to perform these integrals, as described in detail in \cite{schaal_astrophysical_2015}.
The resulting quadrature rule is exact for polynomials of degree $\leq 2d+1$,
which allows to integrate products of two basis functions without any error,
and would evaluate \eqref{eq:wdot-integrals} exactly if the fluxes and source terms were linear in the conserved state.

\subsection{Time integration}
\label{sec:time-integ-sig-vel}

We integrate the weights in time using \eqref{eq:wdot} with the traditional RKDG approach,
using strong-stability preserving (SSP) explicit Runge-Kutta schemes of selectable order between 1 and 4 \citep[see e.g.][]{gottlieb_high_2005}.
As described in \cite{schaal_astrophysical_2015}, we use a \emph{global} timestep $\Delta t$ respecting the constraint of \cite{cockburn_tvb_1989-1}:
\begin{equation}
  \Delta t = \min_K \frac{C}{2d+1} \left( \sum_\alpha \frac{|v_\alpha^K| + c^K}{\Delta x^K} \right)^{-1},
  \label{eq:cfl}
\end{equation}
where $c^K$ is the maximum signal velocity in cell $K$ (that is, $\sum_\alpha (|v_\alpha| + c)$ is an upper bound for the largest eigenvalue of the flux Jacobian).
In absence of magnetic fields, $c$ is the sound speed $c_S = \sqrt{\gamma \press/\rho}$.
In the MHD case, we take this upper bound to be $c = \sqrt{c_S^2 + c_A^2}$ where
$c_A=|\Bfield|/\sqrt{\rho}$ is the local Alfv\'en speed.
With this choice, $c$ is always greater or equal to the fast magnetosonic speed in the cell.

$C < 1$ is the chosen Courant number, which we typically set to $C=0.8$.
Note that the presence of source terms may require further reduction of the timestep;
in particular we discuss the case of Powell source terms for divergence control in Section~\ref{sec:powell-dt}.
Some RK schemes (typically at order 4 or more) may also require reducing the timestep \citep{gottlieb_high_2005}.

For most of the test problems presented in Section~\ref{sec:results}
and in particular the convergence tests presented in \ref{sec:conv-order},
we typically set the time integration order to match the spatial order of the scheme,
to prevent time integration errors from dominating the total error of the solution.
However, very high-order time integration schemes may not be required in practice
for many science applications
\citep[see e.g.][]{velascoromero_planetdiscinteractions_2018};
we elaborate on this point in Section~\ref{sec:discuss-hi-order-time-integ} in the discussion.

\section{Numerical ingredients}
\label{sec:num-ingredients}

In this section, we detail the numerical components required to achieve stability and accuracy with the scheme.

\subsection{Divergence control}
\label{sec:divB}

As noted in Section~\ref{sec:gov-eqs}, the Maxwell equations impose $\divB = 0$ on all physical realizations of the magnetic field, everywhere and at all times.
While the induction equation guarantees that an initially divergence-free magnetic field will remain so in time,
truncation errors in discretized schemes can cause non-zero numerical divergence to appear.
These errors can trigger a non-linear instability in the MHD equations and lead to blow-up of the numerical solution (see \citealt{brackbill_effect_1980,toth__2000}, and also \citealt{kemm_origin_2013} for a detailed mathematical discussion of this instability).
In addition, even if the numerical divergence stays bounded, divergence errors can still result in non-physical perturbations to the flow, such as plasma acceleration along the magnetic field lines.

The issue of divergence control across cells is not specific to DG, and has received a lot of attention in the context of finite difference and finite volume MHD codes, resulting in the development of multiple techniques.
% Projection
Projection methods, introduced by \cite{brackbill_effect_1980} and used in some recent schemes \citep[e.g.][]{derigs_novel_2016} project the magnetic field onto a globally divergence-free representation at every time step.
The main drawback of this technique is that it requires solving a global elliptic Poisson problem at each projection operation, which is expensive and less scalable than purely hyperbolic formulations
as it requires global exchange of information.
% Constrained transport
Another family of methods, constrained transport, keeps the magnetic field divergence-free to machine precision for some careful choice of discretization and update scheme for the induction equation 
\citep[see e.g.][]{evans_simulation_1988,dai_divergence-free_1998,ryu_divergence-free_1998,balsara_staggered_1999,gardiner_unsplit_2005}.
Constrained transport methods have also been extended to non-staggered grids \citep{rossmanith_unstaggeredhighresolution_2006,helzel_unstaggeredconstrainedtransport_2011}
and adaptive meshes \citep{teyssier_kinematicdynamosusing_2006,fromang_high_2006}.
Constrained transport has been very popular with finite difference and finite volume grid codes in astrophysics
\citep[e.g.][]{stone_zeus-2d:_1992,fromang_high_2006,stone_athena:_2008,collins_cosmological_2010,mocz_moving_2016},
because of its suitability for second-order mesh methods,
exact divergence control, and lack of any tunable parameter in the scheme.
Constrained transport schemes have also been extended to higher-order reconstruction methods \citep[see e.g.][]{balsara_divergence-free_2009}, and also to DG, involving either dual discretizations \citep{li_central_2011,li_arbitrary_2012,xu_new_2016,balsara_von_2017,zhao_rungekutta_2017}
or updating a vector potential with its own higher-order DG discretization \citep[e.g.][]{rossmanith_high-order_2013}.
These methods ensure that the magnetic field is exactly globally divergence-free at all times.
The main drawback of constrained transport in a DG setting is its implementation complexity and cost,
requiring significantly more operations and storage to update the magnetic field in a divergence-free way.

For this work, we adopted two widespread divergence control techniques which allow working with cell-centred discretizations
while preserving the hyperbolic character of the equations:
the Powell scheme, based on the addition of a nonconservative source term to the MHD equations,
and hyperbolic divergence cleaning, which dynamically advects and dampens the numerical divergence using an additional scalar field.
In the rest of this section, we describe both methods and detail our implementations.

\subsubsection{Powell source terms}

The so-called Powell scheme, after \cite{powell_solution-adaptive_1999},
follows the insights of \cite{godunov_symmetric_1972},
who pointed out that the conserved system \eqref{eq:cns-eq} of ideal MHD does not formally conserve entropy and is also not Galilean invariant, unless a specific source term proportional to $\divB$ is added.
While identically zero on continuous physical solutions, this source term modifies the nature of the equations.
The extra term may be obtained from deriving the local form of the conserved MHD equations based on integral conservation laws \citep{powell_solution-adaptive_1999},
or from requiring entropy stability \citep{godunov_symmetric_1972,chandrashekar_entropy_2016,winters_affordable_2016,liu_entropystablehigh_2018}.
Defining the column vector
\begin{equation}
  \powellq(\cns) \coloneqq (0, \Bfield, \dotp{\vel}{\Bfield}, \vel)^T,
  \label{eq:powell-q}
\end{equation}
the method introduces an additional source term at the right hand side of the conserved equation \eqref{eq:cns-eq}:
\begin{equation}
  {\bm s}_\text{Powell} = - (\divB) \powellq(\cns).
  \label{eq:powell-src}
\end{equation}
\cite{powell_solution-adaptive_1999} derived the characteristics of the ideal MHD system with this source term,
and showed that the addition of \eqref{eq:powell-src} results in an additional wave to the usual 7 waves,
whose effect is to advect away $\divB/\rho$ with the flow and restore Galilean invariance.

The main advantages of the Powell method are that it can be easily adapted to existing grid schemes,
and does not require setting or tuning any free parameter.
This scheme has been implemented in astrophysical MHD codes, both with adaptive mesh refinement \citep{mignone_plutocodeadaptive_2012} or moving mesh \citep{pakmor_simulations_2013,mocz_discontinuous_2014} grids.
In the DG context, it was also adopted by \cite{warburton_discontinuousgalerkinmethod_1999} for viscous MHD flows.
\cite{pakmor_simulations_2013} have found the Powell method to be more robust and stable than hyperbolic divergence cleaning when used with large dynamic ranges in time and space discretizations in the context of moving mesh simulations with local timestepping.

However, the Powell method comes with important limitations.
Firstly, it does not completely eliminate the divergence, as it advects it away with the flow.
It can therefore result in local accumulation of numerical divergence in the case of standing shocks,
which are among the most challenging problems for static mesh MHD divergence control
\citep{balsara_staggered_1999,toth__2000}.
% Becomes non-conservative
Secondly, after adding the source term \eqref{eq:powell-src}, the numerical scheme is not strictly conservative any more:
while the source term vanishes for exact physical solutions, 
it will locally inject conserved quantities whenever numerical divergence errors are present.
As noted by \cite{toth__2000}, this will result in wrong jump conditions across shock fronts.
In this paper, we therefore will be carefully evaluating these effects in our implementation,
including with a dedicated set of numerical tests in Section~\ref{sec:tests-div}.

In the rest of this section, we describe the details of our Powell implementation,
before covering the simpler hyperbolic cleaning scheme in Section~\ref{sec:dedner}.

\subsubsection{Choice of numerical flux function}
\label{sec:riemann-solver}

In this work, we consistently use the so-called HLLD fluxes of \cite{miyoshi_multistatehllapproximate_2005},
which have gained widespread adoption in astrophysics,
largely because of their robustness, low numerical diffusion, and relative computational inexpensiveness.
In particular, we use the same HLLD fluxes with both Powell and hyperbolic cleaning.

As noted by \cite{powell_solution-adaptive_1999}, the addition of the source term \eqref{eq:powell-src} will modify the characteristics of the ideal MHD system by adding an 8-th so-called divergence wave.
Formally, this would call for modifying the numerical flux function used at cell faces,
since usual 1D Riemann solvers will not propagate jumps in the normal component of the magnetic field.
Special Riemann solvers have been developed for 8-wave schemes
\citep[e.g.][]{powell_approximate_1994,fuchs_approximateriemannsolvers_2011,chandrashekar_entropy_2016,winters_affordable_2016}.
%%%
We have experimented with a number of such numerical fluxes, including the 8-wave entropy-stable flux of \cite{chandrashekar_entropy_2016} and local Lax-Friedrichs 8-wave fluxes.
Note that while our solver of choice, HLLD, is not an 8-wave Riemann solver,
fluxes that do not incorporate the divergence wave have been used successfully with Powell schemes
\citep{warburton_discontinuousgalerkinmethod_1999,mocz_discontinuous_2014,pakmor_magneticfieldformation_2017,pakmor_simulations_2013},
in which case the method reduces to adding a properly discretized source term.

Note however that, despite our HLLD flux not propagating the \nth{8} divergence wave,
we use the full 8-wave formalism when computing local characteristic eigensystems.
We discuss characteristic decomposition in more detail in the description of slope limiting in Section~\ref{sec:limiter-char}.

\subsubsection{LDF bases and discretization of the Powell term}
\label{sec:powell-discretization}

Our Powell scheme requires expanding the magnetic field on locally divergence-free bases for stability.
LDF basis functions have been found by several authors to significantly improve the stability of DG Maxwell and MHD schemes \citep{cockburn_locally_2004,li_locally_2005,yakovlev_locally_2013,zhao_positivity-preserving_2014,karami_halashi_reconstructed_2016},
even though they are not sufficient by themselves for divergence control
\citep{li_locally_2005,yakovlev_locally_2013}.
With this prescription, the magnetic field can be made exactly locally divergence-free inside the cells,
but not \emph{globally} as the normal component of the magnetic field is not guaranteed to be continuous across cell interfaces.
The role of the Powell terms in our scheme is therefore to stabilize the divergence contribution at the faces.
We now describe our choice of Powell term discretization with LDF basis functions,
which our tests have found to be robust and accurate even with a non-diffusive solver like HLLD which does not propagate the divergence wave itself.

As noted by \cite{waagan_positivemusclhancockscheme_2009,waagan_robust_2011,fuchs_approximateriemannsolvers_2011} in the context of finite volume methods,
the exact discretization of the Powell term is critical to the stability of the scheme,
and our experience with DG schemes can only support that statement.
The problem is to find a consistent discretization of the contribution of the Powell source term \eqref{eq:powell-src}
to the right hand side of the DG integral in \eqref{eq:wdot}:
\begin{equation}
  \int_K \dotp{{\bm s}_\text{Powell}}{\bfunc_l} \diffxcell =
  - \int_K (\divB) \left( \dotp{\powellq}{\bfunc_l} \right) \diffxcell,
  \label{eq:powell-src-int}
\end{equation}
where $\powellq$ is defined in \eqref{eq:powell-q}.
Although this term features a spatial derivative, it cannot be cast into conservative form as a flux,
and integration by parts is not helpful.
In addition, while $\divB$ vanishes identically inside $K$ because we use a LDF basis,
the discontinuity of the normal components of $\Bfield$ will contribute surface terms to the integral,
which mandates a careful discretization.

\begin{figure}
  \centering
  \tikzsetnextfilename{tikz-cell-interface}
  \begin{tikzpicture}[xscale=4, yscale=2]
    \tikzstyle{Lstate}=[red]
    \tikzstyle{Rstate}=[blue]
    \tikzstyle{Fstate}=[black]

    \newcommand{\yL}{-0.5};
    \newcommand{\yR}{ 0.5};
    \newcommand{\yF}{ 0.1};
    \newcommand{\xepsL}{0.08};
    \newcommand{\xepsR}{0.0};

    \begin{scope}
      \clip (-1, -0.9) rectangle (1, 0.8);
      \fill[gray!50!white] (-\xepsL,-1) rectangle (0, 1); % regularization region
      \draw[very thick] ( 0,-1) -- (0, 1); % face
    \end{scope}
    \draw[<->] (-\xepsL, -0.7) -- node[midway,above] {$\epsilon$} ++(\xepsL, 0);
    \draw[-latex] (0,-0.8) -- +(0.2,0) node[right] {$\facenormal$}; % face normal

    \draw[-latex,yshift=-0.1] (-1, 0) -- (0.5, 0) node[below] {$x$}; % x axis

    \node[below] at (0,-0.9) {Face $F$};
    \node[below] at (-0.5,-0.9) {Cell $L$};
    \node[below] at ( 0.5,-0.9) {Cell $R$};

    \begin{scope}
      \clip (-1, -1) rectangle (-\xepsL,1);
      \draw[thick,Lstate,shorten >=1pt] (-0.9,-0.8) .. controls (-0.6,-0.0) and (-0.3,-0.6) .. (-\xepsL,\yL)
        node[circle,draw,inner sep=1pt] (uL) {}; % left state
    \end{scope}

    \begin{scope}
      \clip (\xepsR, -1) rectangle (0.8, 1);
      \draw[thick,Rstate,shorten >=1pt] (1.0,0.5) .. controls (0.6,-0.1) and (0.3,0.6) .. (\xepsR,\yR)
        node[circle,draw,inner sep=1pt] (uR) {}; % right state
    \end{scope}

    \node[circle,fill,draw,inner sep=1pt] at (0.0,\yF) (uF) {}; % face state

    \path (uL) node[below left] {$\cns_L$};
    \path (uR) node[above right] {$\cns_R$};
    \path (uF) node[right] {$\cns_F$};

    \draw[xshift=-2pt,Bar->,shorten >=0, shorten <=0] (-\xepsL,\yL) -- node[midway,left] {$\Delta \cns_L$} (-\xepsL,\yF);
    \draw[densely dotted,Lstate,thick] (uL) -- (uF);

  \end{tikzpicture}
  \caption{
    \emph{Discretization of the Powell term at a cell interface},
    from the point of view of the $L$ cell ($\facenormal$ pointing outwards).
    $\cns_L$ and $\cns_R$ are the state vectors immediately left and right of the face $F$, obtained from the respective smooth DG state representations.
    $\cns_F$ is the state exactly at the face, which can be defined as the interface state returned by the Riemann solver.
    The total jump $\Delta \cns = \cns_R - \cns_L$ at the face is split between the left and right cells into $\Delta \cns_L$ and $\Delta \cns_R$ using the face state.
  }
  \label{fig:powell}
\end{figure}
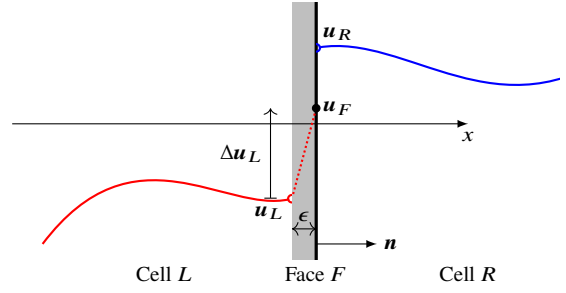

The face discontinuity suggests to split up the volume integral \eqref{eq:powell-src-int} into contributions in the interior
(which will vanish by construction since $\divB=0$ for divergence-free bases),
and contributions close to the faces.
The overall situation is sketched in 1D in Fig.~\ref{fig:powell},
from the point of view of the left cell $L$, close to a face $F$ shared with its neighbour cell $R$.
The grey area represents the volume of cell $L$ ``close'' to the face, for some thin layer thickness $\epsilon \rightarrow 0$.
The conserved state $\cns_L$ is obtained from the smooth DG state representation in cell $L$.
Given some choice of state $\cns_F$ at the interface (which we will discuss below),
the state in the left cell will jump by $\cns_F - \cns_L$ across the layer.
This suggests that, in the notations of Fig.~\ref{fig:powell},
the integral~\eqref{eq:powell-src-int} may be computed as
\begin{equation}
  \int_K \dotp{{\bm s}_\text{Powell}}{\bfunc_l} =
  - \int_{\partial K}
    \left( \dotp{ \powellq^\dagger }{\bfunc_l} \right)
    \dotp{\left(\Bfield_F - \Bfield_L\right)}{\facenormal} \diffS
  \label{eq:powell-src-face}
\end{equation}
for some choice of $\powellq^\dagger$ which determines how we average the non-linear product $(\divB) \powellq$.
If we choose some explicit form of conserved state $\cns$ which interpolates between $\cns_L$ and $\cns_F$ in the layer
(represented for example by the dotted line in Fig.~\ref{fig:powell}),
we may in principle obtain \eqref{eq:powell-src-face} and $\powellq^\dagger$ together directly by integrating \eqref{eq:powell-src-int} on the layer along the $x$ direction,
and taking the limit $\epsilon \rightarrow 0$.
Note however that we may also choose $\powellq^\dagger$ arbitrarily, independently from such considerations.

\paragraph{Interface state $\cns_F$}
For HLL-type Riemann solvers, a natural choice for $\cns_F$ is to take the state from the Riemann fan which contains the interface.
In particular, this choice guarantees that $\cns_F$ is properly upwinded,
which \cite{waagan_positivemusclhancockscheme_2009} argues to be critical to the stability of Powell schemes.
In this work, we use the interface state from the HLLD solver.

\paragraph{Normal component $\Bnormal_F$ at the interface}
\label{sec:normal-bx-choice}
Let $\Bnormal$ denote the component of the magnetic field normal to the face.
One-dimensional Riemann solvers such as HLLD usually assume that $\Bnormal$ is constant across the interface
to satisfy the 1D divergence-free condition.
These solvers therefore cannot prescribe any value $\Bnormal_F$ for $\Bnormal$ at the face whenever $\Bnormal$ is discontinuous.
Consequently, before calling the Riemann solver to obtain the flux and face state,
we decide on a face normal magnetic component $\Bnormal_F$,
and then assign it to the left and right states by setting $\Bnormal_L = \Bnormal_R = \Bnormal_F$.
A straightforward choice is to pick $\Bnormal_F \coloneqq \frac{1}{2} (\Bnormal_L + \Bnormal_R)$
\citep[e.g.][]{dedner_hyperbolic_2002,pakmor_simulations_2013}.
This choice results in a surface term proportional to the jump $\frac{1}{2}(\Bnormal_R - \Bnormal_L)$
which is also found in a number of published Powell term implementations
\citep{waagan_positivemusclhancockscheme_2009,chandrashekar_entropy_2016,liu_entropystablehigh_2018}.
In this work, we found that we obtained slightly smaller divergence errors near shocks by instead averaging the Alfv\'en velocities of the left and right states, and setting
\begin{equation}
  \label{eq:bxface-average-alfven}
  B^n_F \coloneqq \frac{B^n_L\sqrt{\rho_L}^{-1} + B^n_R\sqrt{\rho_R}^{-1}}{\sqrt{\rho_L}^{-1} + \sqrt{\rho_R}^{-1}},
\end{equation}
which we use in the rest of this work.

\paragraph{Choice of $\powellq^\dagger$}
Many prescriptions may be used for $\powellq^\dagger$,
most of which we tested were found to be unstable in our DG scheme, including
$\powellq^\dagger \coloneqq \powellq(\cns_L)$ or
$\powellq^\dagger \coloneqq \powellq(\cns_F)$,
despite the latter benefiting from upwinding from the Riemann solver.
We describe two choices that we have found to be stable across all test problems.

First, as previously noted, we can construct a $\powellq^\dagger$ by regularizing the integral \eqref{eq:powell-src-int}
near the face and taking the limit $\epsilon \rightarrow 0$ for some choice of regularization (see Fig.~\ref{fig:powell}).
For a simple linear interpolation in primitive variables between the states $\cns_L$ and $\cns_F$, this yields
\begin{equation}
  \powellq \dagger = \begin{bmatrix}
    0 \\
  \frac{1}{2}(\Bfield_L + \Bfield_F) \\
  \frac{1}{6}(
    2 \dotp{\Bfield_L}{\vel_L} +
      \dotp{\Bfield_L}{\vel_F} +
      \dotp{\Bfield_F}{\vel_L} +
    2 \dotp{\Bfield_F}{\vel_F} ) \\
  \frac{1}{2}(\vel_L + \vel_F)
  \end{bmatrix}.
  \label{eq:powellq-interp}
\end{equation}
This choice seems to be stable in our tests, and provides overall good results.

Second, we also found that we obtain a stable scheme by simply evaluating $\powellq^\dagger$ at the average conserved state in the cell
\begin{equation}
  \powellq^\dagger \coloneqq \powellq(\left< \cns \right>_L), \quad
  \left< \cns \right>_L \coloneqq \frac{1}{8} \int_{[-1,1]^3} \cns_L(\xcell) \mathrm{d}\xcell.
  \label{eq:powellq-average}
\end{equation}
This prescription fully retains all the high order information about $\Bnormal$ at the face,
which we find to be essential to preserve both the high-order property and stability of the scheme.
However, it discards any local variation of the state within the cell to evaluate the $\powellq$ part of the Powell term.
For some test problems, the choice \eqref{eq:powellq-average} yields slightly better results than \eqref{eq:powellq-interp}.
Because the cell average may be obtained from the lowest order DG weights without any additional computation,
this choice is also computationally inexpensive.
We therefore resort to using \eqref{eq:powellq-average} by default.

For divergence control, the Orszag--Tang vortex problem described in Section~\ref{sec:test-orszag-tang} proved to be a particularly discriminatory test,
as divergence issues will promptly cause distortions or ripples in smooth post-shock regions of the flow,
or result in exponential divergence blow-up.
The problem is exacerbated by the use of non-diffusive Riemann solvers like HLLD.

For completeness, we remark that \cite{janhunen_positive_2000} proposed a combination of an HLL-inspired solver, together with a source term similar to \eqref{eq:powell-src}
but with non-zero components in the induction equation only,
which amounts to taking $\Bfield=0$ in \eqref{eq:powell-q}.
As a result, Janhunen's source term preserves conservation of momentum and energy.
In practice, while we find Janhunen's term to work well with HLLD fluxes for some test problems,
it seems to provide insufficient control of the magnetic field divergence in strong shock situations,
which was also noted by other authors \citep[e.g][]{gaburov_astrophysical_2011}.

\subsubsection{Time step control for Powell term}
\label{sec:powell-dt}

In presence of the Powell source terms, we add an additional constraint to the timestep $\Delta t$ to ensure that the time integration can stably resolve any local change in conserved quantities,
in extreme cases where the Powell injection term would possibly become stiff.

Since the Powell term \eqref{eq:powell-src} contributes to the momentum, energy and magnetic components of the state vector, any of these may in principle be used to limit the timestep.
In practice, we found that using the total energy $E$ provides a robust criterion to control relative changes, since the total energy is always a positive quantity, and the energy contribution is sensitive to all of $\vel$, $\Bfield$ and $\divB$.
Therefore, we limit $\Delta t$ so that the change $\Delta E$ in total energy $E$ during a timestep $\Delta t$ due to the Powell source term \eqref{eq:powell-src} respects:
\begin{equation}
  \frac{\Delta E}{E} \approx \frac{1}{E} \left| \divB \right| \left| \vel \cdot \Bfield \right| \Delta t < \eta
\label{eq:powell_dt}
\end{equation}
for some threshold $\eta$ which we typically take to be $\eta=0.2$.

\subsubsection{Hyperbolic divergence cleaning}
\label{sec:dedner}

As a separate option for divergence control,
we have also implemented the so-called hyperbolic divergence cleaning technique,
proposed by \cite{dedner_hyperbolic_2002},
which introduces an additional scalar dynamical field which couples to $\divB$
and results in hyperbolic advection and parabolic damping of the divergence.
For some suitable choice of the advection velocity and damping time-scale parameters,
divergence cleaning will efficiently attenuate and advect away divergence errors.
Since it is straightforward to implement in existing schemes,
it has been adopted by a number of MHD codes
\citep[e.g.][]{gaburov_astrophysical_2011,mignone_plutocodeadaptive_2012,tricco_constrainedhyperbolicdivergence_2012,hopkins_accurate_2016}
including in a DG context
\citep[e.g.][]{boscheri_high_2014,zanotti_solving_2015,dumbser_simple_2016,kidder_spectre:_2017}.

Our implementation closely follows the so-called generalized Lagrange multiplier (GLM) formulation of \cite{dedner_hyperbolic_2002}.
An additional dynamical scalar field $\psi$ is added to the 8 conserved variables of MHD.
This field couples to the divergence of the magnetic field through a modified induction equation:
\begin{equation}
  \frac{\partial \Bfield}{\partial t} + \bm \nabla \cdot ( \Bfield \otimes \vel - \vel \otimes \Bfield ) + \bm \nabla \psi = 0.
  \label{eq:dedner-induction}
\end{equation}
The field $\psi$ is evolved according to an additional dynamical equation:
\begin{equation}
  \frac{\partial \psi}{\partial t} + \frac{c_h^2}{c_p^2} \psi + c_h^2 \bm \nabla \cdot \Bfield = 0,
  \label{eq:dedner-psi-evol}
\end{equation}
and the resulting coupled GLM system \eqref{eq:dedner-induction}--\eqref{eq:dedner-psi-evol} makes the fluctuations of $\psi$ propagate away from their sources at speed $c_h$
while damping them with a time-scale $\tau = c_p^2/c_h^2$.

In practice, we introduce the extra variable $\psi$ in the DG scheme as an additional scalar Legendre component.
The analytical internal fluxes $\flux$ of Eq.~\eqref{eq:mhd-flux} in the conserved formulation are modified:
equation \eqref{eq:dedner-induction} amounts to an extra diagonal term $\psi \idm_3$ in the fluxes of $\Bfield$.
Eq.~\eqref{eq:dedner-psi-evol} is broken down into two contributions:
the $\bm \nabla \cdot \Bfield$ term yields $c_h^2 \Bfield$ as the internal fluxes for $\psi$,
whereas $\psi/\tau$ is treated as a damping source term on the right hand side of \eqref{eq:cns-eq-ref}.
The numerical fluxes at the faces are also modified:
first, the interface values $\psi^\star$ and $\Bfield_x^\star$ are computed using upwinding of the linear characteristics
of the GLM system, following the 1D derivation of \cite{dedner_hyperbolic_2002}.
The approximate MHD fluxes are then estimated using the HLLD Riemann solver.
The magnetic field at the face may be set to $\Bfield_x^\star$ above,
or chosen to follow the prescription used for Powell terms in \ref{sec:normal-bx-choice};
in practice we found that this choice seems inconsequential and we use \eqref{eq:bxface-average-alfven}.
Finally, the face fluxes for $\Bfield_x$ and $\psi$ are corrected with the linear upwinded fluxes obtained from $\psi^\star$ and $\Bfield_x^\star$.

The parameters $c_p$ and $c_h$ also need to be set.
In our implementation, on grids with uniform cell size $\Delta x$, $c_h$ is chosen by adapting the procedure of \cite{dedner_hyperbolic_2002} to DG and setting
\begin{equation}
  c_h \coloneqq \frac{C'}{2d+1} \frac{\Delta x}{D \Delta t},
\end{equation}
where $D$ is the number of space dimensions,
and $C'$ is related to the CFL number $C$ chosen in \eqref{eq:cfl} and is such that $C < C' < 1$.
The resulting $c_h$ speed is faster than any other signal velocity appearing in the CFL constraint \eqref{eq:cfl},
but still compatible with this same stability condition for the explicit DG scheme.

For non-uniform grids however (i.e. with mesh refinement),
we observed that a time-varying $c_h$ will act as a local source of divergence,
as was demonstrated and studied by \cite{tricco_constrained_2016} in the context of smoothed particle MHD.
In this case, we resort to using a global and time-independent value of $c_h$ manually adapted to each problem.
We typically pick $c_h$ so that it remains greater than
about twice the maximum signal velocity of Section~\ref{sec:time-integ-sig-vel},
across all cells and all timesteps in the simulation,
and this condition is checked at runtime.

The parameter $c_p$ is set following \citet{dedner_hyperbolic_2002} by fixing $c_r \coloneqq c_p^2/c_h = 0.18$.
Note that in some cases, the resulting damping time-scale $c_p^2/c_h^2$ in \eqref{eq:dedner-psi-evol} may be too small to be resolved by the timestep $\Delta t$,
in which case $c_p$ is set so that $c_p^2/c_h^2 \gtrsim 3 \Delta t$.
Unfortunately, all three parameters $c_h, c_p, c_r$ above have dimensions,
which makes this formulation only suitable for well-understood test problems.
This can be reduced to one dimensional parameter $c_h$ and one dimensionless parameter $\alpha$ following \cite{mignone_secondorderunsplitgodunov_2010},
however these authors suggest that $\alpha$ is still slightly resolution-dependent.

To conclude on our divergence control implementations,
note that in our test runs,
we use exclusively either hyperbolic cleaning or Powell terms,
by enabling only the corresponding additional fluxes and source terms of each selected method;
in particular, our hyperbolic cleaning implementation does not use the ``extended GLM'' formulation of \citet{dedner_hyperbolic_2002}.

Finally, while hyperbolic divergence cleaning is implemented in a basis-independent way,
we typically only use it together with the componentwise Legendre basis for the magnetic field:
locally divergence-free bases are not required for hyperbolic cleaning,
are slightly more computationally expensive than pure Legendre bases,
and may in some cases result in slightly degraded convergence orders as discussed in Section~\ref{sec:conv-order}.
As such, our hyperbolic cleaning method produces magnetic fields which are formally
neither locally nor globally exactly divergence-free;
the role of the $\psi$ field is to provide a dynamical mechanism
which evolves the solution towards a divergence-free configuration.

\subsection{Slope limiting}
\label{sec:limiter}

A major challenge for high-order codes is to limit the appearance of ``ringing'' artefacts around discontinuous solution features.
This problem stems from Godunov's theorem: any linear scheme of order 2 or above will introduce spurious extrema in the numerical solution.
When applied to linear problems (i.e. for which the fluxes $\flux_\alpha$ and source terms are linear)
the RKDG scheme described so far is itself linear:
the solution weights $\wgt_l(t+\Delta t)$ are a linear combination of the weights $\wgt_l(t)$.
This problem is exacerbated for non-linear equations which can form shocks from smooth initial conditions, such as the MHD equations.

A typical workaround to this problem is to introduce in the scheme some non-linear procedure, whose role is to detect and attempt to control oscillations by locally modifying the solution.
This so-called \emph{limiting} process is recognized as a major challenge for high-order schemes \citep{jianxian_qiu_runge-kutta_2004,balsara_higherorderaccuratespacetime_2017}.
Many limiters have been proposed in the literature, first in the context of finite volume methods, but also specifically for discontinuous Galerkin schemes.

Conceptually, limiting proceeds in two successive steps \citep{jianxian_qiu_runge-kutta_2004}:
firstly, a detection procedure is used to identify ``troubled cells'' which are potentially subject to oscillations.
Secondly, cells marked by this first procedure then have their weights $\wgt_l$ modified by the limiter.
A high detection sensitivity (low false-negative rate) achieves scheme stability and oscillation reduction,
whereas a high specificity (low false-positive rate) ensures that the solution is not needlessly limited,
which could result in effective order reduction and useless application of the potentially costly limiter procedure.

\subsubsection{TVB slope limiter}
\label{sec:limiter-tvb}
In our current implementation,
we use the widespread so-called total variation bounded (TVB) minmod slope limiter \citep{shu_tvbuniformlyhighorder_1987,cockburn_tvb_1989}.
This limiter will detect troubled cells by comparing the slopes (linear components)
of the solution in the cell with finite difference with averages in neighbouring cell.
When triggered, the limiting procedure downgrades the scheme to second and sometimes first order locally in the cell,
altering the slopes so that the limited solution locally respects some upper bound on total variation.
In this work, this limiter serves as a starting point
to deal with the presence of shocks in test problems.
Note that more sophisticated DG limiters which preserve higher-order information at shocks have been developed,
and we discuss possible extensions and future improvements in Section~\ref{sec:discuss-limiters}.

We first illustrate the TVB slope limiter for some scalar quantity $u$ in 1D.
In a 1D reference cell $[-1, 1]$ with coordinate $\xi$, $u$ is expanded on a Legendre basis following \eqref{eq:u-exp}.
We may define the cell average $\avg{u}$ and cell slope $\slope{u}{\xi}$ of $u$ by projecting $u$ onto the space of piecewise linear functions of the form
\begin{equation}
  \xi \mapsto \avg{u} + \slope{u}{\xi} \cdot \xi.
\end{equation}
Note that since $\xi \mapsto 1$ and $\xi \mapsto \xi$ are Legendre polynomials,
the projection simply identifies $\avg{u}$ and $\slope{u}{\xi}$ with the first two weights in the basis expansion of $u$
(up to scale factors depending only on the chosen normalization for Legendre polynomials).

The idea of the TVB detection procedure is to compare the slope $\slope{u}{\xi}^i$ in cell $i$ to finite difference approximations of the slope based on averages $\avg{u}$ in the neighbouring left and right cells $i-1$ and $i+1$:
\begin{align}
  \Delta_L u \coloneqq (\avg{u}^{i+1} - \avg{u}^i) / \Delta \xi, \\
  \Delta_R u \coloneqq (\avg{u}^{i} - \avg{u}^{i-1}) / \Delta \xi,
\end{align}
where $\Delta \xi = 2$ is the size of the reference cell $[-1,1]$.
In the case of a linear solution $u = Ax + B$ in physical coordinates $x$, we have $\slope{u}{\xi} = \Delta_L u = \Delta_R u = \frac{\Delta x}{2} A$
using \eqref{eq:x-xi}.
The TVB limiter first checks the absolute slope $|\slope{u}{\xi}|$ to avoid limiting close to local extrema,
where oscillations are unlikely to appear and higher order information needs to be retained.
If $u$ is smooth and admits an upper bound on its second derivative
(i.e. $|\partial^2 u / \partial x^2| \leq 2M$ for some $M$)
then within one cell of a local extremum, $|\partial u / \partial x| \leq 2M \Delta x$.
Therefore, we avoid limiting cells for which $|\partial u / \partial x| \leq 2M \Delta x$,
or equivalently using \eqref{eq:x-xi}, $|\slope{u}{\xi}| \approx |\partial u / \partial \xi| \leq M \Delta x^2$.
For cells whose slope is above this threshold, TVB limiting applies a traditional total value diminishing limiter,
based for example on the minmod function:
\begin{align}
  \mathrm{minmod}(a, b, c) & \coloneqq \left\{ %
    \begin{tabular}{lr}
      $\min(a,b,c)$ & $a, b, c > 0$, \\
      $\max(a,b,c)$ & $a, b, c < 0$, \\
      0 & otherwise.
  \end{tabular} \right.
\end{align}
The new TVB limited slope $\slopeshort$ is computed as
\begin{align}
  \slopeshort \coloneqq \left\{ %
    \begin{tabular}{lr}
      $\slope{u}{\xi}$ & if $|\slope{u}{\xi}| \leq M \Delta x^2$, \\
      $\mathrm{minmod}(\slope{u}{\xi},\beta \Delta_L u,\beta \Delta_R u)$ & otherwise,
  \end{tabular} \right.
  \label{eq:tvb-limiter}
\end{align}
where $\beta \in [1,2]$ is a parameter that controls the aggressiveness
of the total variation diminishing (TVD) part of the limiter
\citep[see e.g.][]{leveque_finitevolumemethods_2002}.
The detection procedure computes $\slopeshort$ and compares it to the original solution slope $\slope{u}{\xi}$.
If $|\slopeshort - \slope{u}{\xi}| > \epsilon$ for some small threshold $\epsilon$, the limiter is marked as triggered.

For triggered cells, the limiting step assigns $\slopeshort$ as the local solution slope
by modifying the corresponding first-degree weight,
and sets all second-degree or higher weights of the solution to 0.
For limited cells, whenever all of $\Delta_L u$, $\slope{u}{\xi}$ and $\Delta_R u$ have the same sign,
the resulting limited weights have polynomial degree 1, and the scheme degrades to locally second-order accurate.
However, whenever $\Delta_L u$, $\slope{u}{\xi}$ or $\Delta_R u$ have conflicting signs (e.g. near extrema, or at oscillations around strong shocks),
the minmod function will assign 0 to $\slopeshort$;
the limited solution becomes constant in the cell, and the scheme locally becomes only first-order.
Note that the cell average $\avg{u}$ is never modified by the limiter;
this ensures that the limiting process stays conservative.
In all of this work, we take $\epsilon = 10^{-8}$,
and we use $\beta=2$, which corresponds to a monotonized central limiter\footnotemark.
\footnotetext{Note that using $\beta=1$ with the TVB limiter is not a suitable choice in practice.
  As resolution is increased,
  the $M\Delta x^2$ threshold will protect only $O(1)$ cells around local extrema from the minmod function,
  and as a result the latter will end up being invoked almost everywhere in the domain.
  If $\beta=1$, then $\mathrm{minmod}(a,b,c) \neq a$ almost everywhere,
  which will end up triggering the limiter on almost all cells of the domain,
  degrading the scheme to second order at best.
}

Equation \eqref{eq:tvb-limiter} prescribes a limiter whose minmod function gets applied almost everywhere,
except very close to extrema as defined by the parameter $M$.
For a fixed $\Delta x$, one can always find a suitable value of $M$,
but across resolutions, we find that we obtain more consistent results if we let $M$ scale with $\Delta x$,
in a prescription similar to \cite{schaal_astrophysical_2015}.
We therefore define $\tilde M \coloneqq M \Delta x$, and choose to keep $\tilde M$ a constant instead of $M$ in \eqref{eq:tvb-limiter}.
This has the effect of making the limiter weaker at higher resolutions,
which we found to be important for divergence control in the Powell scheme.
We come back to this non-intuitive aspect of the limiter in Section~\ref{sec:discuss-limiter-powell-interaction} in the discussion.

In 2D and 3D, the limiter is simply applied in each space direction independently,
acting only on one of the 2 or 3 directional slopes at a time.

\subsubsection{Characteristic limiting}
\label{sec:limiter-char}
To apply this scalar slope limiter to systems of equations,
we can in principle apply the limiter to each scalar conserved variable successively,
downgrading the local cell accuracy to second order or less whenever one of the components triggered the limiter.

Another option is to compute a local decomposition into characteristics,
applying the limiter on characteristic variables instead.
In this approach, the slopes $\Delta_L \cns, \slope{\cns}{\xi}, \Delta_R \cns$ are first transformed to characteristic slopes,
limited using the TVB procedure, and the resulting limited slope $\slopeshort$ is transformed back to conserved variables.
For a given spatial direction along which to perform the 1D limiting process,
the characteristic variables $\Delta \chrv$ are obtained from the conserved differences (or slopes) $\Delta \cns$ with
\begin{equation}
  \Delta \chrv = \Lchrm(\avg{\cns}) \bm \cdot \Delta \cns,
\end{equation}
where $\Lchrm(\avg{\cns})$ is the matrix of left eigenvectors of the conserved 1D MHD equations along the chosen direction,
linearized around the average state $\avg{\cns}$ in the central cell.
After limiting $\Delta \chrv$ as described in \ref{sec:limiter-tvb},
we obtain the limited conserved variables using the matrix $\Rchrm(\cns) = \Lchrm(\cns)^{-1}$ of right MHD eigenvectors:
\begin{equation}
  \Delta \cns = \Rchrm(\avg{\cns}) \bm \cdot \Delta \chrv.
\end{equation}
Suitable expressions for numerical evaluation of $\Lchrm$ and $\Rchrm$ may be found e.g. in \citet{powell_solution-adaptive_1999} or \citet{stone_athena:_2008}.
\citet{jiang_high-order_1999} conveniently summarize how to obtain eigenvectors for the divergence wave of the Powell scheme from conserved variables,
and \citet{dedner_hyperbolic_2002} cover the extension of the characteristic matrices for hyperbolic divergence cleaning.

Characteristic limiting is well motivated by the description of local variations as a superposition of local linearized physical waves,
and is generally recognized as yielding better results than conserved variable limiting \citep{jianxian_qiu_runge-kutta_2004,balsara_higherorderaccuratespacetime_2017}.
Our numerical experiments are consistent with these earlier results,
and we therefore systematically resort to limiting characteristic variables.
In particular, we find that conserved variable limiting can result in post-shock oscillations and noise,
readily visible for example on the Orszag--Tang vortex test problem,
whereas characteristic limiting results both in sharp shocks and noise-free solutions in smooth regions of the flow.
Note that it is also possible to apply the limiting process component-wise to the primitive variables,
which can be useful in particular to enforce positivity of the pressure.
In our case, we use a separate DG limiter for positivity as described in Section~\ref{sec:limiter-pos}.
We have not investigated primitive variable limiting in this work;
characteristic limiting is generally regarded as
better physically justified than component-wise conservative or primitive limiting,
while possessing superior entropy properties
\citep[see e.g.][]{cockburn_tvb_1989,balsara_higherorderaccuratespacetime_2017}.

For each independent limiting space direction $\alpha$,
we compute the 1D characteristics along direction $\alpha$ with the appropriate 1D matrices $\Lchrm$ and $\Rchrm$,
and apply the scalar limiting described in \ref{sec:limiter-tvb} to each characteristic variable independently.

The main drawback of characteristic limiting is the expensive construction and application of the matrices $\Lchrm$ and $\Rchrm$ for each cell;
in practice the cost can be amortized by processing multiple characteristic matrices within a same inner loop,
which allows the efficient use of vector CPU instructions.

\subsubsection{Choice of limiter threshold}
We conclude by noting that suitably choosing the parameters $M$ or $\tilde M$ is central to the good performance of this limiter.
Unfortunately, irrespective of the choice of limiter threshold scaling prescription, neither $M$ nor $\tilde M$ are dimensionless, and their optimal value depends on the initial conditions, spatial resolution, and choice of units.

For simple scalar problems with smooth initial conditions,
$M$ may be interpreted as an upper bound on second derivatives in the initial conditions \citep{cockburn_tvb_1989}.
Finding good a priori choices of $M$ for systems is more complicated \citep{jianxian_qiu_runge-kutta_2004}.
In the case of conserved variable limiting, each variable formally has different units, and it is unclear whether a single numerical value of $M$ may apply to all components meaningfully.
In the case of characteristic limiting, the normalization of characteristic variables depends on the arbitrary normalization of the eigenvectors,
making $M$ dependent on this choice as well.
$\tilde M$ is related to admissible gradients in the solution,
and is subject to the exact same shortcomings of units and normalization.

Despite these issues, we use this simple limiter based on $\tilde M$ as a starting point,
and discuss some possible promising alternatives in Section~\ref{sec:discuss-limiters},
including limiters which do not require setting a dimensional parameter.

\subsubsection{Slope limiting with LDF bases}
\label{sec:limiting-div0}
At the end of the limiting process, whenever the slope limiter was triggered and has modified the weights in the cell,
we are about to reset the slopes of the fields in the cell, and clear the higher-order moments in the weights.
When using LDF bases, some additional caution is required:
after the TVB slope limiter step operates in each space direction $\beta$ independently,
the limited slopes $\Bslope{\alpha}{\beta}$ of each component $\Bfield_\alpha$ of the magnetic field
may not be divergence-free, i.e. we may have
\begin{equation}
  \delta \coloneqq \sum_\alpha \Bslope{\alpha}{\alpha} \neq 0.
\end{equation}
If $\delta \neq 0$, then the resulting slopes are not representable on a LDF basis.
One could simply think of projecting back new slopes onto the divergence-free basis using $L^2$ projection;
however, this procedure has a number of flaws in the general case where $\delta \neq 0$.
First, this $L^2$ projection is not total variation diminishing, and may reintroduce local extrema in the magnetic field that were just taken away by the limiting process.
In particular, it may generate non-zero slopes of $\Bfield$ in directions along which the magnetic field was initially uniform, thereby breaking symmetries by coupling of spatial directions.
Finally, $L^2$ projection may create higher-order contributions (degree 3 and above) in the divergence-free weights,
which is not desirable in a limited cell;
this particular issue is discussed in more detail in Appendix~\ref{sec:apdx-div0-proj}.

To solve this problem, we use a simple non-linear procedure applicable to any number of dimensions.
We start from limited slopes $\Bslope{\alpha}{\beta}$ obtained from any slope limiting procedure;
in general those will have $\delta \neq 0$.
Note that the off-diagonal slopes $\Bslope{\alpha}{\beta}$ with $\alpha \neq \beta$ do not contribute to $\delta$, and therefore do not require any correction.
Let $\slopeshort_\alpha \coloneqq \Bslope{\alpha}{\alpha}$ be the limited diagonal slopes before divergence correction.
Our goal is to derive corrected slopes $\tilde \slopeshort_\alpha$ verifying $\tilde \delta = \sum_\alpha \tilde \slopeshort_\alpha = 0$.

We start by separating the divergence $\delta$ into positive and negative contributions $\delta^+$ and $\delta^-$:
\begin{equation}
  \delta^\pm \coloneqq \sum_\alpha \max(\pm \slopeshort_\alpha, 0),
\end{equation}
such that $\delta = \delta^+ - \delta^-$, with $\delta^\pm \geq 0$.

Suppose the slopes $\Bslope{\alpha}{\beta}$ have divergence $\delta > 0$; then $\delta^+$ is in excess compared to $\delta^-$ and we have $\delta^+ > \delta^- \geq 0$.
In this case, we rescale the slopes which contribute to $\delta^+$ (i.e. diagonal slopes which are positive) by the appropriate factor to exactly cancel out the divergence.
We set:
\begin{equation}
  \text{For } \delta > 0, \quad \tilde \slopeshort_\alpha \coloneqq
  \left\{
    \begin{array}{ll}
      \slopeshort_\alpha & \text{if } \slopeshort_\alpha \leq 0, \\
      (\delta^-/\delta^+)\slopeshort_\alpha & \text{if } \slopeshort_\alpha > 0,
    \end{array}
  \right.
\end{equation}
so that we obtain $\tilde \delta^+ = \tilde \delta^-$.
Similarly, in case of negative divergence, we set:
\begin{equation}
  \text{For } \delta < 0, \quad \tilde \slopeshort_\alpha \coloneqq
  \left\{
    \begin{array}{ll}
      \slopeshort_\alpha & \text{if } \slopeshort_\alpha \geq 0, \\
      (\delta^+/\delta^-)\slopeshort_\alpha & \text{if } \slopeshort_\alpha < 0.
    \end{array}
  \right.
\end{equation}
It is easy to check that with this prescription, for any initial value of $\delta$, we obtain $\tilde \delta = 0$.
Because
$\delta > 0 \Rightarrow 0 \leq \delta^- / \delta^+ < 1$
and
$\delta < 0 \Rightarrow 0 \leq \delta^+ / \delta^- < 1$,
we always have $|\tilde \slopeshort_\alpha| \leq |\slopeshort_\alpha|$, and the procedure is total variation diminishing.

Finally, we assign the newly obtained divergence-free slopes $\tilde \slopeshort_\alpha$
to the $\Bslope{\alpha}{\alpha}$,
and we obtain the limited magnetic field weights by $L^2$ projection of the resulting second-order solution
$\Bfield_\alpha(\xcell) = \avg{\Bfield}_\alpha + \sum_\beta \Bslope{\alpha}{\beta} \xcell_\beta$ onto the LDF basis functions.
Since this second-order solution is now divergence-free,
this last projection is exact and does not suffer from any of the issues mentioned at the beginning of this section.

\subsection{Enforcing density and pressure positivity}
\label{sec:limiter-pos}

In the conserved variables formulation, the thermal pressure is derived from the total energy by subtracting the kinetic and magnetic energy terms.
Under strong shock conditions, this can result in unphysical negative thermal pressure $\press$ at quadrature points.
This is also true for the density which, although represented exactly in the conserved variables, can still suffer from high-order oscillations in rarefied regions.

Schemes have been developed to try to maintain positivity of pressure and density in the context of higher-order finite volume methods.
A possible solution revolves around rewriting the conserved system as a conservation law for some appropriately defined entropy variables.
This solution was adopted by \cite{ryu_cosmological_1993} in the context of the Euler equations for cosmological simulations, and later extended to MHD by \cite{balsara_maintaining_1999}.
Combining this idea with appropriate Riemann solvers for the numerical fluxes, one can then construct finite volume schemes with some positivity properties.
\cite{chandrashekar_entropy_2016} developed an entropy-stable finite volume scheme for MHD with a prescribed numerical flux and Powell term,
and other authors have further developed this approach, see e.g. \cite{winters_affordable_2016,derigs_novel_2016}.

In practice, positivity can however generally only be proven for a limited set of schemes, in the absence of general source terms and relying on selective numerical flux prescriptions.
In addition, as noted by \cite{balsara_maintaining_1999}, discretization or round-off errors can still contribute to causing negative states even with positivity-preserving strategies in place.

For this work, we follow the approach used for hydrodynamics in \cite{schaal_astrophysical_2015} and adopt the general framework of \cite{zhang_positivity-preserving_2010} for positivity limiting in a DG context.

Our positivity limiter proceeds in two steps to modify the cell weights $\wgt$.
In a first step, the cell averages (represented by \nth{0}-degree weights $\bar{\wgt}$) are checked for positive density and pressure.
If positivity is satisfied, then the cell weights $\wgt$ are left unchanged.
If the cell averages violate positivity, then $\wgt$ is modified by setting the average density and/or pressure to predefined floor values $\epsilon_\rho$ and $\epsilon_\press$.
This operation is not conservative: it modifies the total amount of mass and energy in the simulation
by injecting conserved quantities to satisfy positivity of the cell averages if needed.
It should therefore be viewed as a last resort to keep the simulation running.
As a diagnostic, we keep track throughout the whole simulation run
of the total amount of each conserved quantity (mass and total energy) injected, if any, by this procedure.
We find in practice that cell averages never need any positivity correction
when using the Powell scheme in any of the test problems presented in this paper,
but can in some cases require correction when using hyperbolic divergence cleaning
with strong shocks in very low plasma-$\beta$ situations.

Once positivity of the cell average state $\bar{\wgt}$ is guaranteed, we follow the general idea of \cite{zhang_positivity-preserving_2010} by computing a scaling factor $\tau \in [0, 1]$ such that the cell state defined by the weights
\begin{equation}
  \tilde{\wgt}(\tau) \coloneqq \tau \wgt + (1-\tau) \bar{\wgt}
  \label{eq:wtilde}
\end{equation}
has positive density and pressure at all \emph{cell} and \emph{face} quadrature points.
This prescription effectively scales the amplitude of the spatial variations of the cell state between $\tau=1$ (which yields $\tilde{\wgt}=\wgt$ and recovers the full unlimited state) and $\tau=0$ (corresponding to $\tilde{\wgt} = \bar{\wgt}$, i.e. a piecewise-constant solution in the cell, which is positive everywhere by construction).
Ideally, we would like to find the maximal (i.e. least impacting) value of $\tau$ such that we have both:
\begin{align}
  \rho(\tilde{\wgt}(\tau)) \geq \epsilon_\rho \label{eq:pos_rho} \\
  \press(\tilde{\wgt}(\tau)) \geq \epsilon_\press \label{eq:pos_press}
\end{align}
at each individual quadrature point of the cell.

Since the density is a linear function of the weights, solving \eqref{eq:pos_rho} for $\tau$ at a given quadrature point $q$ with local density $\rho_q$ simply yields:
\begin{align}
  \tau_q = \frac{\bar{\rho} - \epsilon_\rho}{\bar{\rho} - \rho_q} \label{eq:tau_rho}.
\end{align}
Taking $\tau$ to be the smallest $\tau_q$ over all quadrature points $q$ which require positivity limiting (i.e. where $\rho_q < \epsilon_\rho$) guarantees that \eqref{eq:pos_rho} is fulfilled.

The situation for the pressure positivity \eqref{eq:pos_press} is more complicated, because $\press$ is not a linear function of the conserved quantities $\cns$ and therefore of the weights $\wgt$.
In the case of hydrodynamics, solving \eqref{eq:pos_press} requires finding the roots of a quadratic polynomial at each non-positive quadrature point, which is the method adopted by \cite{zhang_positivity-preserving_2010,schaal_astrophysical_2015}.
For MHD however, this approach requires cubic root finding due to the magnetic energy term in the total energy, which is both costly and challenging to implement in a numerically robust way.
\cite{wang_robust_2012} noted that we can easily obtain a value of $\tau$ which satisfies \eqref{eq:pos_press} by exploiting the fact that $\press(\cns)$ is a concave function of the conserved state $\cns$, i.e.:
\begin{align}
  \forall \tau \in [0, 1], \quad \press(\tau \cns + (1-\tau) \bar{\cns}) \geq \tau \press(\cns) + (1-\tau) \press(\bar{\cns}),
\end{align}
which can be easily checked by noting that the eigenvalues of the Hessian of $\cns \mapsto \press(\cns)$ are all negative.
Therefore, choosing
\begin{align}
  \tau_q = \frac{\bar{\press} - \epsilon_\press}{\bar{\press} - \press_q} \label{eq:tau_press}
\end{align}
guarantees that \eqref{eq:pos_press} holds at quadrature point $q$.
Even though this $\tau$ is not necessarily maximal, this prescription retains the space-varying properties of the solution while ensuring positivity,
and we found it to be very efficient and much more stable numerically than iterative root finding.

We can now pick a single value of $\tau$ which satisfies simultaneously \eqref{eq:pos_rho} and \eqref{eq:pos_press} at all quadrature points, by taking the smallest of all the $\tau_q$ prescribed by both \eqref{eq:tau_rho} and \eqref{eq:tau_press}.
In the case of a cell which does not need any limiting because its conserved state is positive at all quadrature points, we set $\tau = 1$.
Finally, we use this single $\tau$ to update the weights in the cell by setting $\wgt \leftarrow \tilde{\wgt}(\tau)$ using \eqref{eq:wtilde} whenever we end up with $\tau < 1$.

Finally, note that our practical implementation differs from \cite{zhang_positivity-preserving_2010,schaal_astrophysical_2015} in two additional important details.
First, instead of introducing a separate set of Gauss-Legendre-Lobatto points for positivity limiting,
we enforce positivity over all face and volume Gauss-Legendre quadrature points%
---the exact same points used in the computation of the face and volume terms in the DG scheme.
This avoids issues related to round-off errors,
which may arise from enforcing positivity and evaluating the conserved state at different quad points,
to which MHD seems particularly prone.
In addition, \cite{zhang_positivity-preserving_2010} suggest reducing the Courant
number whenever the positivity limiting procedure is enabled.
In practice, we find that we do not need to modify the timestep criterion for our test problems,
as long as the limiting is applied consistently at each of the Runge-Kutta substeps
before computing any conserved state at quadrature points.

\subsection{Adaptive mesh refinement}

In order to capture a large dynamic range for astrophysical applications,
our code provides tree-based spatial adaptive mesh refinement,
in which each cell may be refined or de-refined independently based on local criteria.
A refinement operation consists in a local splitting of the parent cell into 8 cubic children cells (in 3D),
which results in an octree-structured grid.

The design of the AMR algorithm broadly follows the adaptive mesh implementation of RAMSES described in \citet{teyssier_cosmologicalhydrodynamicsadaptive_2002}.
The DG framework provides a clean setting for defining prolongation (refinement) and restriction (de-refinement) operations,
through the use of local $L^2$ projections onto basis functions.
Our MHD implementation is identical to the pure hydrodynamics implementation discussed in detail in \cite{schaal_astrophysical_2015}.
For LDF bases, the AMR prolongation and restriction operations
follow the same $L^2$ projection procedure as for the Legendre basis as described in \citet{schaal_astrophysical_2015},
but using the natural $L^2$ inner product between vector fields $[-1,1]^3 \rightarrow \mathbb{R}^3$:
\begin{align}
  \innerp{\bm u}{\bm v} \coloneqq \int_{[-1,1]^3} \dotp{\bm u}{\bm v} \, \diffxcell.
\end{align}
For refinement, the weights of a child cell are obtained by projecting the relevant octant
of the parent cell onto the child's LDF basis.
Since the solution in the parent cell is locally divergence-free everywhere in the cell,
the solution in each child cell will also be represented exactly on the LDF basis.
Conversely, the restriction operation projects the solution formed by all children cells
onto the LDF basis of the parent cell.
These operations are linear, and may therefore be represented as
fixed pre-computed matrices operating on the weights,
albeit with different prolongation and restriction matrices than for Legendre basis functions.

Due to the non-uniform nature of AMR grids,
some precautions are required for solution limiting, in particular to ensure positivity;
we refer the reader to \citet{schaal_astrophysical_2015} for more details,
as the introduction of magnetic fields does not alter this particular aspect of the method.

For cell-based AMR,
the decision to refine or derefine a cell is based on a local refinement criterion:
cells which feature rapid variations of the solution are split,
in order to locally introduce additional spatial resolution.
\citet{schaal_astrophysical_2015} used the linear slopes (degree one polynomials) from the DG weights
to decide whether to split a cell.
In some instances, we found that
polynomials modes of degree $d \geq 2$ can actually contribute more
than the linear modes to the total variation inside a cell.
For this reason, we choose to compute the refinement criterion based on all available high-order information.
To this end, we measure the amount of local variation of the solution in a cell
using a so-called \emph{smoothness indicator}, an important ingredient of WENO schemes
\citep[see e.g.][]{jiang_efficient_1996,shu_essentiallynonoscillatoryweighted_1998,balsara_monotonicitypreservingweighted_2000,zhong_simpleweightedessentially_2013}.
For a polynomial $u(\xlab)$ of degree $d$ defined for $\xlab$ in cell $K$,
we determine its smoothness $S_\alpha(u)$ along each direction $\alpha \in \{1, 2, 3\}$
using the 1D indicator of e.g. \citet{zhong_simpleweightedessentially_2013},
computed in the reference cell $[-1,1]^3$:
\begin{align}
  S_\alpha(u) \coloneqq \sum_{k=1}^{d} 2^{2k-1} \int_{[-1,1]^3}
    \left( \frac{\partial^k}{\partial \xi_\alpha^k} u(\xcell) \right)^2 \, \diffxcell.
\end{align}
$S_\alpha(u)$ may be computed as a function the weights of $u$ in the cell,
with different expressions for the Legendre and LDF basis functions
which are best derived using a symbolic computation package.

Refinement and derefinement is performed by specifying a set of fields $u$ to monitor for refinement,
together with a threshold smoothness $\tilde S$.
The refinement criterion is then evaluated on a cell-by-cell basis:
\begin{itemize}
  \item If a leaf (unsplit) cell has $S_\alpha(u) > 2 \tilde S$
    for any refinement field $u$ or direction $\alpha$, it is marked for refinement,
  \item If a split cell has $S_\alpha(u) < \frac{1}{2} \tilde S$
    for all refinement fields $u$ and directions $\alpha$, the cell is marked for derefinement.
\end{itemize}

We provide two specific tests of adaptive mesh refinement in our code,
for the Orszag--Tang vortex problem (Section~\ref{sec:test-orszag-tang}, Fig.~\ref{fig:test-orszag-tang-amr}),
and for the MHD rotor problem (Section~\ref{sec:test-rotor}, Fig~\ref{fig:test-rotor-amr}).
For these test problems, we set $\tilde S=0.03$
and refine on the density and magnetic field components.
In the relevant test problem sections,
we show maps of the solution, mesh and magnetic field divergence
for both the Powell and cleaning schemes,
and discuss the results in more detail,
including the reduction in the number of cells
and wall time offered by the adaptive grid.
Note that for our implementation, at a fixed smallest cell size,
the gains in memory (number of cells) and wall time will usually be very similar,
because we use global timesteps, and the global $\Delta t$ is driven by the finest cells due to the CFL \eqref{eq:cfl}.
With local timesteps, it becomes possible to advance the coarser AMR cells
with a larger $\Delta t$ than the finer levels, while still respecting the CFL condition everywhere.
We discuss options for future extensions to local timestepping in Section~\ref{sec:local-timestepping}.

\section{Results}
\label{sec:results}

\subsection{General test problem setup}

In the following section, we present test problems run using our DG schemes at various orders.
Specific care has been taken to ensure that the test problems are run with consistent and homogeneous settings,
without problem-specific fine tuning,

In all test problems, we use HLLD as the approximate Riemann solver.
Except for the convergence order tests,
positivity limiting is enabled with density and pressure floors $\epsilon_\rho = \epsilon_\press = 10^{-12}$,
and we limit the slopes using the characteristic slope limiter,
for which the modified threshold parameter $\tilde M$ is always set to $\tilde M=5$.
While this sometimes results in sub-optimal limiting,
we feel that it achieves a good compromise and presents an honest picture of the capabilities of the code
across a range of test problems without any limiter fine-tuning.

In the following, ``DG-$(d+1)$'' designates the DG method with degree $\leq d$ basis polynomials%%
---whose spatial convergence order is typically $d+1$.
Unless otherwise specified, the results are presented for degree $d=2$ polynomials,
i.e. the \nth{3} order scheme DG-3,
and using the Powell scheme for divergence control.
By default, we match the order of the Runge-Kutta time integration to the spatial scheme order $d+1$,
up to a maximal order of RK4.
Comparisons with hyperbolic cleaning are shown whenever they are informative or show relevant differences.
Details of the divergence control schemes are discussed in Section~\ref{sec:divB}.

Most of the test problems shown are computed on 2D or 3D Cartesian grids,
for which the resolution level $\ell$ corresponds to $2^\ell$ grid points per dimension.
We also illustrate the use of our scheme with adaptive mesh refinement in Fig.~\ref{fig:test-orszag-tang-amr}.

\subsection{Convergence order tests}
\label{sec:conv-order}

We first present tests of the spatial convergence order of the code.
Carefully measuring convergence of higher-order MHD codes is surprisingly challenging,
as smooth MHD test problems present numerical subtleties that are revealed by the very low numerical diffusion of higher-order methods.
We rely on widely-used smooth MHD test problems for convergence assessment:
the so-called isodensity vortex in 2D,
and non-linear circularly polarized Alfv\'en waves in 2D and 3D.

For the smooth convergence order tests of this Section~\ref{sec:conv-order}, we disable the slope and positivity limiters,
to ensure that the limiters do not interfere with effective order measurement.
Note that given a problem with a smooth solution and a desired resolution, the slope limiter setting $M$ (or equivalently $\tilde M$) may always be set in such a way that the limiter will never trigger.

\subsubsection{Computing errors against analytical solution}

To study the convergence properties of the code, we compute the $L^p$ error between a numerical scalar function $f$ and its reference solution $f_\mathrm{ref}$ over the simulation volume $V$ as
\begin{equation}
  \left\| f - f_\mathrm{ref} \right\|_p = \left[ \frac{1}{V} \int_V \left| f(\xlab) - f_\mathrm{ref}(\xlab) \right|^p d\xlab \right]^\frac{1}{p}.
  \label{eq:lp_norm}
\end{equation}
The integral is computed using Gaussian quadrature over each cell.
As described in \cite{schaal_astrophysical_2015}, we use a higher number of points for this quadrature rule ($d+3$ instead of $d+1$ used in the numerical scheme)
to account for the fact that $\left| f(x) - f_\mathrm{ref}(x) \right|^p$ will generally not be a polynomial of degree $\leq 2d$.
This simple prescription ensures that the error calculated from \eqref{eq:lp_norm} in our convergence tests
will not be dominated by errors from the quadrature in the norm itself.

\subsubsection{Isodensity MHD vortex in 2D}
\label{sec:conv-order-vortex}

We first consider the so-called isodensity MHD vortex of \cite{balsara_second-order-accurate_2004}.
This problem follows the evolution of a stationary magnetized vortex crossing a periodic domain $[-10, 10]^2$,
advected with a background flow until final time $t_f = 20$,
after which it will have returned to its initial location at the centre of the domain.
We broadly follow the setup of \cite{li_locally_2005}.
We set $\gamma = 5/3$, and initialize the unperturbed background flow with
$\rho=1$, $\press=1$, velocity $\vel=\vel_0=\sqrt{2}(\cos \alpha, \sin \alpha, 0)$ and $\Bfield=0$.
Most authors including \citet{li_locally_2005} set $\alpha=45^\circ$, however we choose to test both $\alpha=45^\circ$ and $\alpha=30^\circ$ to test 2D configurations not aligned with the grid.
Letting $r^2 = x^2 + y^2$, the velocity and magnetic field are perturbed according to
\begin{align}
\delta \vel    &= \frac{\eta}{2\pi} \bm \nabla \bm \times \left(0, 0, e^{(1-r^2)/2}\right), \\
\delta \Bfield &= \frac{\xi }{2\pi} \bm \nabla \bm \times \left(0, 0, e^{(1-r^2)/2}\right).
\end{align}
Dynamical equilibrium is achieved by correcting the pressure following
\begin{align}
  \delta \press &= \left[ \xi^2(1-r^2) - \eta^2 \right] \frac{1}{8\pi^2} e^{1-r^2}.
\end{align}
We use $\eta = \xi = 1$.
The vortex is simply advected in an equilibrium configuration, so the solution at any time $t$ is found by
$\cns(\xlab, t) = \cns(\xlab - \vel_0 t, 0)$.

Note that it is important to take a large enough domain, so that the perturbations are negligible at the domain border when setting up initial conditions,
otherwise waves will appear at the periodic boundaries.
This issue was studied in the context of higher-order Euler codes by \cite{spiegel_surveyisentropiceuler_2015} for the similar isentropic vortex test.

\begin{figure*}
  \begin{tabular}{cc}
    \hspace{1cm} LDF basis with Powell terms &
    \hspace{1cm} Legendre basis with hyperbolic cleaning \\
    \includegraphics[width=0.45\textwidth]{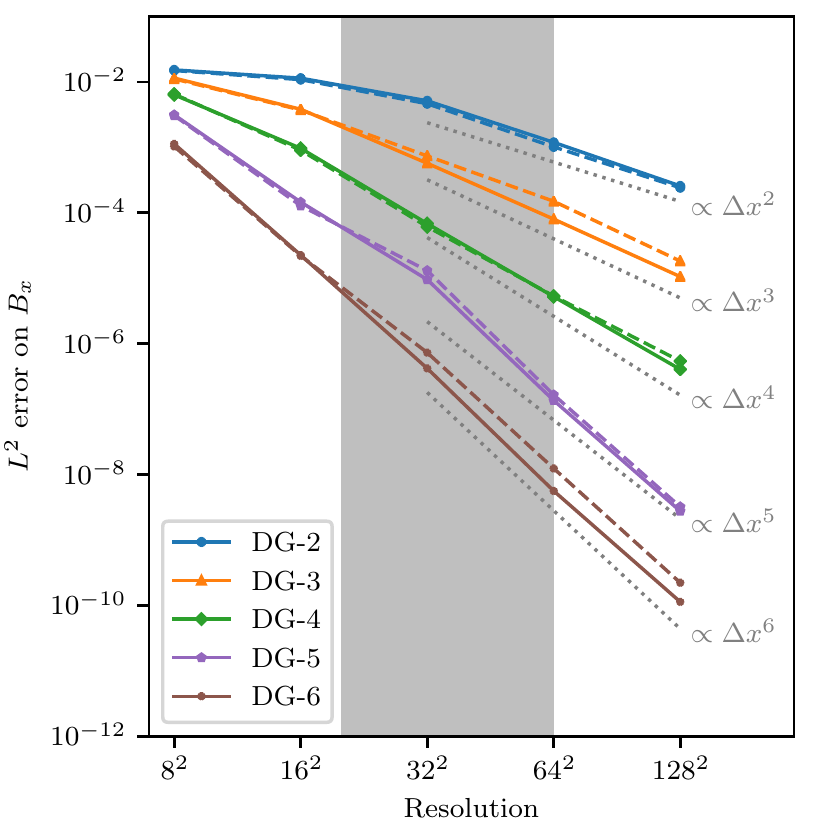} &
    \includegraphics[width=0.45\textwidth]{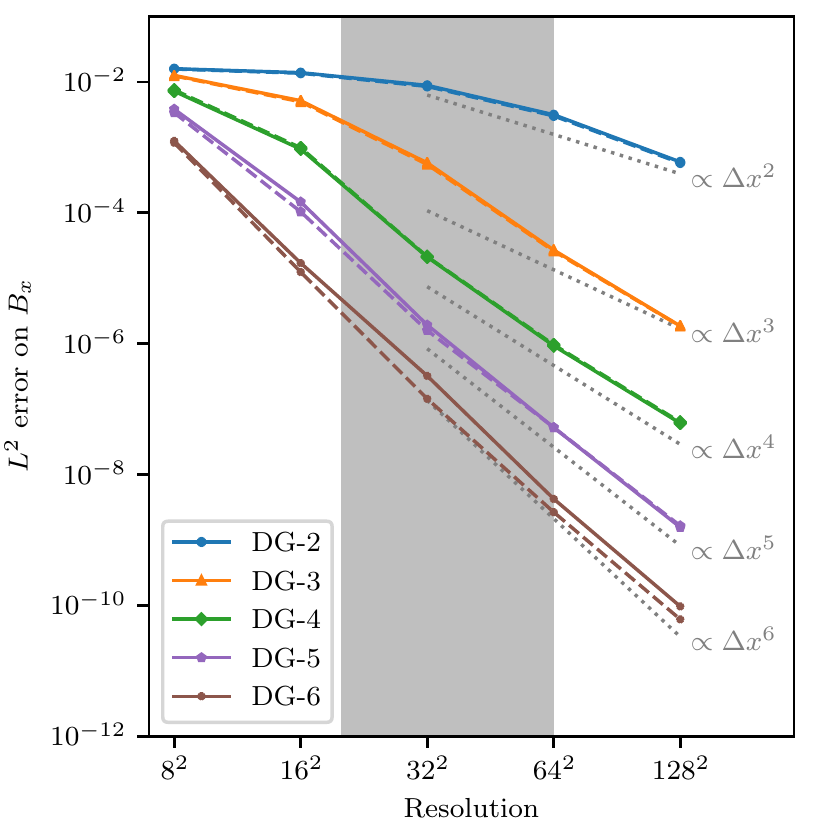} \\
    \includegraphics[width=0.45\textwidth]{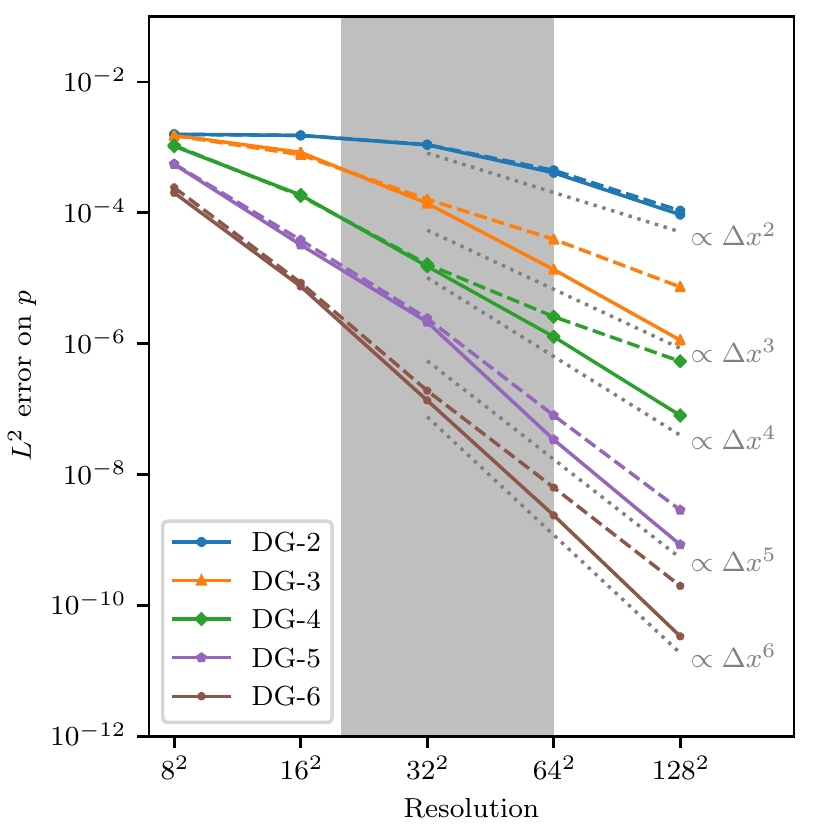} &
    \includegraphics[width=0.45\textwidth]{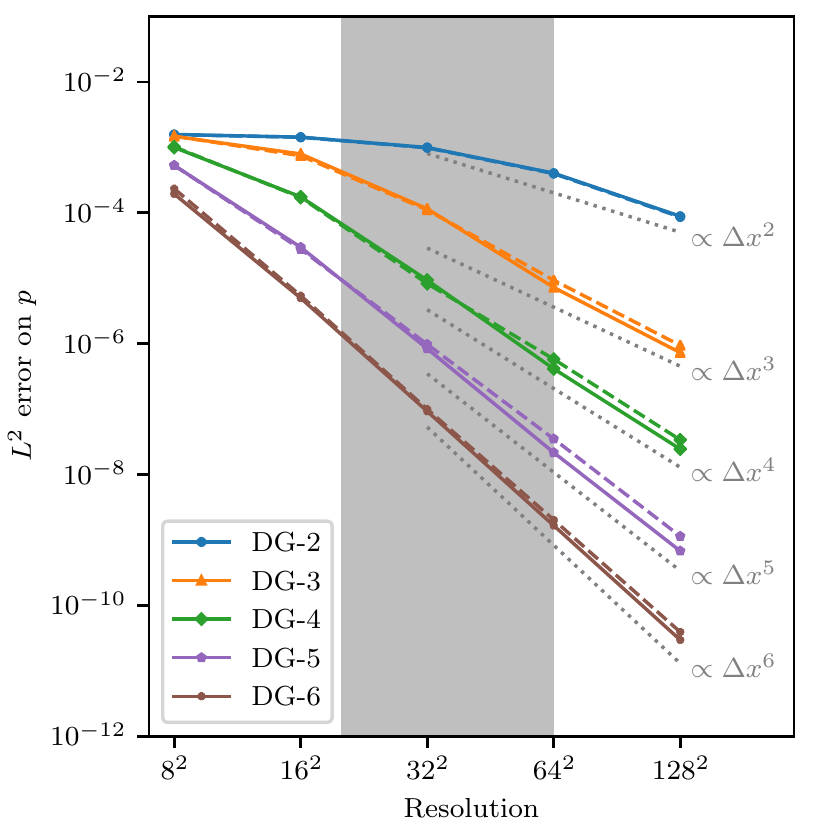} \\
  \end{tabular}
  \caption{
    \emph{Convergence of errors in the MHD isodensity vortex problem}.
    Solution $L^2$ errors are plotted for $B_x$ (top row) and pressure (bottom row),
    for both the LDF basis with Powell terms (left column),
    and Legendre basis with hyperbolic cleaning (right column).
    Errors are measured at time $t=20$ after the vortex has crossed the whole computational domain.
    Dotted lines show theoretical slopes for convergence orders 2 to 6.
    Solid and dashed lines correspond to errors for an advection angle $\alpha=45^\circ$ and $\alpha=30^\circ$ respectively.
    With the Legendre basis, the errors are insensitive to the vortex advection angle.
    With the LDF basis, the effective convergence order generally degrades slightly with $\alpha=30^\circ$,
    in particular for the pressure.
    The shaded area corresponds to a range of resolutions for which the vortex is resolved but not over-resolved (see discussion in the text).
  }
  \label{fig:conv-smvrtx-2d}
\end{figure*}

Fig.~\ref{fig:conv-smvrtx-2d} presents the $L^2$ solution errors
after the vortex has crossed the box at time $t=20$.
Results are shown for the $x$ component of the magnetic field and pressure (top and bottom rows),
and for the Powell scheme with LDF basis and Legendre basis with hyperbolic cleaning (left and right columns).
Solid and dashed lines correspond to errors for an advection angle $\alpha=45^\circ$ and $\alpha=30^\circ$ respectively.
The grey shaded area shows the approximate range of problem resolutions across which
the vortex structure is resolved with at least 2 cells (lower resolution limit),
but not over-resolved,
in the sense that the pressure fluctuation $\delta \press$ changes by 1\% or more over at least one cell (upper resolution limit).
This is the regime most interesting for science applications,
where the spatial resolution is dynamically adapted to the feature size of interest.

\begin{figure*}
  \begin{tabular}{cc}
    \hspace{1cm} LDF basis with Powell terms &
    \hspace{1cm} Legendre basis with hyperbolic cleaning \\
    \includegraphics{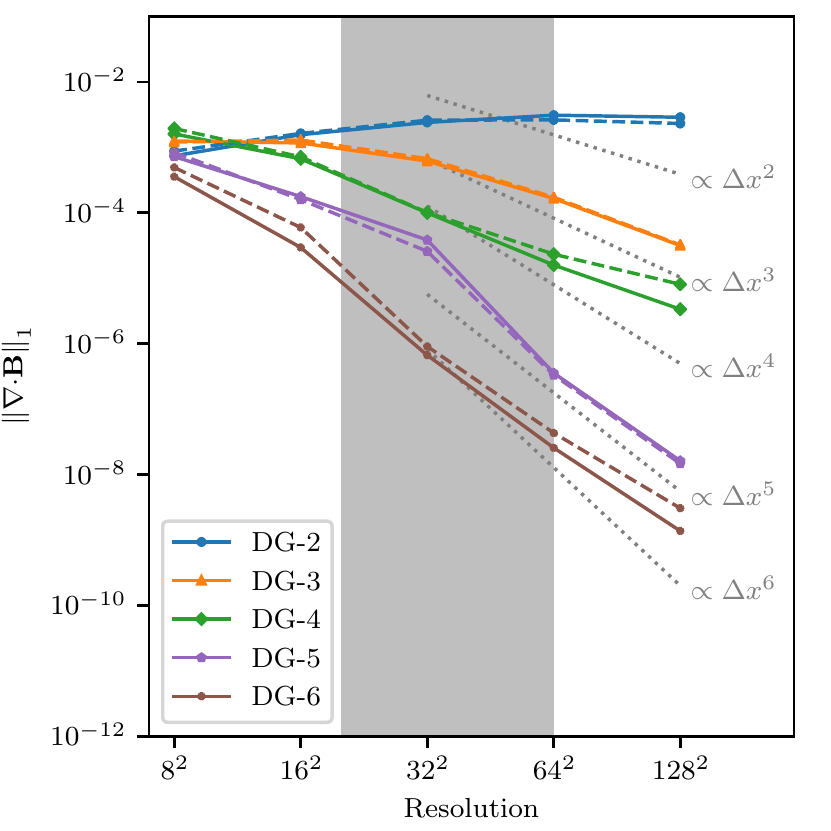} &
    \includegraphics{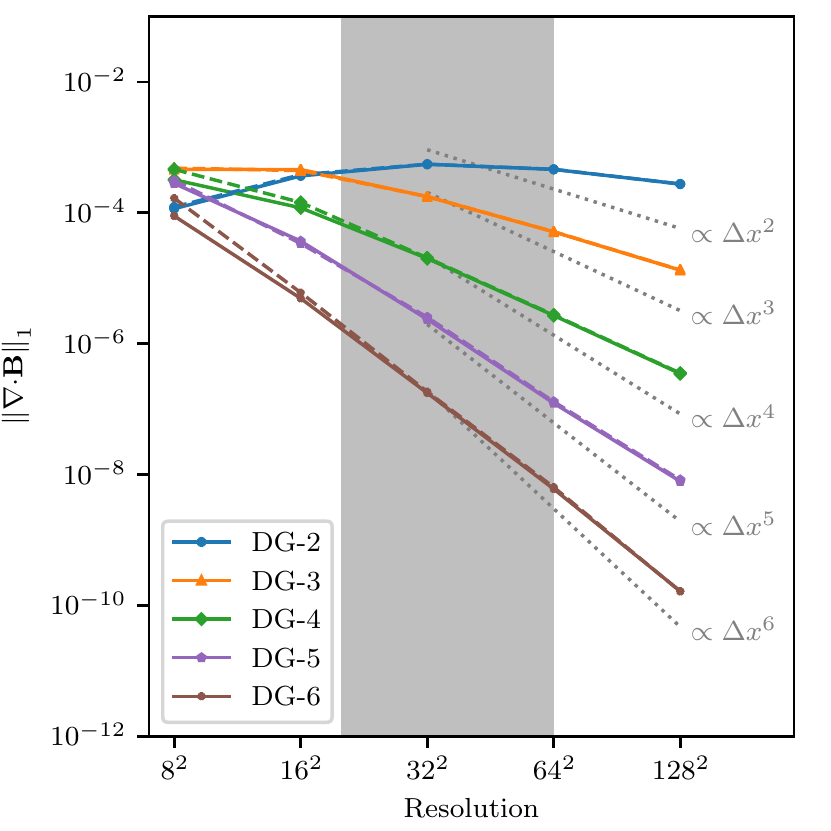}
  \end{tabular}
  \caption{\emph{Convergence of $\divB$ in the MHD isodensity vortex problem}.
    The norm of the divergence of the magnetic field
    (defined as in Eq.~\ref{eq:divb-total})
    is shown at $t=20$
    for the LDF basis with Powell terms (left panel),
    and Legendre basis with hyperbolic cleaning (right panel).
    Similarly to Fig.~\ref{fig:conv-smvrtx-2d},
    dotted lines show theoretical slopes for convergence orders 2 to 6,
    solid and dashed lines correspond to $\alpha=45^\circ$ and $\alpha=30^\circ$ respectively,
    and the shaded area corresponds to a range of resolutions for which the vortex is barely resolved.
  }
  \label{fig:conv-smvrtx-2d-div}
\end{figure*}

We find that the code achieves the expected convergence order overall.
When using LDF basis functions however, the effective order of convergence is
degraded with $\alpha=30^\circ$ in the case of over-resolved solutions.
This effect is mostly visible on the pressure, in the lower left panel of Fig.~\ref{fig:conv-smvrtx-2d}.
We speculate that this effect is caused by some instability related to the LDF bases,
whose discussion we postpone to the end of this section.

% Convergence of the divergence
We first comment on the evolution of the global numerical divergence present in the solution
as a function of scheme order and grid resolution.
Fig.~\ref{fig:conv-smvrtx-2d-div} presents the norm of the magnetic field divergence
for the smooth vortex problem at final time $t=20$,
for both the Powell and hyperbolic cleaning schemes.
The norm of the divergence is computed using Eq.~\ref{eq:divb-total},
which is sensitive to the divergence inside the cells,
as well as to discontinuities of the normal component of $\Bfield$ across cell faces.
We find that the divergence asymptotically converges to zero with resolution for the higher orders,
however, the convergence order of $\divB$ is generally slower than scheme order:
for the cleaning scheme, the effective order of convergence is about one order less
than the spatial order of the scheme---as one would expect for a first derivative of a field.
For the Powell scheme, the convergence rate suffers some additional degradation,
and presents a dependence on the angle $\alpha$ similar to the pressure errors of Fig.~\ref{fig:conv-smvrtx-2d}.
In addition, the convergence of $\divB$ is very sensitive to the resolved character of the solution for this smooth problem.
For resolved smooth solutions (right of the grey band in Fig.~\ref{fig:conv-smvrtx-2d-div}),
higher order schemes provide a more accurate representation of the solution,
thereby reducing the discontinuity of the normal component of the magnetic field and
allowing the global divergence of $\Bfield$ to converge to zero with order and resolution.
For unresolved solutions on the other hand (left of the grey band in Fig.~\ref{fig:conv-smvrtx-2d-div}),
additional resolution does not translate into reduction of the divergence,
particularly at low scheme orders.
For the adopted stringent definition of $\left\|\divB\right\|_1$ given by Eq.~\ref{eq:divb-total},
the Powell convergence rates for $\divB$ are generally slightly slower than with hyperbolic cleaning on this problem.
Interestingly, we found that the norm of the more lenient ``signed divergence'' definition of Eq.~\ref{eq:divb-gauss}
asymptotically converges, for both Powell and cleaning methods, at a rate matching the spatial order of the scheme.

We now discuss the convergence order degradation observed with the Powell scheme.
It is tempting to point at the lack of divergence damping with the Powell scheme as a possible culprit, because
for such schemes, divergence errors are known not to converge away with resolution for discontinuous solutions \citep{toth__2000}.
\citet{dumbser_unified_2008} also suggested that high-order methods alone are not enough to achieve optimal convergence without some form of divergence cleaning.
However, we found that a similar convergence degradation also appears when using hyperbolic cleaning on top of LDF bases%
---whereas hyperbolic cleaning with Legendre bases is immune, as Fig.~\ref{fig:conv-smvrtx-2d} demonstrates.
This suggests that the effect could be connected to the LDF bases, rather than caused by the Powell treatment alone.

We suspect that this deterioration is caused by the joint use of locally divergence-free basis functions,
together with the continuous treatment of the normal component of $\Bfield$ in the chosen HLLD Riemann solver.
While the detailed mechanics of this effect remain unclear at this point,
we speculate that projection of source terms (in particular at the faces) onto LDF bases could in some cases
contaminate high order modes (see Appendix~\ref{sec:apdx-div0-proj}).
Potential ways to alleviate this issue with HLLD fluxes could be to rely on an 8-wave version of this Riemann solver,
for example the one developed by \cite{fuchs_approximateriemannsolvers_2011},
or to combine different flux functions based on local smoothness of the solution
\citep[see e.g.][]{derigs_novel_2016}.
We leave these investigations to future work.

At this point, we also wish to stress that this convergence order degradation does not appear to be of great practical importance at this stage,
because the errors are still well behaved and the beneficial effects of order convergence are still present,
especially in the shaded area where the problem is marginally resolved:
at resolutions around $32^2$, order convergence helps capture the vortex much more efficiently than spatial convergence,
with both choices of divergence control.
It is of course possible that this effect is only the early-time manifestation of a more serious long-term instability,
however we have seen no indication of such problems in our test runs so far.
We believe these results demonstrate the necessity and value of testing multidimensional schemes in non-grid-aligned configurations (i.e. $\alpha=30^\circ$ in this case).

\subsubsection{Circularly polarized Alfv\'en wave problem in 2D}

Circularly polarized Alfv\'en waves are simple, exact smooth analytic solutions of the MHD equations for any wave perturbation amplitude.
They are therefore particularly suitable to study the convergence order of the scheme,
as well as its amount of numerical dispersion and dissipation.
They consist of plane waves in which the magnetic field and velocity oscillate in phase in a circular polarization perpendicular to the propagation direction.

We broadly follow the travelling wave setup proposed by \cite{toth__2000}.
Using periodic boundary conditions on a square domain $[0,L]^2$, we construct an Alfv\'en plane wave of unit wavelength,
propagating at angle $\alpha > 0$ with the $x$ axis.
In order to respect periodicity, we pick $\alpha$ and $L$ such that the box extends for exactly 1 wavelength in the $y$ direction, and $m$ wavelengths in the $x$ direction (with $m>0$ an integer).
This implies $\alpha = \arctan(1/m)$ and $L = 1/\sin \alpha$.
We pick $m=2$ which yields $\alpha \approx 26.6^\circ, L \approx 2.24$.
This ensures that the problem is truly two-dimensional.
Calling $(\parallel, \perp, z)$ the rotated $(x,y,z)$ frame so that the wave propagates along the $\parallel$ direction,
the phase of the wave at any point $(x, y)$ is given by $\phi = 2\pi x_\parallel$, with $x_\parallel = x \cos \alpha + y \sin \alpha$.
The magnetic field is initialized as
$(B_\parallel, B_\perp, B_z) = (1, \epsilon \sin \phi, \epsilon \cos \phi)$, and
the initial velocity is set to 
$(v_\parallel, v_\perp, v_z) = (v_0, \epsilon \sin \phi, \epsilon \cos \phi)$.
We set $\epsilon=0.1$ and take $v_0=0$, which produces waves travelling along the $\parallel$ direction towards negative $x_\parallel$ at the Alfv\'en velocity $v_A = B_\parallel / \sqrt{\rho} = 1$.
Density and pressure are uniform with $\rho=1, \press=0.1$, and we take $\gamma = 5/3$.
Note that the magnetic pressure is uniform thanks to the $B_\parallel$ and $B_\perp$ components being in quadrature, and exact pressure equilibrium is achieved.

The analytic solution to this problem at any time $t$ for the complete state vector $\cns$ is simply
$\cns(x,y,t) = \cns_0( x - (v_0-v_A) t \cos \alpha, y - (v_0-v_A) t \sin \alpha)$
where the signs of $-v_A$ reflect the fact that the wave travels towards negative $x_\parallel$.

Unfortunately, although it is an exact solution of non-linear MHD, this setup is subject to parametric instabilities,
which are easily triggered by low dissipation high-order schemes
\citep[see e.g. the discussion in][and references therein]{balsara_efficient_2009}.
In practice, this restricts convergence order tests to low resolutions,
before the instability starts picking up and dominating the very low errors of high order schemes.
We present results with up to $64$ grid points per dimension, which is comparable to the maximum resolution shown by \cite{balsara_efficient_2009}.

\begin{figure*}
  \begin{tabular}{cc}
    \hspace{1cm} LDF basis with Powell terms &
    \hspace{1cm} Legendre basis with hyperbolic cleaning \\
    \includegraphics[width=0.45\textwidth]{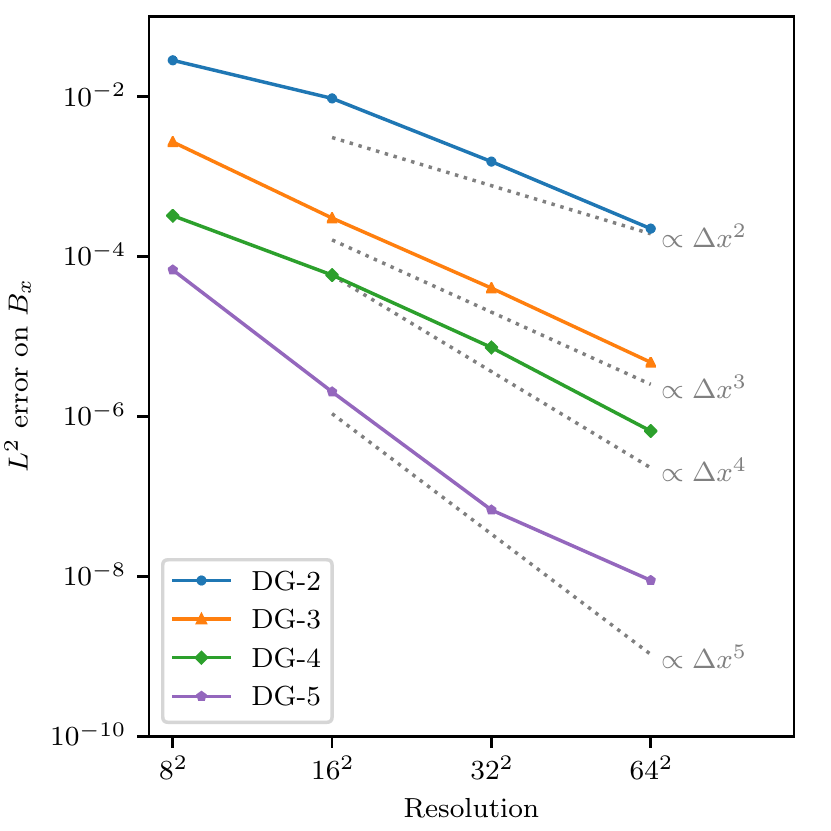} &
    \includegraphics[width=0.45\textwidth]{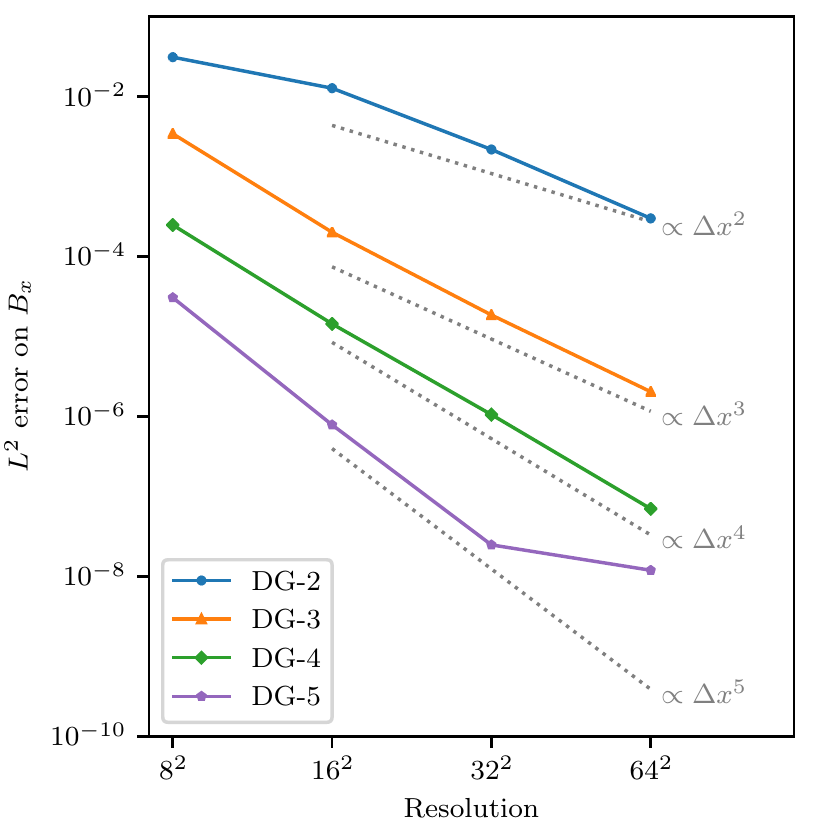}
  \end{tabular}
  \caption{
    \emph{Convergence of errors in the 2D Alfv\'en wave problem}.
    Solution $L^2$ errors on the $x$ component of the magnetic field are shown
    for the LDF basis with Powell terms (left),
    and Legendre basis with hyperbolic cleaning (right).
  Errors are computed after the wave has crossed five wavelengths, at time $t=5$.
  At the lowest resolutions, the DG-2 and DG-3 schemes do not accurately capture the wave fronts,
  which explains their apparent superconvergence.
  At error levels $\sim 10^{-8}$, the solution is polluted by the parametric instability of Alfv\'en waves discussed in the text.
  The effective order of convergence is degraded for the DG-4 scheme with LDF bases, probably because of projection effects.
  We refer to the text for a discussion of these last two effects.
  }
  \label{fig:conv-alfven-2d}
\end{figure*}

Fig.~\ref{fig:conv-alfven-2d} presents the $L^2$ solution errors for the $x$ component of the magnetic field,
after the wave has crossed one wavelength at time $t=1$.
The convergence order generally follows the theoretical order of convergence for both divergence control schemes.
Note that at $64^2$ resolution, the error of the \nth{5}~order scheme starts levelling off at around $10^{-8}$,
which we attribute to the triggering of the aforementioned parametric instability;
\citet{balsara_efficient_2009} found evidence for this effect in their \nth{4}~order code at comparable resolutions in their 3D Alfv\'en wave problem.
Note also that our Powell scheme---which relies on LDF bases---%
suffers from a slightly degraded convergence rate for the fourth-order DG-4 scheme compared to hyperbolic cleaning.
We deem that this effect could be related to the observed deterioration of the convergence order for the isodensity MHD vortex of Section~\ref{sec:conv-order-vortex}.

\subsubsection{Circularly polarized Alfv\'en wave problem in 3D}

To test convergence order of the code in 3D,
we run a comparable Alfv\'en wave setup, partially based on \cite{balsara_efficient_2009}.
For this problem, the domain is $[0, 1]^3$ with periodic boundary conditions in all three directions.
We take a uniform background density $\rho=1$ and thermal pressure $\press=100$.

A rotated coordinate system $(x', y', z')$ is setup so that the $x'$ direction is aligned with the cube diagonal,
following the exact transformation described in \cite{balsara_efficient_2009}.
Letting $\phi$ be the phase angle of the wave, the velocity field in the primed reference frame is
$\vel' = (1, \epsilon \cos \phi, \epsilon \sin \phi)$,
while the magnetic field is
$\Bfield' = (1, -\epsilon \cos \phi, -\epsilon \sin \phi)$,
for which we pick $\epsilon = 0.02$.
The unprimed velocity and magnetic fields are obtained by transforming back to unprimed coordinates.

The resulting wave travels at speed 2 along $x'$,
since it has Alfv\'en velocity 1 and is also advected with the flow with velocity 1 along $x'$ due to our choice of $\vel'$.
The phase angle $\phi$ is therefore given by:
\begin{equation}
  \phi = \frac{2\pi}{\lambda} (x' - 2t).
\end{equation}
Choosing the wavelength $\lambda = 1/\sqrt{3}$ ensures that the problem is periodic in all 3 directions $x$, $y$ and $z$,
since whenever we cross one box length in one of these directions, $x'$ shifts by $1/\sqrt{3}$.
With this choice of $\lambda$, the cube diagonal crosses 3 wave periods.
The problem is run until time $t_f = \sqrt{3}/2$, after which the 3 wave fronts will have crossed the whole domain once.

\begin{figure*}
  \begin{tabular}{cc}
    \hspace{1cm} LDF basis with Powell terms &
    \hspace{1cm} Legendre basis with hyperbolic cleaning \\
    \includegraphics[width=0.45\textwidth]{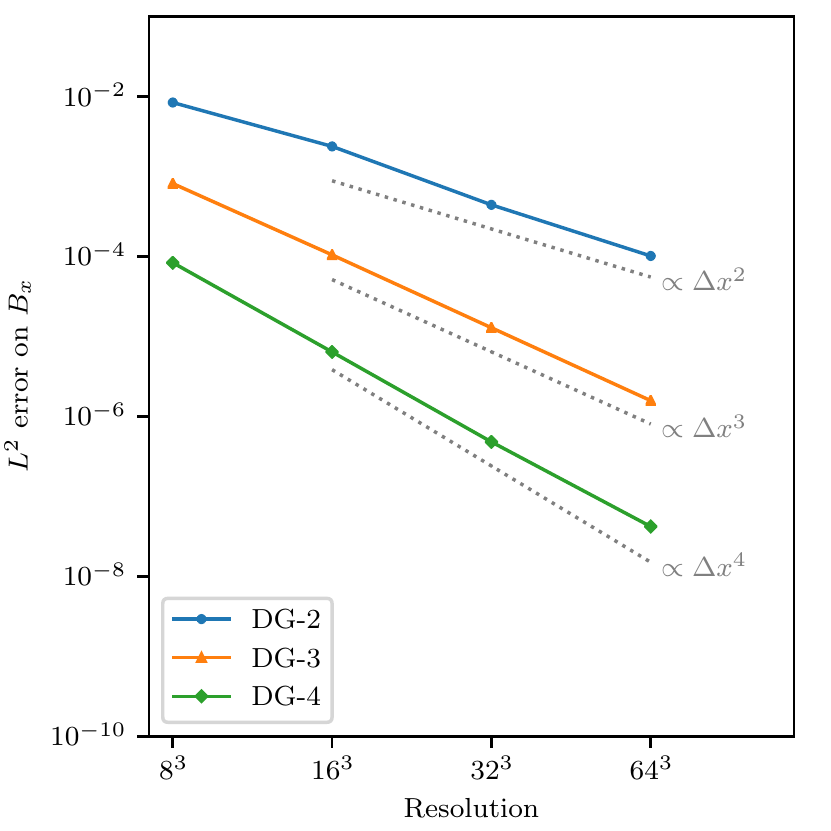} &
    \includegraphics[width=0.45\textwidth]{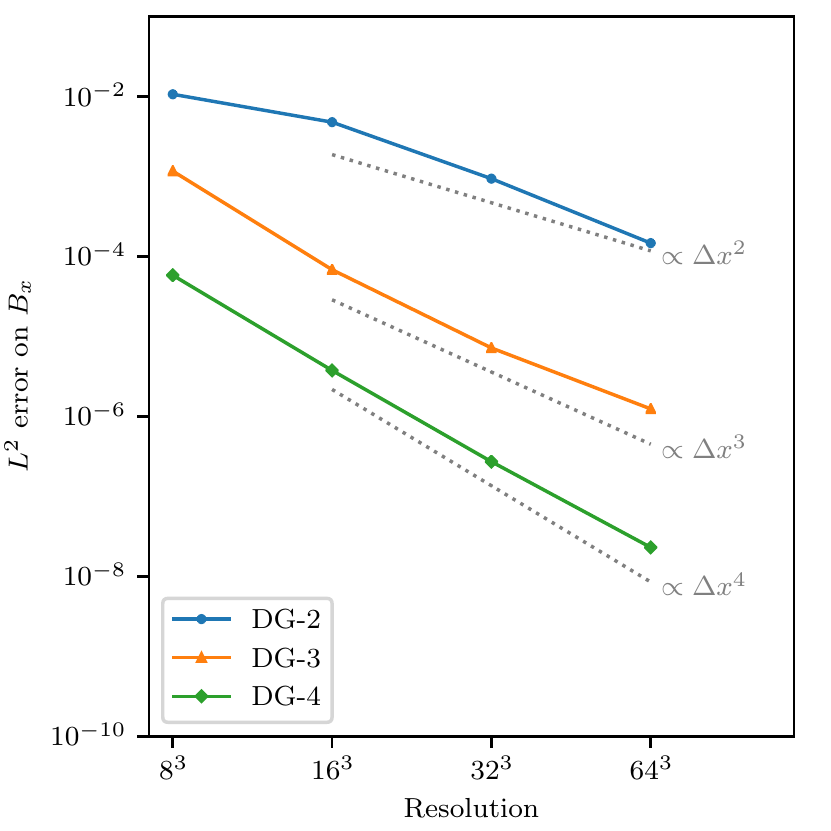}
  \end{tabular}
  \caption{
    \emph{Convergence of errors in the 3D Alfv\'en wave problem}.
    Solution $L^2$ errors on the $x$ component of the magnetic field are shown
    for the LDF basis with Powell terms (left),
    and Legendre basis with hyperbolic cleaning (right).
    Errors are computed after 3 wave fronts have crossed the box domain, at time $t_f = \sqrt{3}/2$.
  }
  \label{fig:conv-alfven-3d}
\end{figure*}

Fig.~\ref{fig:conv-alfven-3d} shows the convergence of the errors on the $\Bfield_x$ component
with the Powell and hyperbolic cleaning schemes for scheme orders 2 to 4.
Both schemes follow overall the expected order of convergence.

\subsection{Shock problems}

We now turn to test problems in 1D, 2D and 3D designed to test the shock capturing properties of the scheme.

\subsubsection{Brio-Wu shock tube}

The famous shock tube problem introduced by \citet{brio_upwind_1988} has now become a classic of shock tests for MHD codes.
For this test, we take the computational domain to be $[0, 1]$ with fixed boundary conditions at $x=0$ and $x=1$.
In the whole domain, the flow is initially at rest ($\vel=0$) and $(B_x, B_z)=(0.75, 0)$.
The initial primitive variables are discontinuous at $x=0.5$, with the left and right states given by
$(\rho, \press, B_y)_L = (1, 1, 1)$
and
$(\rho, \press, B_y)_R = (0.125, 0.1, -1)$
respectively.
We set $\gamma=2$, and run the simulation until final time $t_f=0.1$.
Note that we actually run our 1D test problems in 2D, with perfectly $y$-independent initial conditions and periodic boundary conditions in $y$,
in order to test the numerical stability of the 1D setup.
We check that no significant $y$ dependence of the solution has developed at the final time.

\begin{figure*}
  \includegraphics[width=\textwidth]{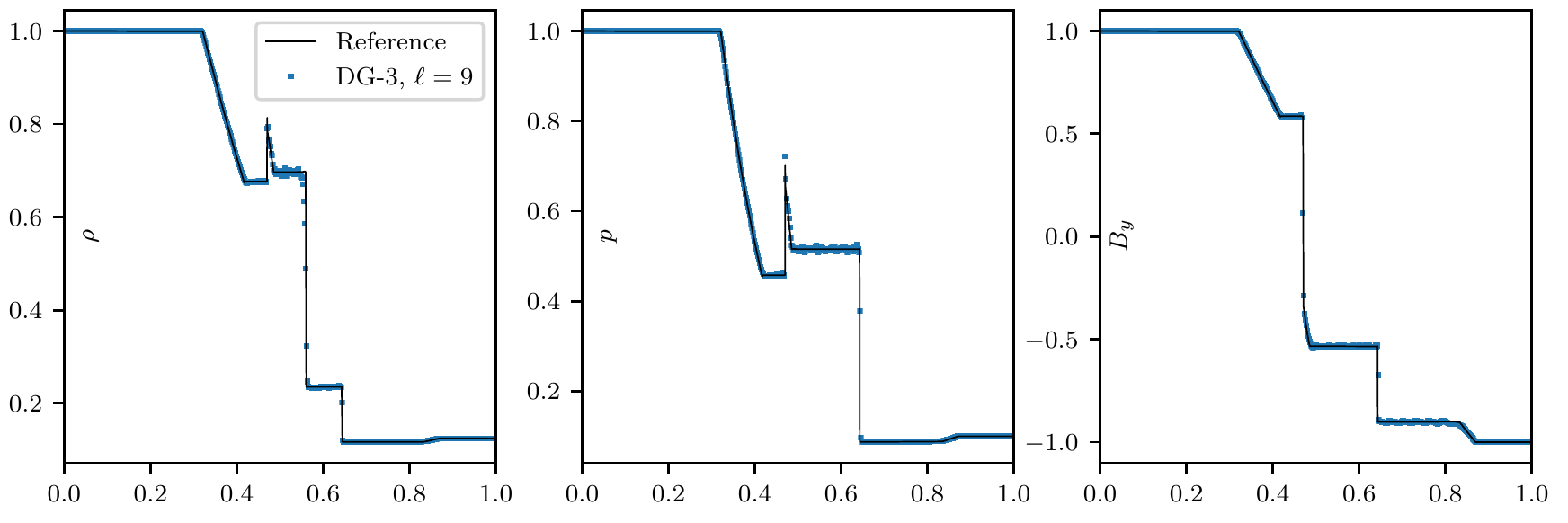}
  \caption{ \emph{Brio-Wu MHD shock tube problem}.
    The density, pressure and $y$ component of the magnetic field are shown at $t=0.1$
    for a grid with 512 cells (resolution level $\ell=9$).
    The reference solution was computed using {\athena} with the third-order Roe solver on $10^4$ mesh points.
    All the MHD waves are correctly captured and shocks are captured within 1--2 cells without overshooting.
    Some oscillations are visible, probably due to the simple limiting procedure employed (see text).
  }
  \label{fig:test-brio-wu}
\end{figure*}

Fig.~\ref{fig:test-brio-wu} presents the density, pressure, and $y$ component of the magnetic field in the Brio-Wu shock tube test problem at final time $t_f=0.1$,
using the third-order DG-3 scheme at resolution level $\ell=9$ (corresponding to 512 cells).
The reference solution was obtained using {\athena} \citep{stone_athena:_2008} with the third-order Roe solver on $10^4$ mesh points.
The DG-3 scheme captures all the MHD waves correctly, and the shocks are resolved with 1--2 cells,
whereas the contact discontinuity is resolved within about 4 cells.
The limiter is efficient at preventing overshoots around shocks,
although some oscillations are visible around the contact discontinuity and after the right rarefaction wave,
probably due to a rather lenient global limiter setting used across all of our tests.
A more finely tuned or more sophisticated limiter could help reducing these oscillations.
We note however that we observed identical oscillations of similar amplitude and wavelength
with {\athena} with HLLD and third-order reconstruction at the same resolution of $512$ cells.

\subsubsection{1D MHD Shu-Osher test problem}
The so-called hydrodynamical Shu-Osher test problem, introduced by \cite{shu_efficientimplementationessentially_1989}, follows the interaction of a supersonic shockwave with smooth density perturbations.
It tests the scheme's ability to resolve small-scale features in the presence of strong shocks.
An MHD version of the test was proposed by \cite{susanto_highorderfinitevolumeschemes_2014},
and used in particular by \cite{derigs_novel_2016} to test their MHD scheme in FLASH.
We follow the setup of these two papers.
The computational domain is $[-5, 5]$ with reflective boundary conditions.
At $t=0$, the shock interface is located at $x_0 = -4$.
In the region $x \leq x_0$, a smooth supersonic flow is initialized with primitive state given by
$( \rho, \vel_x, \vel_y, \vel_z, \press , \Bfield_x, \Bfield_y, \Bfield_z) =
 (  3.5, 5.8846, 1.1198,      0, 42.0267,         1,    3.6359,         0)$.
In the rest of the domain $x > x_0$, smooth stationary perturbations are setup with primitive state
$( 1 + 0.2\sin 5x, 0, 0,  0, 1, 1, 1, 0)$.
The flow is evolved until final time $t_f = 0.7$.

\begin{figure*}
  \includegraphics[width=0.66\textwidth]{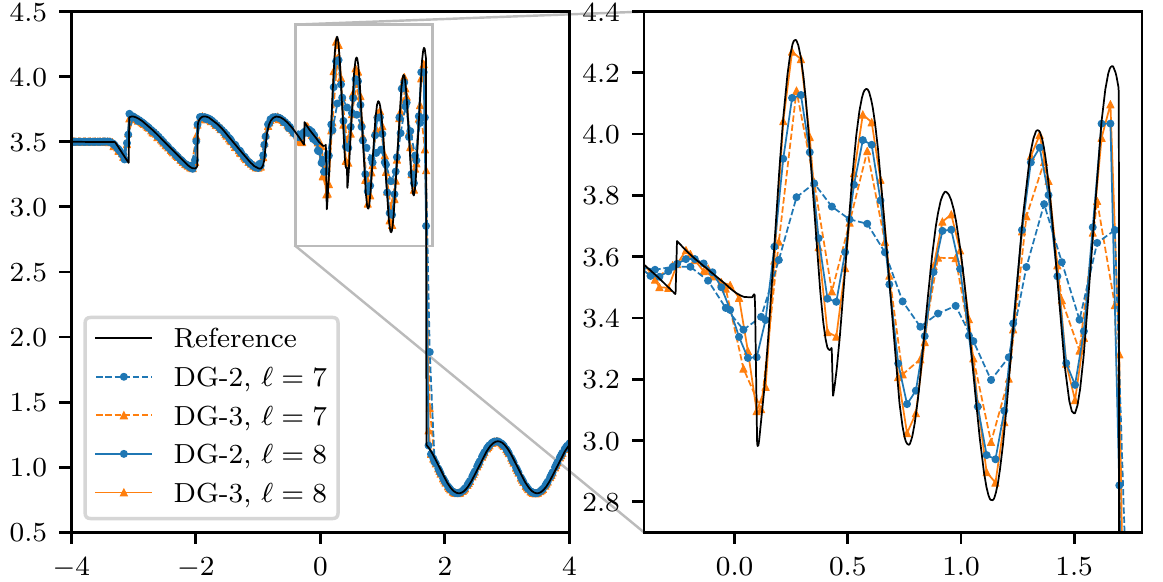}
  \caption{\emph{MHD Shu-Osher shock tube test problem}.
    The density profile is shown at final time $t_f=0.7$.
    This figure can be directly compared to \protect\cite{derigs_novel_2016}.
    The reference solution was computed using {\athena} with the third-order Roe solver on $10^4$ mesh points.
    The right panel is a zoom on the region shown framed in the left panel.
    The numerical solution for second-order (DG-2) and third-order (DG-3) schemes are shown,
    for resolution levels $\ell=7$ (128 grid points) and $\ell=8$ (256 points).
  }
  \label{fig:test-shu-osher}
\end{figure*}

The density profile at $t_f$ is shown in Fig.~\ref{fig:test-shu-osher},
for resolution levels 7 (128 cells) and 8 (256 cells).
To compare the results and focus on the impact of the spatial discretization order,
we use the RK3 SSP time integrator for all runs of this test problem,
but we have checked that the results are unchanged if we use RK2 for the DG-2 schemes.
The zoom-in panel on the right shows that going from second to third order greatly improves the capturing of the strong oscillations in the interaction region, in particular at the lower resolution level 7.
This test problem demonstrates the ability of higher-order methods to capture finer features in the flow with lower numerical diffusion.
In particular, our third-order scheme is very competitive for this test problem:
\citet{chakravarthy_simplehybridfinite_2015} show that they can capture the oscillations with 400 grid points (about 14 points per oscillation period, see their Figure~9),
\citet{derigs_novel_2016} demonstrate good capturing with 256 points (about 9 per period, see their Figure~6),
and with the DG-3 scheme, we obtain results comparable to this latter work with only 128 total grid points (i.e. 4 to 5 points per period, see the corresponding DG-3 $\ell=7$ line in Fig.~\ref{fig:test-shu-osher}).

\subsubsection{Orszag--Tang vortex problem}
\label{sec:test-orszag-tang}

We now consider the Orszag--Tang vortex problem, a widely-used test problem for MHD.
The vortex starts from a smooth initial field configuration, and quickly forms shocks before transitioning to 2D MHD turbulence.

For this problem, our computational domain is $[0, 1]^2$ and we use $\gamma = 5/3$.
The initial density and pressure are uniform, with $\rho = \frac{25}{36\pi}$ and $\press = \frac{5}{12\pi}$.
The initial gas velocity field is $\vel = (-\sin(2\pi y), \sin(2\pi x), 0)$,
and the initial magnetic field is $\Bfield = (-B_0 \sin(2\pi y), B_0 \sin(4\pi x), 0)$,
with $B_0 = 1/\sqrt{4\pi}$.

\begin{figure*}
  \includegraphics[width=\textwidth]{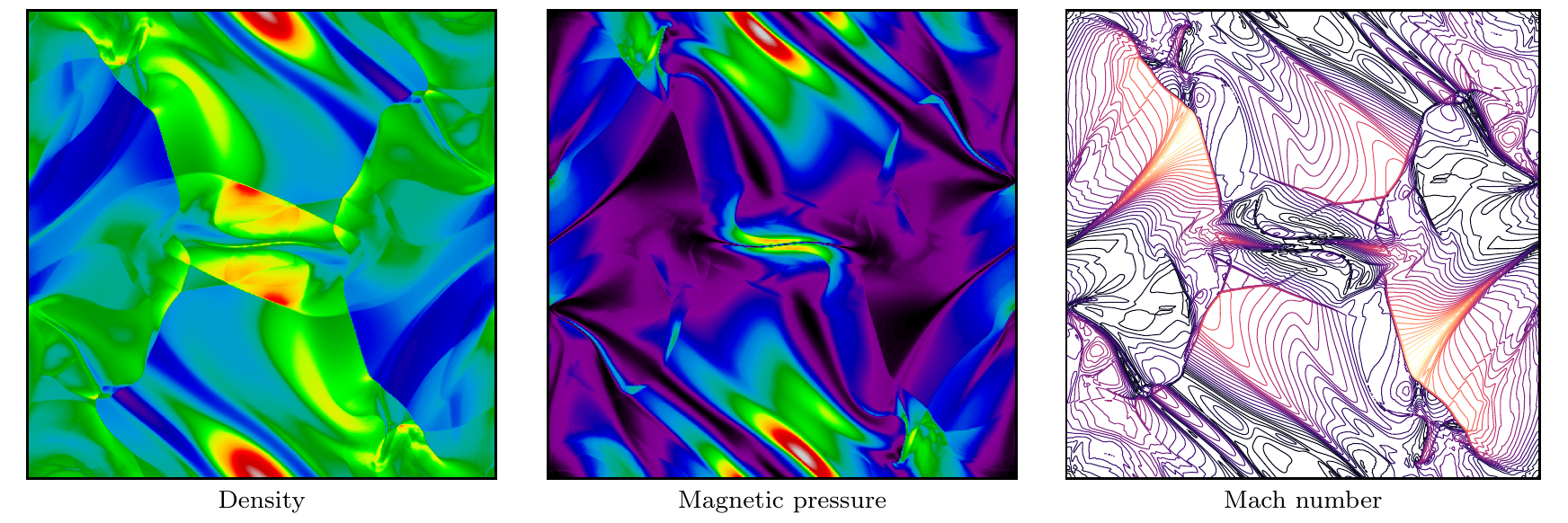}
  \caption{\emph{Orszag--Tang vortex test problem at $t=0.5$}.
    The density, pressure and Mach number are shown on a $512^2$ grid,
  computed using the third-order DG scheme with the Powell method.}
  \label{fig:test-orszag-tang}
\end{figure*}

The solution at $t=0.5$ is presented in Fig.~\ref{fig:test-orszag-tang}.
This is the well-recognizable picture of the Orszag--Tang vortex, presented by many authors discussing MHD codes.
Note that we obtain both sharp shocks and smooth, noise-free flow with resolved features between the shocks.
During our experiments we found that details of both the limiting procedure and the Powell term discretization could strongly impact the shape of the low-density smooth regions.

\begin{figure*}
  \includegraphics[width=\textwidth]{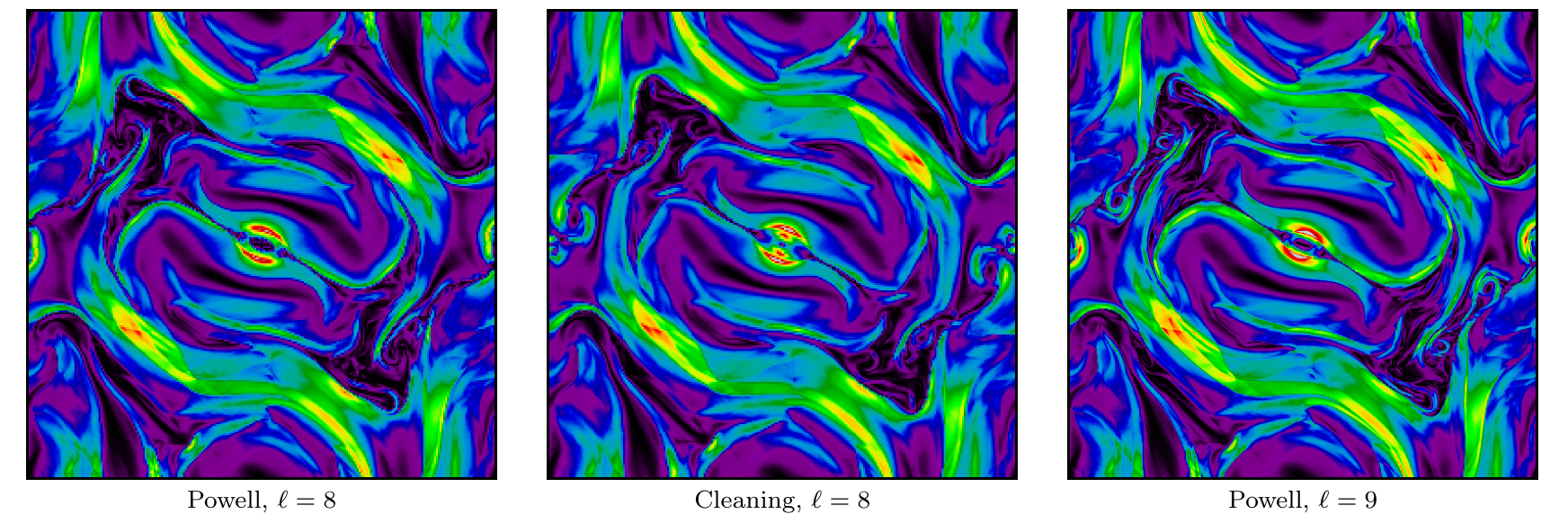}
  \caption{\emph{Orszag--Tang vortex test problem at $t=1$}.
    The magnetic pressure is shown for the third-order DG scheme
    with the Powell and divergence cleaning methods on a $256^2$ grid (left and centre, respectively),
    and for the Powell method on a $512^2$ grid (right).
    All resulting magnetic field configurations are in good agreement.
  }
  \label{fig:test-orszag-tang-t1}
\end{figure*}

While the Orszag--Tang vortex is a classic MHD test, many authors do not discuss how their codes capture the evolution beyond $t=0.5$.
During our experiments, we found that while many limiters and Powell term prescriptions work equivalently well until $t=0.5$,
long time integration of this problem can be challenging, in particular for divergence errors, as noted by \cite{balsara_total_1998}.
We found that running the problem for longer times could discriminate well between otherwise seemingly equivalent discretizations.
In particular, at times $t \in [0.75, 0.85]$, complex shock interactions occur near the centre of the box which can easily cause divergence blow-ups.
In Fig.~\ref{fig:test-orszag-tang-t1}, we show maps of the magnetic pressure at $t=1$
with both divergence control methods at level 8 resolution,
and for the Powell method at level 9 resolution.
While some differences can be we observed in the central vortex
and in the finer detailed structure of the folds,
the resulting magnetic field configurations are in good agreement,
even well into the transition into 2D turbulence.

\begin{figure*}
  \includegraphics[width=\textwidth]{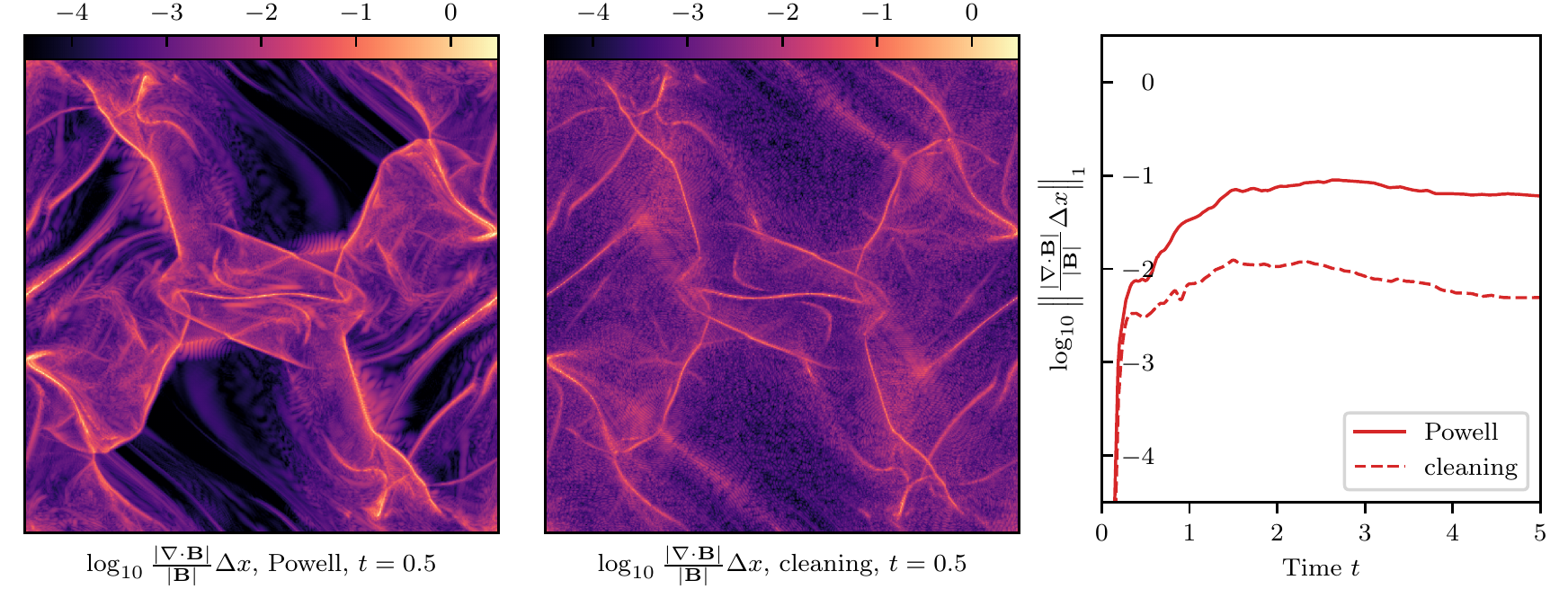}
  \caption{
    \emph{Magnetic field divergence in the Orszag--Tang test problem}.
    Maps of the normalized magnetic field divergence (see Appendix~\ref{sec:apdx-divergence-normalized})
    are shown for the DG-3 scheme at $512^2$ resolution
    using the Powell (left) and divergence cleaning (centre) schemes.
  The right panel shows the time evolution of the $L^1$ norm of the magnetic field divergence for both methods.}
  \label{fig:test-orszag-tang-div}
\end{figure*}

Fig.~\ref{fig:test-orszag-tang-div} shows maps of the normalized divergence
using the DG-3 Powell and cleaning schemes at $512^2$ resolution,
as defined in Appendix~\ref{sec:apdx-divergence-normalized}.
The divergence is concentrated around shocks, with both methods forming similar patterns.
Because hyperbolic cleaning propagates the divergence through the simulation box,
divergence waves quickly fill the whole domain,
producing the fairly uniform background clearly visible in the centre panel.
The right panel of Fig.~\ref{fig:test-orszag-tang-div}
presents the time evolution of the divergence with both methods up to $t=5$,
demonstrating that $\divB$ remains under control for long run times with both schemes.
The Powell method results in divergence levels about one order of magnitude greater than hyperbolic cleaning,
in broad overall agreement with existing comparisons across different test problems
\citep[see e.g.][]{tricco_constrainedhyperbolicdivergence_2012,hopkins_accurate_2016,derigs_idealglmmhdentropy_2018}.

\begin{figure}
  \includegraphics[width=\columnwidth]{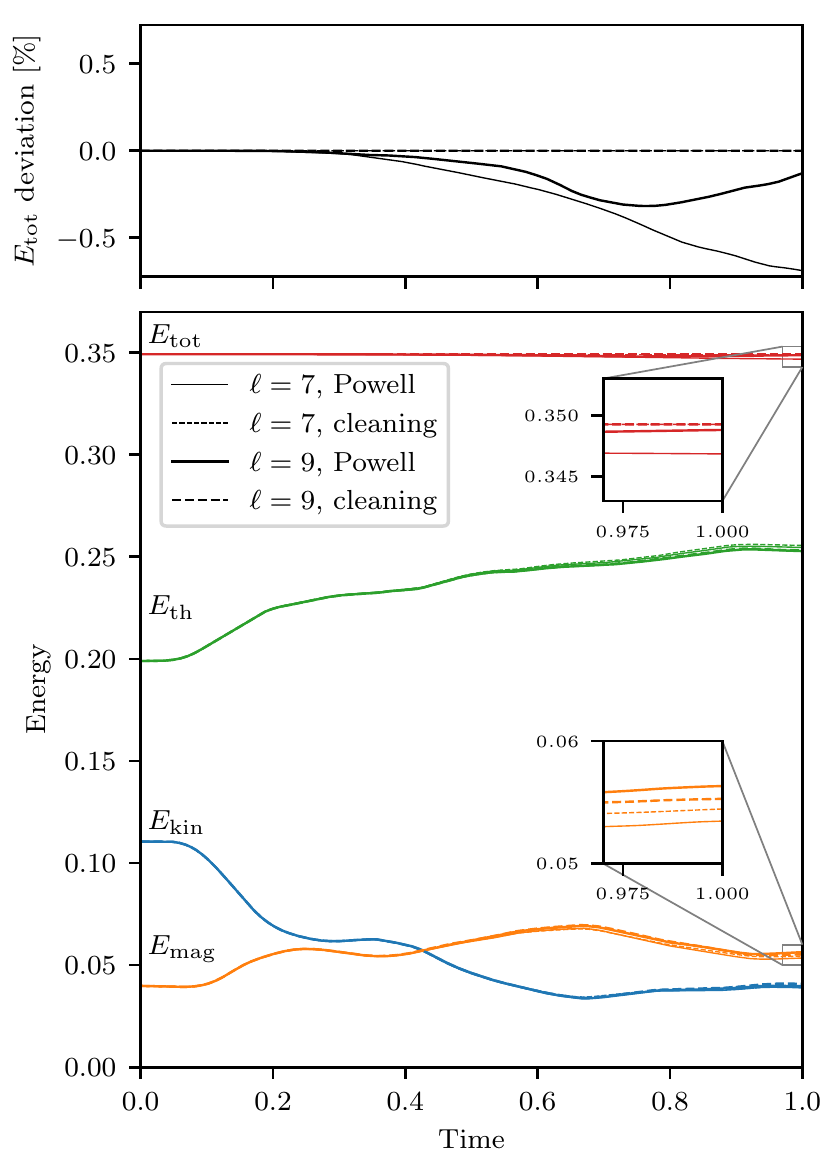}
  \caption{
    \emph{Evolution of energies in the Orszag--Tang vortex problem}, for the \nth{3} order DG-3 scheme,
    at resolution levels 7 ($128^2$) and 9 ($512^2$) using the Powell and hyperbolic cleaning schemes.
    The bottom panel shows the time evolution of the different energies integrated over the whole domain,
    while the top panel shows the deviation in the total energy $\Etot$ from its initial value
    due to non-conservative terms for the Powell scheme.
  }
  \label{fig:test-orszag-tang-energy}
\end{figure}

Because the Orszag--Tang test gives rise to many complex flow features including regions of potentially severe numerical divergence,
we also use this test problem to assess the impact of non-conservative Powell terms on scheme conservation.
We find that in general the total energy is the conserved quantity most directly impacted by the Powell term.
Fig.~\ref{fig:test-orszag-tang-energy} shows the time evolution of the magnetic, kinetic, thermal and total energies
($\Emag$, $\Ekin$, $\Etherm$, $\Etot$ respectively) for the DG-3 scheme over $0 \leq t \leq 1$.
In the bottom panel, the energies are shown for resolution levels $\ell=7$ ($128^2$) and $\ell=9$ ($512^2$) and for both divergence control schemes.
The energy histories are in good overall agreement across methods and resolutions,
although differences develop over time.
The top panel shows the slight deviation of the total energy $\Etot$ from its initial value due to non-conservative source terms with the Powell scheme.
Over the whole time evolution, $\Etot$ has deviated by at most $0.6\%$ for $\ell=7$, and $0.3\%$ for $\ell=9$.
Increasing the resolution seems to contribute to reduce the deviation of $\Etot$,
even though it is known that divergence errors in general do not locally converge away with resolution \citep{toth__2000}.
Note that the deviation is not evolving monotonically,
as such, it is difficult to appreciate the long-term impact of non-conservative source terms on the solution.
Note also that the effect of the divergence control scheme on the magnetic energy (lower inset)
is comparable in magnitude to the effect of resolution,
so the detailed differences in energies may not all be attributed to the difference in divergence treatment.

We point out that total conservation is not a sufficient test of the quality of the solution against divergence errors,
because energy can be created and destroyed \emph{locally} in the flow with different signs,
depending on the sign of various quantities in the source term \eqref{eq:powell-src}.
While the conservation results are encouraging overall,
we intend to perform more careful comparisons when applying both schemes to physical problems.
Driven turbulence problems in particular may turn out to be more sensitive to non-conservation issues,
because energy is continually injected in the simulation domain.

\begin{figure*}
  \includegraphics[width=\textwidth,trim={0 7mm 0 0mm},clip]{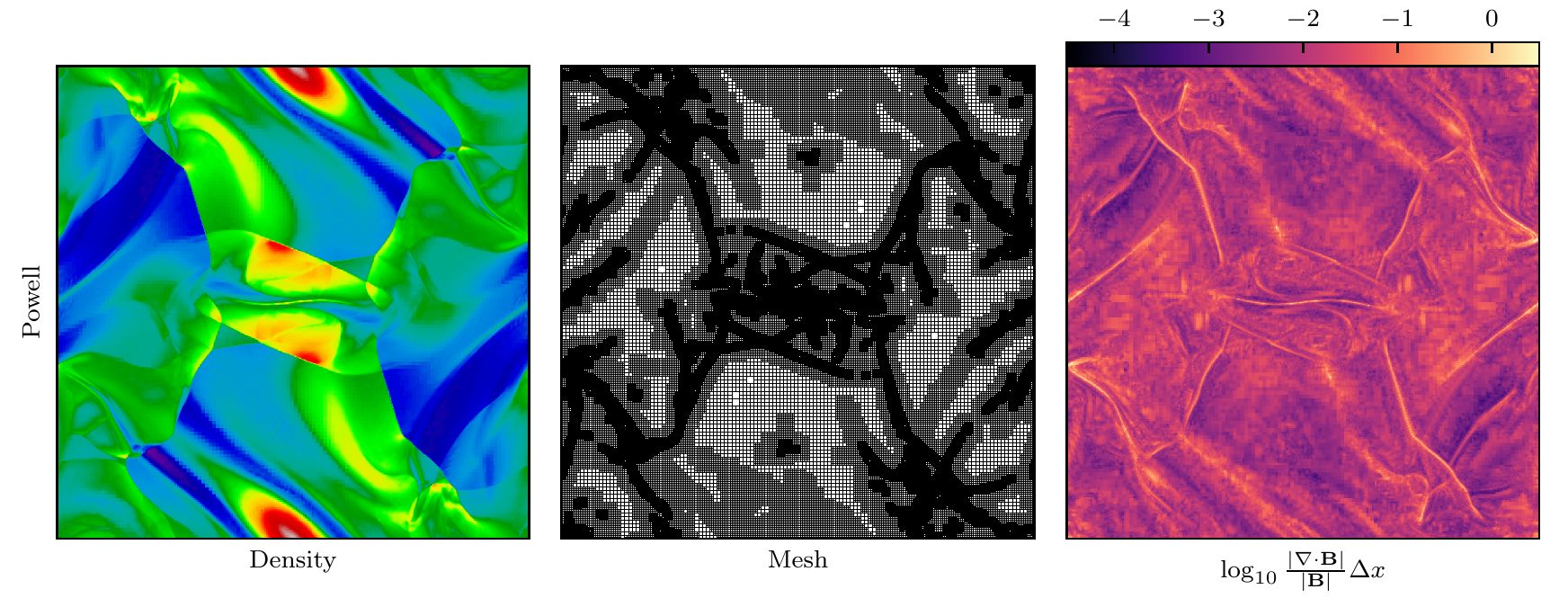}
  \includegraphics[width=\textwidth,trim={0 2mm 0 4mm},clip]{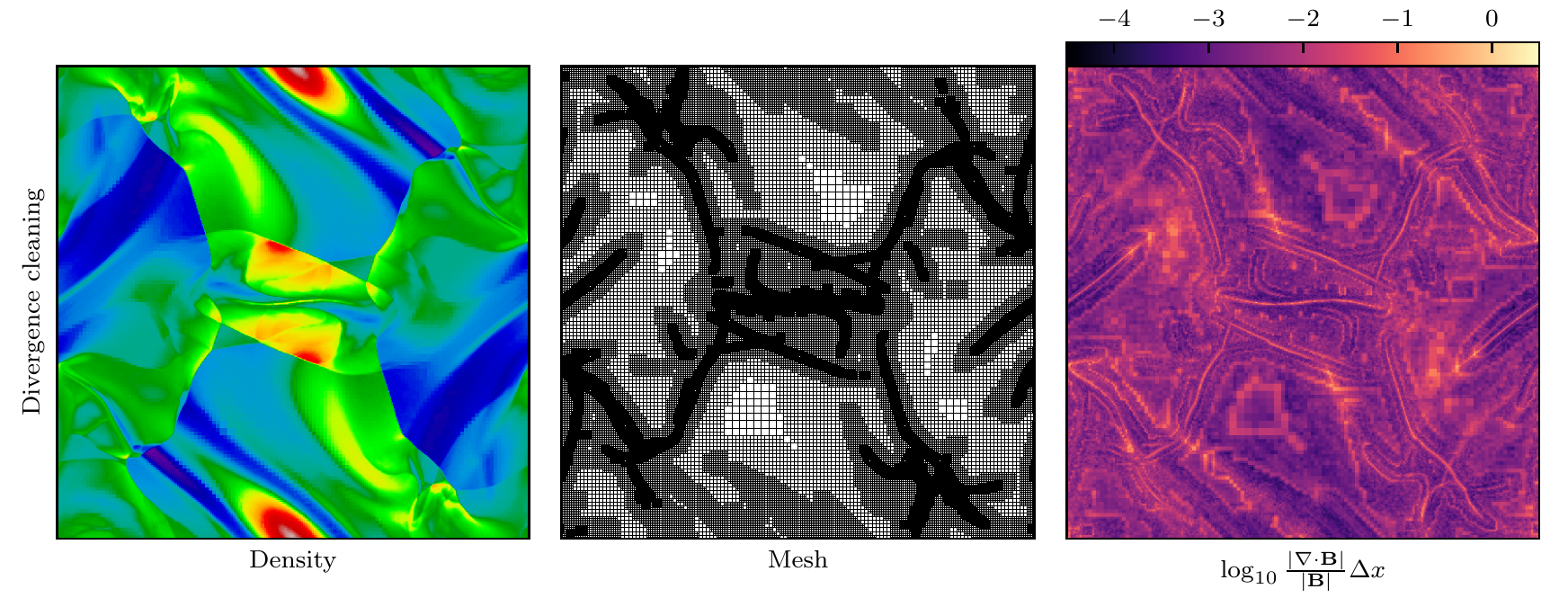}
  \caption{\emph{Orszag--Tang vortex problem with adaptive mesh refinement} at $t=0.5$,
    from resolution levels 6 to 9 ($64^2$ to $512^2$) using the DG-3 Powell (top) and divergence cleaning (bottom) schemes.
    The columns show the density field (left),
    the geometry of the AMR grid (centre),
    and the normalized magnetic field divergence (right).
    The refinement is based on the density and magnetic field components.
    The density map may be directly compared to the $512^2$ Cartesian run of Fig~\ref{fig:test-orszag-tang}.
  }
  \label{fig:test-orszag-tang-amr}
\end{figure*}

We conclude this section with Fig.~\ref{fig:test-orszag-tang-amr},
which shows the same Orszag--Tang problem run up to $t=0.5$
with adaptive resolution from $64^2$ to $512^2$,
using the DG-3 Powell and divergence cleaning schemes.
The solution may be directly compared to the $512^2$ Cartesian run of Fig.~\ref{fig:test-orszag-tang}.
The AMR solutions obtained using both divergence control methods are very similar,
and both in very good agreement with the Cartesian solution.
On the Powell density map, some slight ripple-like noise is visible
in a few localized smooth regions in the vicinity of shocks.
This is likely the result of the interaction of the divergence control method with AMR,
as those features are absent from both the Cartesian Powell
and AMR divergence cleaning runs.

The adaptive mesh provides both memory and wall time reduction for this problem.
With the Powell scheme, at $t=0.5$, the AMR grid of Fig.~\ref{fig:test-orszag-tang-amr}
features $2.4\times$ fewer cells than the $512^2$ Cartesian grid
of Fig.~\ref{fig:test-orszag-tang} of equivalent maximum resolution.
The same AMR run also reaches $t=0.5$ about $3.0\times$ faster:
in addition to requiring fewer cells, the adaptive mesh
allows computing the early stages of the evolution with a larger timestep,
because the initial Orszag--Tang flow is very smooth
and therefore well captured everywhere by coarse cells.

\subsubsection{Advected Orszag--Tang vortex}
\label{sec:test-orszag-tang-adv}
In this problem, we test the code's Galilean invariance by
running the Orszag--Tang vortex test case up to $t_f = 0.5$,
giving the fluid an additional uniform global bulk velocity $\vel_0 = (10, 10, 0)$.
This corresponds to a hypersonic advection with initial Mach number $\approx 14$.
At time $t_f$, the flow will have crossed the box 5 times in the $x$ and $y$ directions,
and end up in a configuration identical to the non-advected problem at the same time.

\begin{figure*}
  \tikzsetnextfilename{tikz-montage-adv-orszag-tang}
  \begin{tikzpicture}
    \node (row1) {
        \includegraphics[width=16cm,trim={0 4mm 0 0},clip]{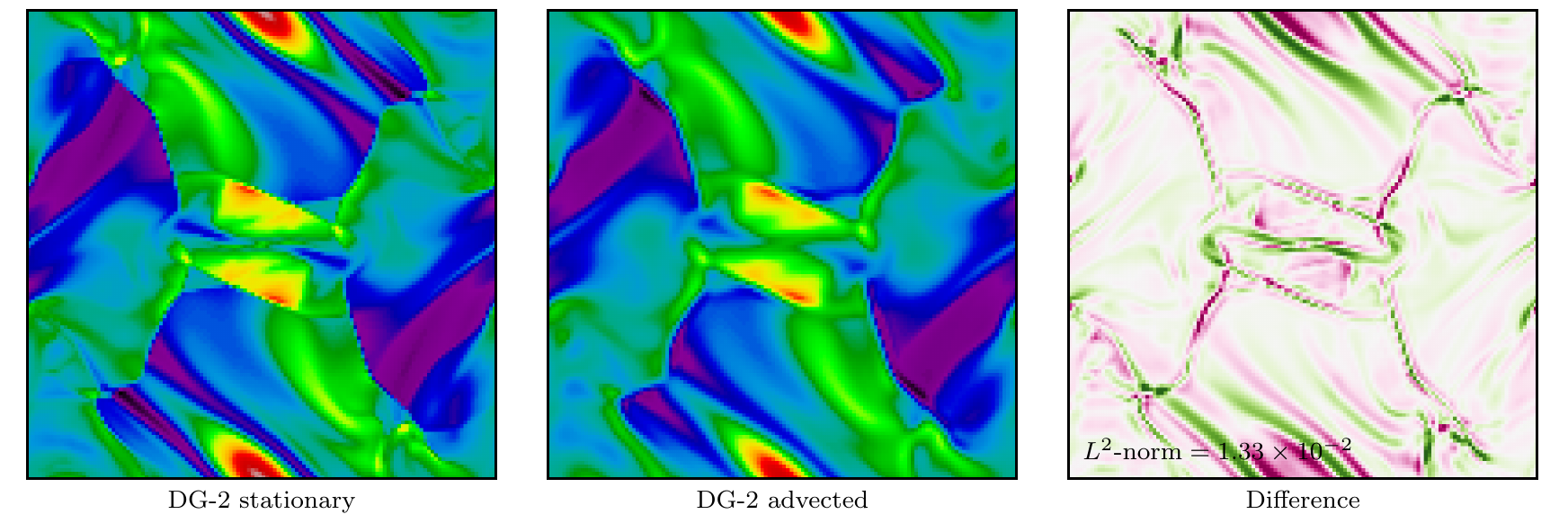}} %
        node[rotate=90] at (row1.west) {\normalsize DG-2, $128^2$, $N_\text{DoF} = 3.8 \times 10^5$} ;
    \node[below=2mm of row1] (row2) {
        \includegraphics[width=16cm,trim={0 4mm 0 0},clip]{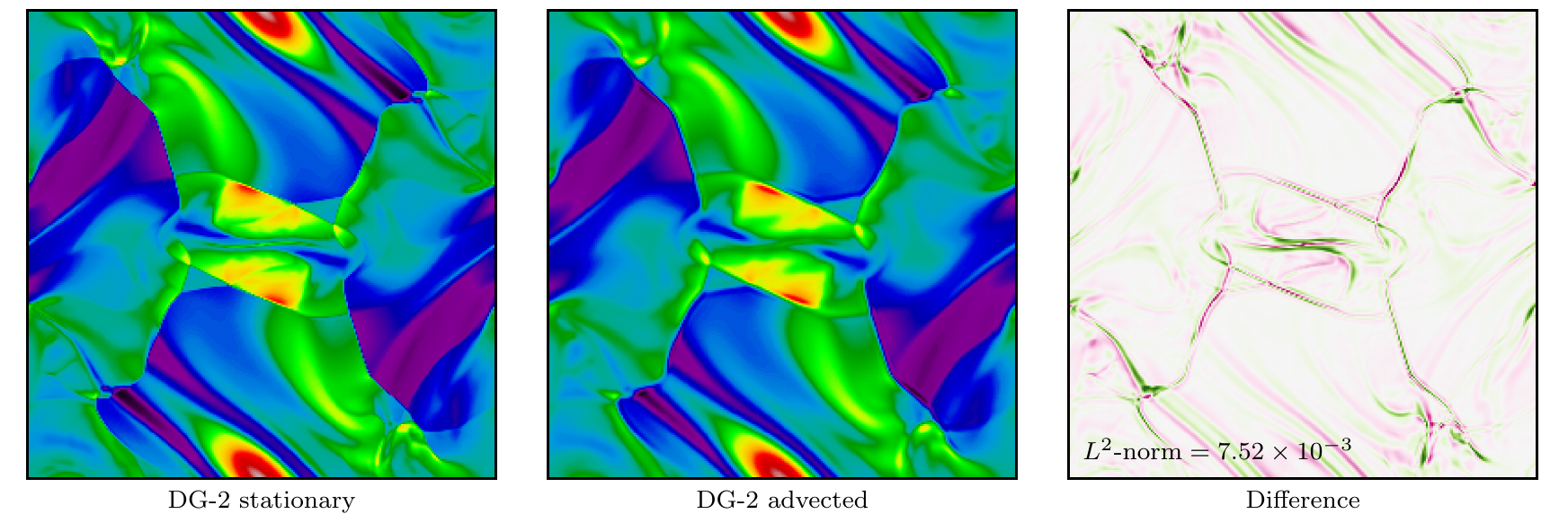}} %
        node[rotate=90] at (row2.west) {\normalsize DG-2, $256^2$, $N_\text{DoF} = 1.5 \times 10^6$} ;
    \node[below=2mm of row2] (row3) {
        \includegraphics[width=16cm,trim={0 4mm 0 0},clip]{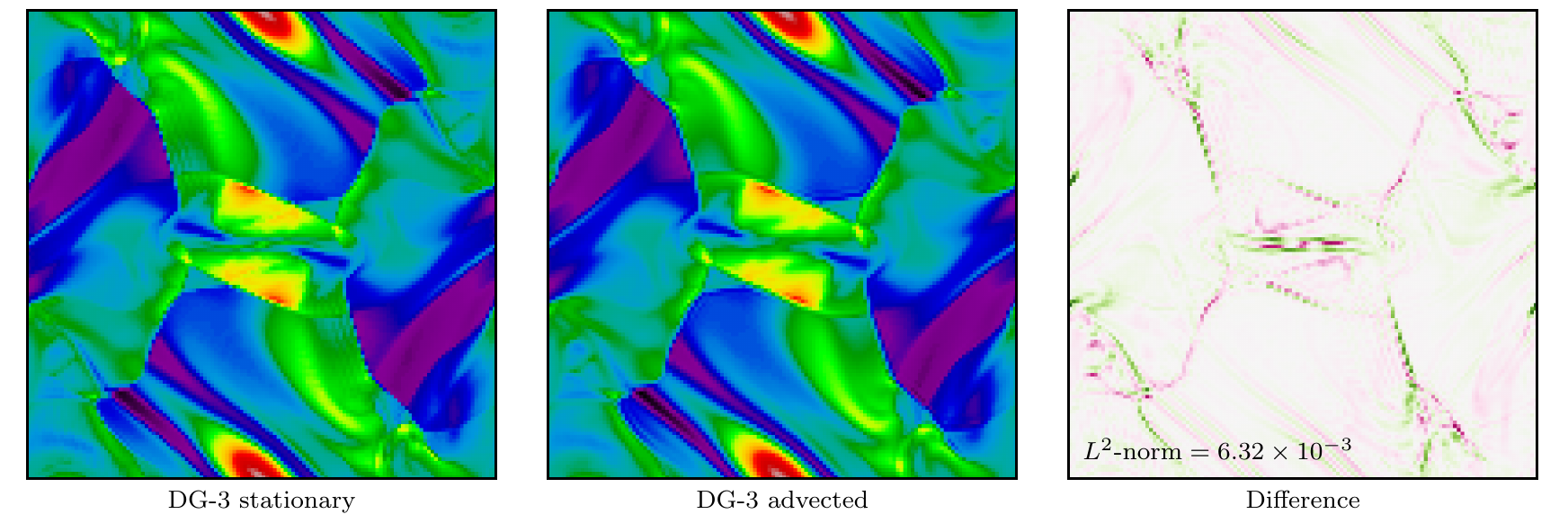}} %
      node[rotate=90] at (row3.west) {\normalsize DG-3, $128^2$, $N_\text{DoF} = 7.4 \times 10^5$} ;
    \path (row1.north west)
      ++(2.8cm, 2mm) node {\normalsize Non-advected, $\vel_0 = (0, 0, 0)$}
      ++(5.2cm, 0mm) node {\normalsize Advected, $\vel_0 = (10, 10, 0)$}
      ++(5.2cm, 0mm) node {\normalsize Difference};
  \end{tikzpicture}
  \caption{\emph{Advected Orszag--Tang vortex problem.}
    We show the density cell averages at $t=0.5$ for the non-advected Orszag--Tang vortex problem (left column), the advected problem (centre column), and the difference between the two (right column),
    for various combinations of scheme order and resolutions (rows).
    The top row is run using second-order DG at $128^2$,
    the middle row increases the resolution of the second-order method to $256^2$,
    while the bottom row uses the third-order scheme at the lower $128^2$ resolution.
    The colour scales are identical across all density maps and all difference maps.
    In the row labels, $N_\text{DoF}$ describes the total number of scalar degrees of freedom in the problem.
    This test shows the ability of higher order methods to reduce advection errors and preserve finer structures in the flow,
    even with limiting in the presence of MHD shocks, at a smaller cost in terms of degrees of freedom.
}
  \label{fig:test-orszag-tang-adv}
\end{figure*}

We present the results for advected and non-advected problems for second-order DG-2 and third-order {DG-3} in Fig.~\ref{fig:test-orszag-tang-adv}.
For this test, all runs use the same RK3 SSP time integrator.
Comparing the density contours of the non-advected and advected DG-2 solutions at resolution $128^2$ (first row),
we notice that the supersonic bulk advection velocity increases numerical diffusion significantly.
In the advected solution, the shocks are smoothed out, the density valleys are shallower,
and some of the complex structures, e.g. in the lower-right corner of the domain, are eroded.
This diffusion is clearly visible on the difference map on the right column,
which shows the deviation between the advected and the non-advected maps,
together with the $L^2$ norm of the difference.
For each run configuration, the total number of degrees of freedom $N_\text{DoF}$ is indicated at the left of the corresponding row.

The middle row of Fig.~\ref{fig:test-orszag-tang-adv} shows the same second-order scheme,
this time with a spatial resolution of $256^2$.
This increased resolution results in four times as many degrees of freedom,
and is able to significantly improve the advection errors in smooth regions of the flow.
However, the $L^2$ norm of the advection error is only reduced by a factor of about $1.8$,
at the cost of a $4\times$ increase in the number of degrees of freedom,
and $8\times$ increase in the time-to-solution due to the CFL condition.

In the bottom row, the resolution was taken back to $128^2$, but the order was increased to DG-3.
Compared to DG-2 at the same resolution, this represents only a $1.95\times$ increase in the number of degrees of freedom,
and a measured $3.2\times$ increase in the time-to-solution.
With the \nth{3} order algorithm, advection errors in smooth regions of the flow are almost completely eliminated,
and errors are also reduced close to shocks,
resulting in the lowest $L^2$ error norm and best overall preservation of Galilean invariance across the three runs.

This test shows that the reduction of advection errors offered by higher orders,
which was discussed by e.g. \cite{robertson_computationaleulerianhydrodynamics_2010}
and demonstrated with DG by \cite{schaal_astrophysical_2015} for Euler hydrodynamics problems,
carries over to complex MHD flows in presence of shocks,
even with our relatively simple slope limiting prescription.
It also shows that in some cases, increasing the order can result in notable improvements,
and at a lesser expense in terms of degrees of freedom than increasing the resolution:
in this case, DG-3 $128^2$ requires only about half as many degrees of freedom as DG-2 $256^2$,
runs more than twice as fast, and results in smaller errors.

\subsubsection{MHD rotor problem}
\label{sec:test-rotor}

We now look at the so-called 2D MHD rotor problem introduced by \cite{balsara_staggered_1999},
for which we use the more stringent ``first rotor problem'' variant of \cite{toth__2000}.
In this setup, a dense disk of fluid rotates within a static fluid background,
with a gradual velocity tapering layer between the disk edge and the ambient fluid.
An initially uniform magnetic field is present,
which winds up with the disk rotation and contains the dense rotating region through magnetic field tension.
The computational domain is $[0,1]^2$.
Initial pressure and magnetic fields are uniform in the whole domain,
with $\press=1$ and $\Bfield = (5/\sqrt{4\pi}, 0, 0)$.
The central rotating disk is defined by $r < r_0$ where $r^2 = {(x-0.5)^2 + (y-0.5)^2}$, and $r_0 = 0.1$.
Inside the disk, $\rho = 10$,
and the disk rotates rigidly with $(\vel_x, \vel_y) = (0.5-y, x-0.5)v_0/r_0$
with $v_0 = 2$.
Beyond $r > r_1$ lies the background fluid,
which has density $\rho=1$ and is at rest: $\vel_x = \vel_y = 0$.
In the annulus $r_0 \leq r \leq r_1=0.115$,
the taper region linearly interpolates between the disk and the background, with
$(\vel_x, \vel_y) = (0.5-y, x-0.5) v_0 f / r_0$
and $\rho = 1 + 9f$,
where $f = (r_1-r)/(r_1-r_0)$ is the tapering function.
The simulation runs until $t_f = 0.15$.
We use periodic boundary conditions, but note that the perturbations will not have reached the domain boundaries at $t_f$,
so reflective or inflow boundary conditions are also suitable.

\begin{figure*}
  \includegraphics[width=\textwidth]{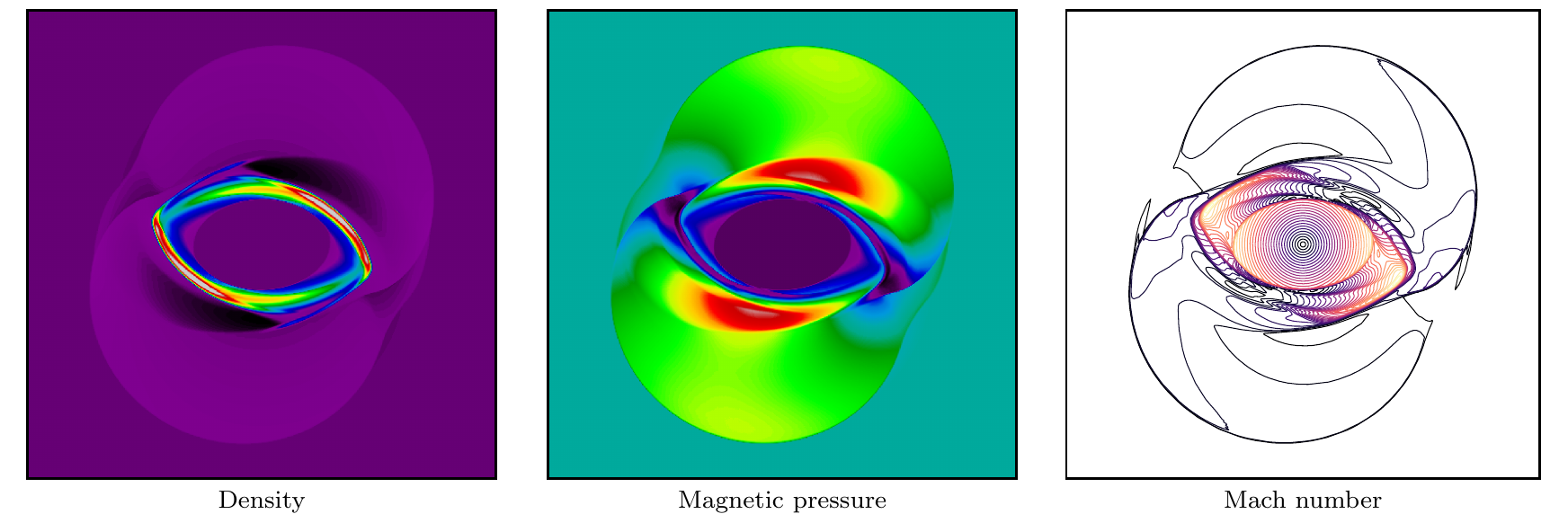}
  \caption{
    \emph{Magnetic rotor test problem}.
    The density, pressure and Mach number contours in the 2D magnetic adiabatic rotor test are shown,
  on a $512^2$ grid using the third-order Powell scheme.}
  \label{fig:test-rotor}
\end{figure*}

\begin{figure}
  \includegraphics[width=\columnwidth]{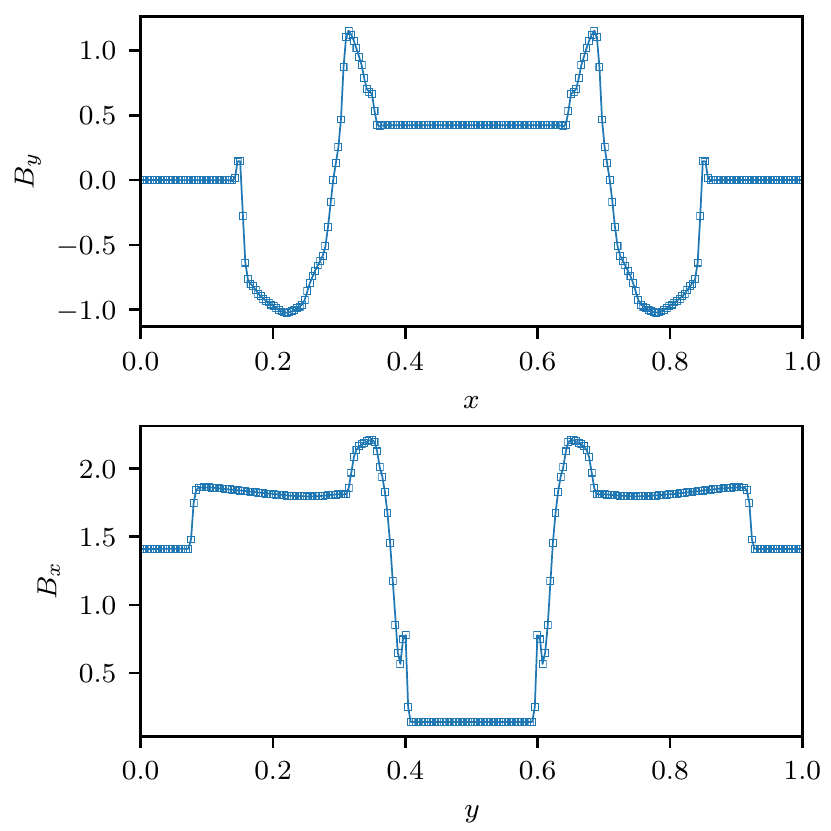}
  \caption{\emph{Slices of the magnetic field in the MHD rotor test}, along $y=0.5$ (top) and $x=0.5$ (bottom),
    computed using DG-3 with Powell terms.
    The slices can be compared to \protect\cite{stone_athena:_2008}.}
  \label{fig:test-rotor-slices}
\end{figure}

The density, magnetic pressure and Mach number contours are presented in Fig.~\ref{fig:test-rotor} for the Powell scheme.
We note that the central contours show a very good conservation of the circular rotation pattern,
which was found by \cite{toth__2000} to be challenging for some MHD schemes.
Also, there are no distortions at the outskirts of the central almond-shaped disk region,
which suggests that the magnetic field divergence is well controlled \citep[see][]{li_locally_2005}.
Fig.~\ref{fig:test-rotor-slices} presents slices\footnotemark of the magnetic field at $t_f$ across horizontal and vertical cuts through the centre of the box,
and can be compared to fig.~26 of \cite{stone_athena:_2008}
with which they show good agreement.
\footnotetext{Note that since we use an even number of grid points per dimension,
we here simply approximate the slices as the average of the two rows (or columns) of cells
immediately adjacent to the box centre line.}

\begin{figure*}
  \includegraphics[width=\textwidth]{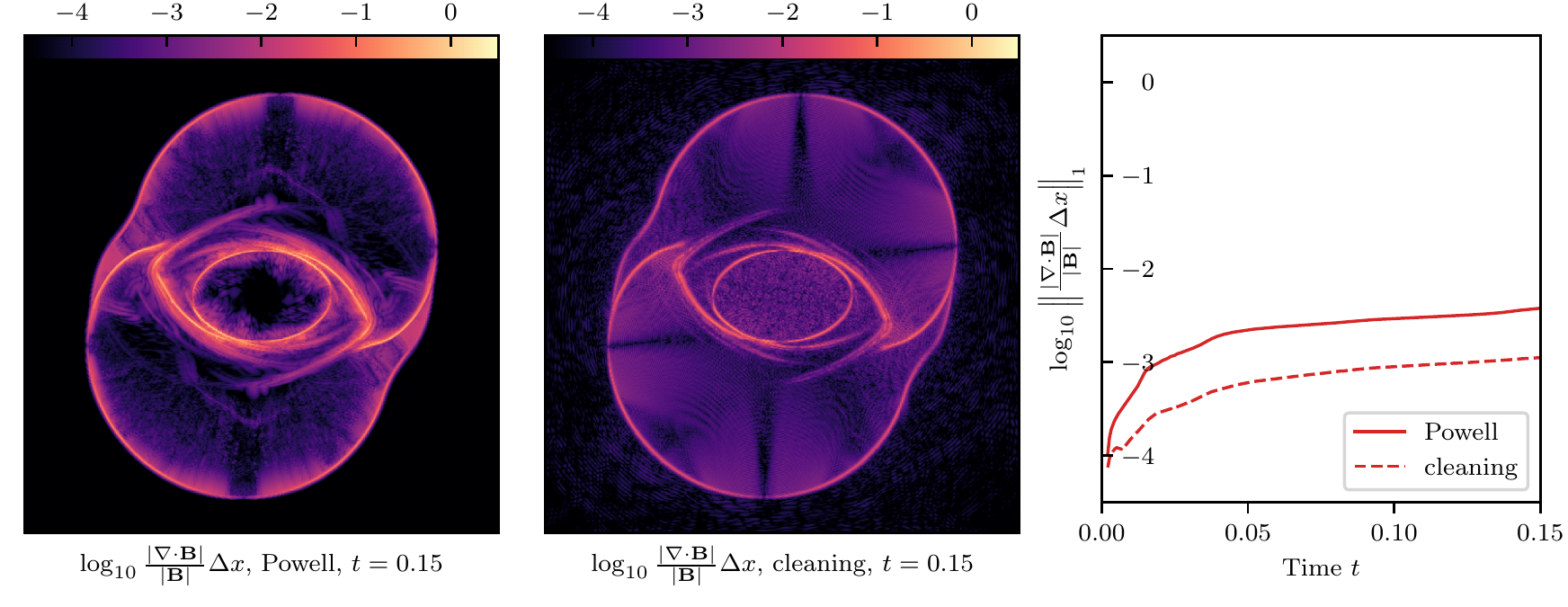}
  \caption{
    \emph{Magnetic field divergence in the rotor test problem}.
    Maps of the normalized magnetic field divergence are shown for the DG-3 scheme at $512^2$ resolution
    using the Powell (left) and divergence cleaning (centre) schemes.
  The right panel shows the time evolution of the $L^1$ norm of the magnetic field divergence for both methods.}
  \label{fig:test-rotor-div}
\end{figure*}

Finally, Fig.~\ref{fig:test-rotor-div} shows maps of the normalized magnetic field divergence
for the Powell and cleaning schemes.
Similarly to the case of the Orszag--Tang vortex presented in Fig.~\ref{fig:test-orszag-tang-div},
$\divB$ is mostly concentrated in regions around shocks.
Hyperbolic cleaning propagates the divergence faster than MHD waves,
creating ripples in the magnetic field which propagate away from the centre in the background fluid.
The time evolution of the divergence is also shown on the right panel;
both the Powell and cleaning schemes follow a very similar overall time evolution.

\begin{figure*}
  \includegraphics[width=\textwidth,trim={0 7mm 0 0mm},clip]{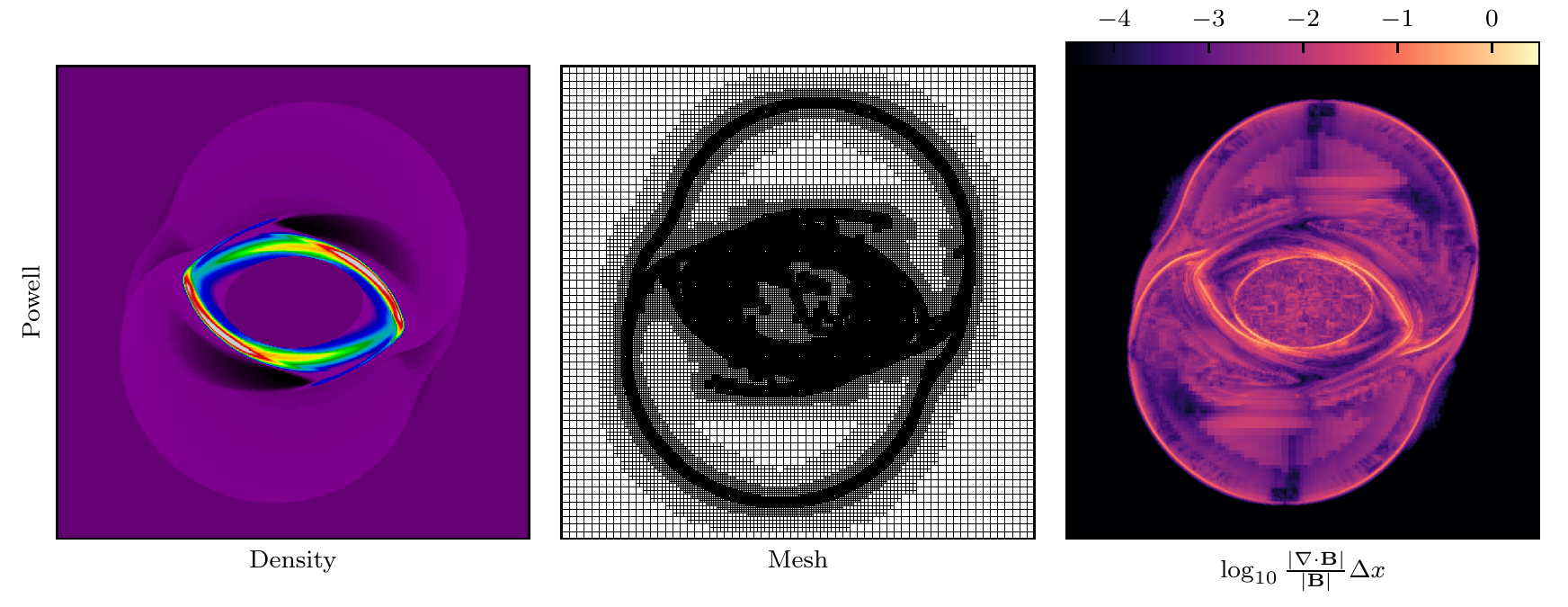}
  \includegraphics[width=\textwidth,trim={0 2mm 0 4mm},clip]{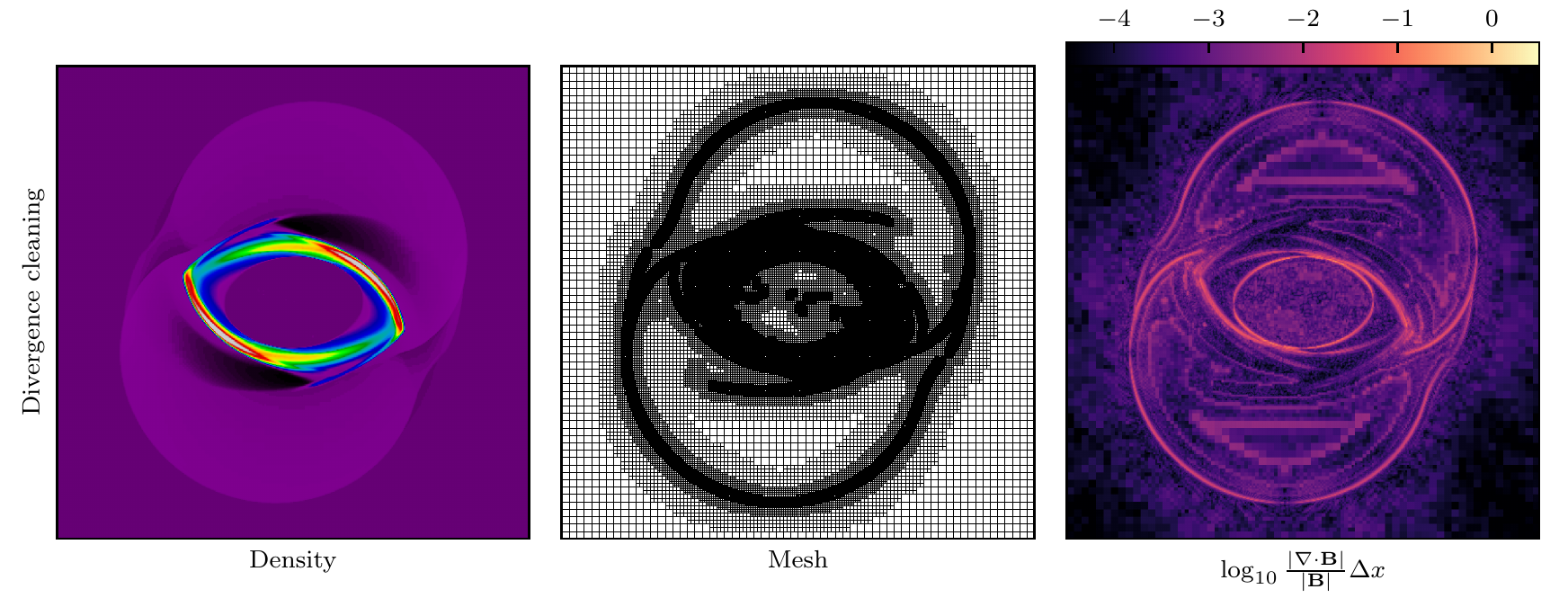}
  \caption{\emph{MHD rotor problem with adaptive mesh refinement} at $t=0.15$,
    from resolution levels 6 to 9 ($64^2$ to $512^2$) using the DG-3 Powell (top) and divergence cleaning (bottom) schemes.
    The columns show the density field (left),
    the geometry of the AMR grid (centre),
    and the normalized magnetic field divergence (right).
    The refinement is based on the density and magnetic field components.
    The density map may be directly compared to the $512^2$ Cartesian run of Fig~\ref{fig:test-rotor}.
  }
  \label{fig:test-rotor-amr}
\end{figure*}

In Fig.~\ref{fig:test-rotor-amr},
we show the MHD rotor problem with adaptive mesh refinement ($64^2$ to $512^2$),
using the DG-3 Powell and divergence cleaning schemes.
Here again, the solution may be directly compared to the $512^2$ Cartesian run of Fig.~\ref{fig:test-rotor},
and like for the Orszag--Tang vortex,
we find that the AMR solutions are in very good agreement with each other,
as well as with the Cartesian solution.
This solution can also be compared to Figures~6 and 7 from \citet{derigs_idealglmmhdentropy_2018},
which also display an AMR solution for this problem spanning the same resolution range.
Their ``no GLM'' method
is a Powell-type method (extra source terms, no divergence cleaning),
which produces divergence artefacts (see their Figure~7);
we observe no such damage in our Powell solution.

On this problem, we also see memory and time-to-solution gains with adaptive mesh refinement.
With the Powell scheme, at $t=0.15$, the AMR rotor grid of Fig.~\ref{fig:test-rotor-amr}
features $3.5\times$ fewer cells than the $512^2$ Cartesian grid
of equivalent maximum resolution of Fig.~\ref{fig:test-rotor}.
That same point in time is reached $3.4\times$ faster with the AMR simulation compared to the Cartesian mesh,
illustrating the important performance gains and memory savings that AMR can bring about already in 2D.

\subsubsection{MHD blast wave in 2D}
This test follows the evolution of MHD discontinuities in a strongly magnetized configuration.
For this test, we adapt the version of \cite{stone_athena:_2008} to a square periodic domain $[0,1]^2$.
The background fluid has zero velocity, $\rho = 1, \press = 0.1$, and magnetic field $\Bfield_0 = (B_0 \cos \alpha, B_0 \sin \alpha, 0)$ with $B_0 = 1, \alpha=45^{\circ}$. Within a radius $r_0=0.1$ of the centre of the box, we set $\press=10$, keeping all other primitive variables at their background values. We use $\gamma=5/3$, and choose $\alpha=45$ degrees.
The simulation is run until a final time of $t_f = 0.2$.

\begin{figure*}
  \includegraphics[width=\textwidth]{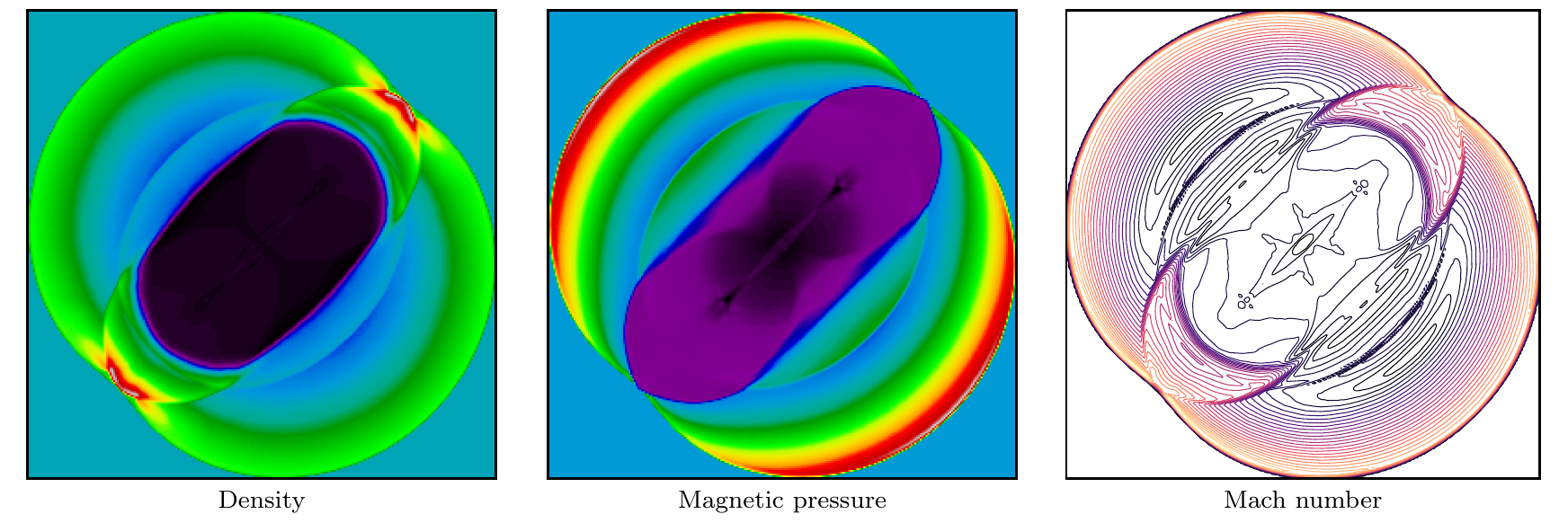}
  \caption{
    \emph{Two-dimensional MHD blast test problem}.
    The density, magnetic pressure and Mach number contours are shown on a $256^2$ grid using the third-order Powell scheme.}
  \label{fig:test-mhd-blast}
\end{figure*}

\begin{figure}
  \includegraphics[width=\columnwidth]{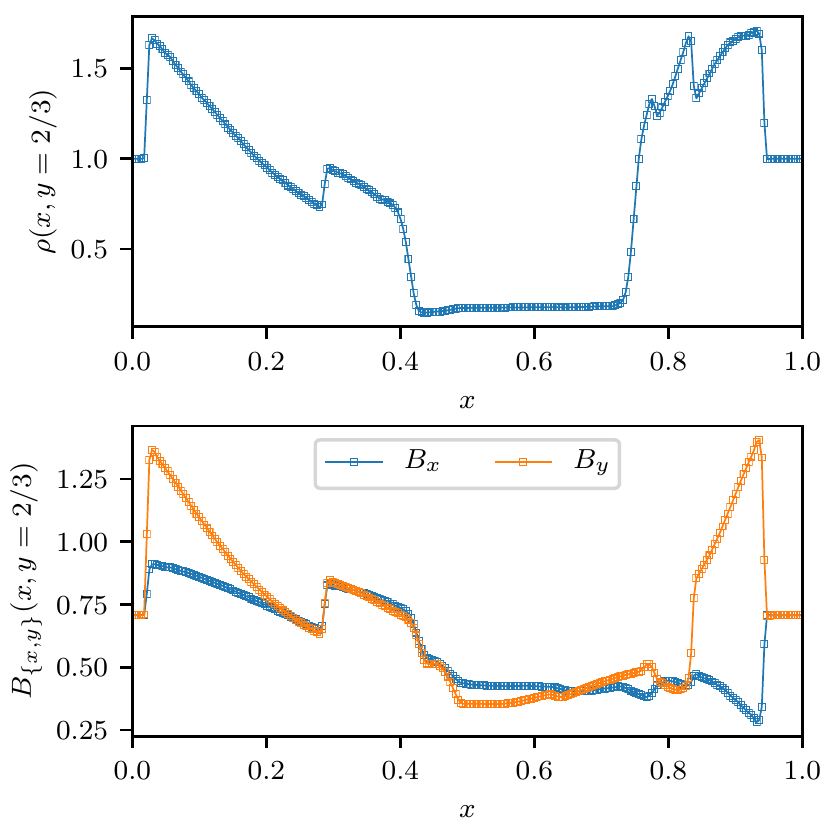}
  \caption{\emph{Slices across $y=2/3$ in the MHD blast wave test problem}, following \protect\cite{hopkins_accurate_2016},
    shown here for the DG-3 Powell scheme.
    We do not observe any of the dramatic errors noted by these authors on their tests with Powell terms.
  }
  \label{fig:test-mhd-blast-slices}
\end{figure}

Note that when running this test with $\Bfield_0$ aligned with the grid with our scheme,
small localized finger-like features develop at the shock front around the axis of symmetry parallel to $\Bfield_0$,
which we attribute to the so-called ``carbuncle'' instability \citep{robinet_shockwaveinstability_2000}.
This instability manifests itself in the case of strong shocks exactly aligned with the grid directions,
when using non-diffusive Riemann solvers.
In few highly specific simulation configurations where this problem may appear, it may be resolved by a modification of the numerical fluxes \citep[see e.g.][]{robinet_shockwaveinstability_2000,stone_athena:_2008};
however here we do not modify the scheme to counter this peculiar instability.

We also successfully run this test with $B_0 = 10$, with an initial plasma $\beta$ of $2 \times 10^{-2}$,
which demonstrates the robustness of our implementation in highly magnetized shock configurations.
In all cases, we find that the symmetry of the problem is very well preserved,
as can be seen on the Mach number contours in Fig.~\ref{fig:test-mhd-blast}.
The shocks are sharp, while post-shock regions remain smooth.

To investigate the potential impact of the Powell method on magnetic field divergence on this test problem,
we also plot slices of the magnetic field in Fig.~\ref{fig:test-mhd-blast-slices} following \cite{hopkins_accurate_2016}.
Unlike these authors, we find no evidence of errors in the components of the magnetic field with the Powell scheme for this test problem in our scheme.

\subsubsection{MHD blast wave in 3D}
To test the shock-capturing and positivity preserving performance of the code in 3D,
we adopt the 3D blast wave setup of \cite{balsara_efficient_2009}.
The computation domain is $[0,1]^3$ with periodic boundary conditions.
The background fluid is initially at rest with respect to the grid, with
$\vel=0$, $\rho=1$, and a uniform magnetic field $\Bfield = (B_0, B_0, B_0)$.
To match the setup for the slices presented in \cite{balsara_efficient_2009}, we take
$B_0 = 100 \sqrt{3}^{-1} \sqrt{4\pi}^{-1}$ in our system of units\footnotemark.
\footnotetext{Note that \cite{balsara_efficient_2009} mention $1000 \sqrt{3}^{-1}$ for $B_0$ in the text,
but their figure seems to correspond to $100 \sqrt{3}^{-1}$ instead.}
The pressure is set to $\press=0.1$ in the background,
and within a central ball of radius $r_0 = 0.1$ we set $\press=1000$ to initialize the blast.
This creates a near-infinite shock strength with a pressure ratio of $10^4$,
in a strongly magnetized background with a plasma-$\beta$ of $7.5\times 10^{-4}.$
We take $\gamma=1.4$, and run the simulation until final time $t_f=0.01$.

\begin{figure*}
  \includegraphics[width=\textwidth]{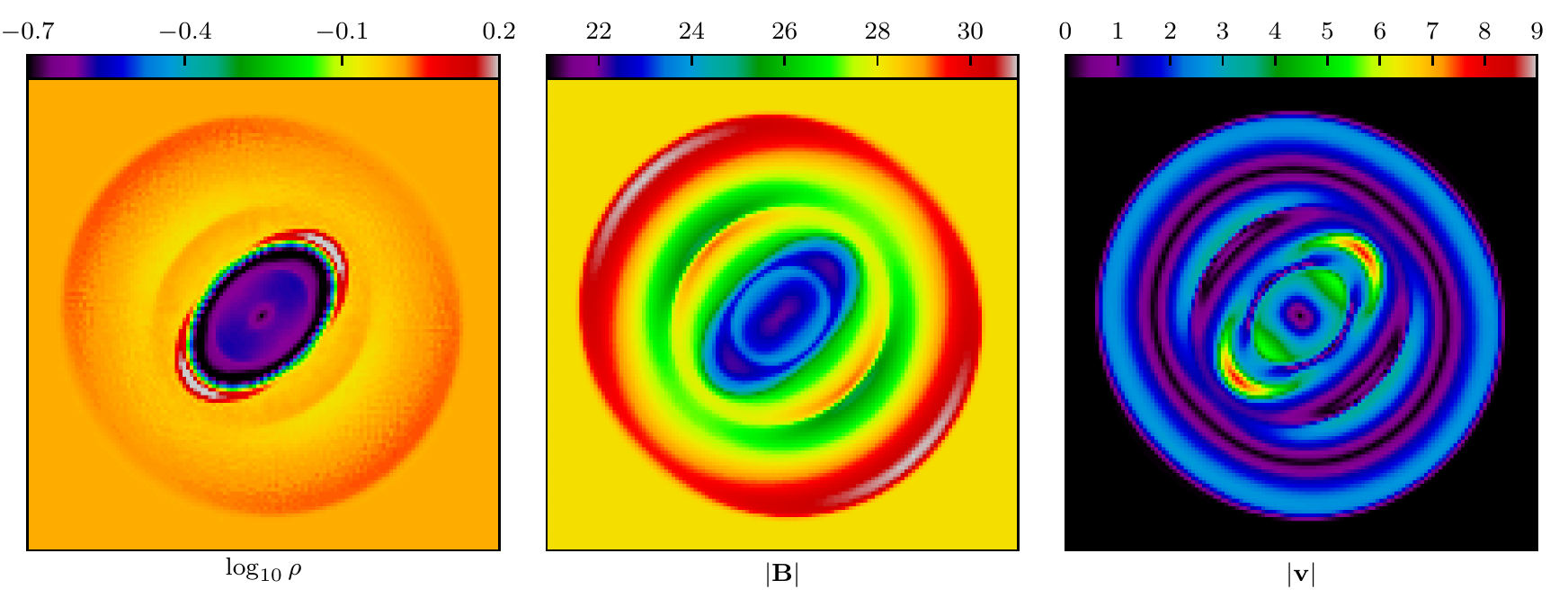}
  \caption{\emph{MHD blast wave in 3D}. Slices in the $(x,y)$ plane showing log density, magnitude of velocity, and magnitude of magnetic field at resolution $128^3$ with the third-order DG method.
  }
  \label{fig:test-blast-3d}
\end{figure*}

Fig.~\ref{fig:test-blast-3d} shows slices at constant $z$ through the centre of the blast at $t_f$,
for the third-order DG-3 method at resolution $128^3$.
Note that for slicing purposes, care must be taken to place the blast centre exactly at the centre of a cell,
which we achieve at $128^3$ resolution by shifting the centre of the high pressure ball by half a cell in every direction.
The slices of Fig.~\ref{fig:test-blast-3d} may be compared to {fig.~11} in \cite{balsara_efficient_2009},
which uses a higher resolution of $151^3$.

The DG Powell scheme is able to maintain positivity of the pressure and density in the whole domain,
while finely capturing the very strong discontinuities and resolving the complex structures in the velocity.
Note that no oscillations are visible around discontinuities.
Some numerical noise is present in the post-shock regions for the density and pressure fields,
but disappears with a more aggressive limiting threshold.
This test shows the robustness and shock-capturing behaviour of the Powell scheme for three-dimensional problems
involving very strong magnetized shocks.

This problem proves particularly challenging for our implementation of hyperbolic cleaning.
In such a low plasma-$\beta$ setup, changes in the magnetic field caused by hyperbolic cleaning
propagate fast in the ambient medium at the cleaning speed $c_h$,
and result in fluctuations of the magnetic pressure which locally cause negative thermal pressures
far away and ahead of the MHD blast wave front.
These negative pressures are successfully but aggressively corrected by our positivity limiter,
and although the computation does not crash,
the gas temperature information is destroyed,
which in turn damages the hydrodynamical solution in the background medium,
far ahead of the fast MHD shocks (depending on the exact choice of $c_h$).
Similar issues have been noted and worked on by a number of authors
\citep[e.g.][]{mignone_secondorderunsplitgodunov_2010,tricco_constrainedhyperbolicdivergence_2012,susanto_highorderfinitevolumeschemes_2014,tricco_constrained_2016};
we come back to this issue and possible solutions in the discussion.

\subsection{Divergence control problems}
\label{sec:tests-div}

We now turn to test problems more specifically aimed at evaluating the efficiency of the divergence control,
with a focus on the Powell scheme.
We have already discussed some aspects related to $\divB$ in some of the previous test problems;
we now show the stability of the Powell scheme
and some consequences of its non-conservative source terms.

On a general note,
like \cite{balsara_staggered_1999} we found that problems with strong moving shocks,
such as the blast or rotor problems,
are not necessarily the most stringent tests of divergence control.
Instead, colliding shocks whose convergence front is at rest with respect to the grid%
---such as the rotated shock tube of \cite{toth__2000} described in \ref{sec:test-rotated-tube},
or some shock-shock interactions in the Orszag--Tang vortex of Section~\ref{sec:test-orszag-tang}---%
proved to be much more challenging tests of the divergence control scheme and its stability in particular.

In addition to shock interactions, we found that smooth problems such as the simple advection of a magnetic field loop
can also be unstable with inappropriate discretizations of the Powell term,
and their smooth character makes it easier to follow the development of the divergence instability.
Without Powell source terms, the instability usually grows faster with scheme order,
as the numerical diffusion at lower orders helps slow down the divergence runaway.

\subsubsection{Loop advection}
\label{sec:test-loop}

This test follows the advection of a magnetic field loop after \cite{gardiner_unsplit_2005}.
On the periodic domain $[-1,1]^2$, the background fluid has $\rho=1, \press=1$,
and a global advection velocity $(\vel_x, \vel_y) = (2, 1)$
so that the ambient flow is not aligned with grid directions.
Letting $r$ be the radial distance to the centre of the box,
the magnetic field is initialized from a vector potential $\bm A = (0, 0, A_z(r))$ with $\Bfield = \bm \nabla \times \bm A$.
To define a magnetic field loop of radius $r_0=0.3$, we set $A_z(r) = \max(0, A_0 (r_0 - r))$.
Taking $A_0 = 10^{-3}$, we obtain a very weakly magnetized configuration with a plasma $\beta$ of order $10^6$,
in which the magnetic field is essentially a passive scalar.
For this field configuration, the MHD current vanishes everywhere, except at $r=0$,
and $r=r_0$ where the corresponding current line and return current tube are singular.

The aim of the test is to verify that the current loop is advected without deformation or noise,
and to monitor the time evolution and dissipation rate of the total magnetic energy,
following \cite{gardiner_unsplit_2005,stone_athena:_2008}.

\begin{figure*}
  \includegraphics[width=\textwidth]{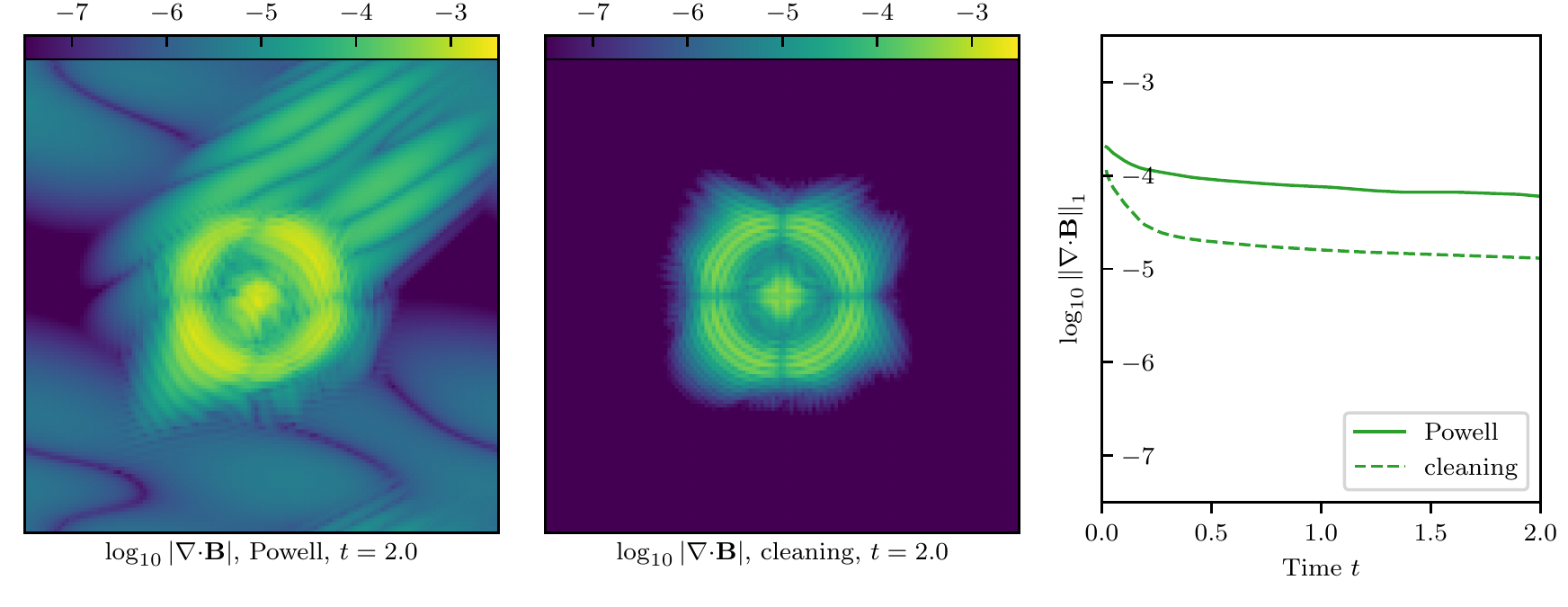}
  \caption{
    \emph{Magnetic field divergence in the loop advection test problem}.
    Maps of the non-normalized magnetic field divergence are shown for the DG-3 scheme at $128^2$ resolution
    using the Powell (left) and divergence cleaning (centre) schemes.
    The right panel shows the time evolution of the $L^1$ norm of the magnetic field divergence for both methods.
    Because the magnetic field vanishes outside of the loop,
    we do not normalize by the local magnetic field strength,
    and instead plot the absolute divergence in code units,
    given by equations \eqref{eq:divb-cell} and \eqref{eq:divb-total} for the maps and time evolution respectively.
  }
  \label{fig:test-loop-div}
\end{figure*}

As discussed in \cite{gardiner_unsplit_2005},
the linearized dynamics of the magnetic field involves the diagonal derivatives $\partial \Bfield_x / \partial x$,
and this test is therefore sensitive to the $\divB$ treatment,
although this is mostly an issue for directionally split methods.
In any case however, the linearized loop advection setup constitutes a so-called resonant hyperbolic problem,
as discussed in \cite{kemm_origin_2013},
and therefore constitutes a good test for the growth of divergence instabilities.
We found this test to be particularly unstable without divergence control,
even with a smooth non-singular loop configuration.
Fig.~\ref{fig:test-loop-div} shows divergence maps for the loop advection problem
for both the Powell and cleaning schemes,
together with the time evolution of the divergence.
With both schemes, the numerical divergence is well under control.

\begin{figure*}
  \includegraphics[width=\textwidth]{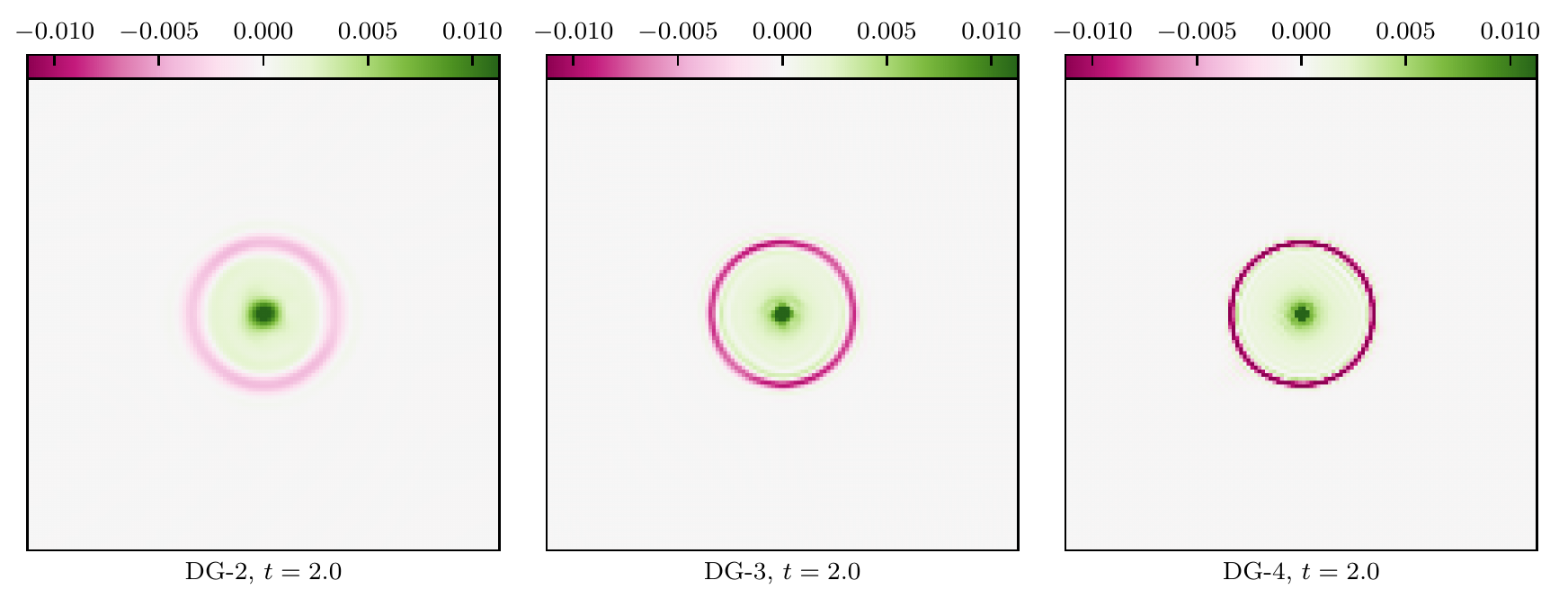}
  \caption{
    \emph{MHD current in the loop advection problem}.
    The average $z$ component of current density $\bm j = \bm \nabla \times \Bfield$ in each cell
    is shown for the Powell DG-2, DG-3 and DG-4 schemes at resolution $128^2$,
    at $t=2$ after two horizontal domain crossings.
    The code preserves the shape of the current loop very well at all orders,
    and the test demonstrates the qualities or higher order schemes for reducing numerical diffusion.
  }
  \label{fig:test-loop}
\end{figure*}

Fig.~\ref{fig:test-loop} shows the $z$ component of the cell average of the current density $\bm j = \bm \nabla \times \Bfield$
at final time $t=2$ after two horizontal domain crossings, at resolution $128^2$, for orders 2, 3 and 4.
For this comparison, we use the RK3 SSP time integrator across all spatial orders.
The current density is a stringent diagnostic since, being a derivative of the magnetic field,
it is very sensitive to noise and local fluctuations.
The scheme preserves the exact circular shape of the current loop at all orders,
with very little noise and oscillations in the current at all orders.
In addition, the reduction in numerical diffusion and advection errors is clearly noticeable as order is increased.
We also see that the current loop gets increasingly resolved at higher orders,
showing an increase in effective resolution even for singular features.

\begin{figure}
  \includegraphics[width=\columnwidth]{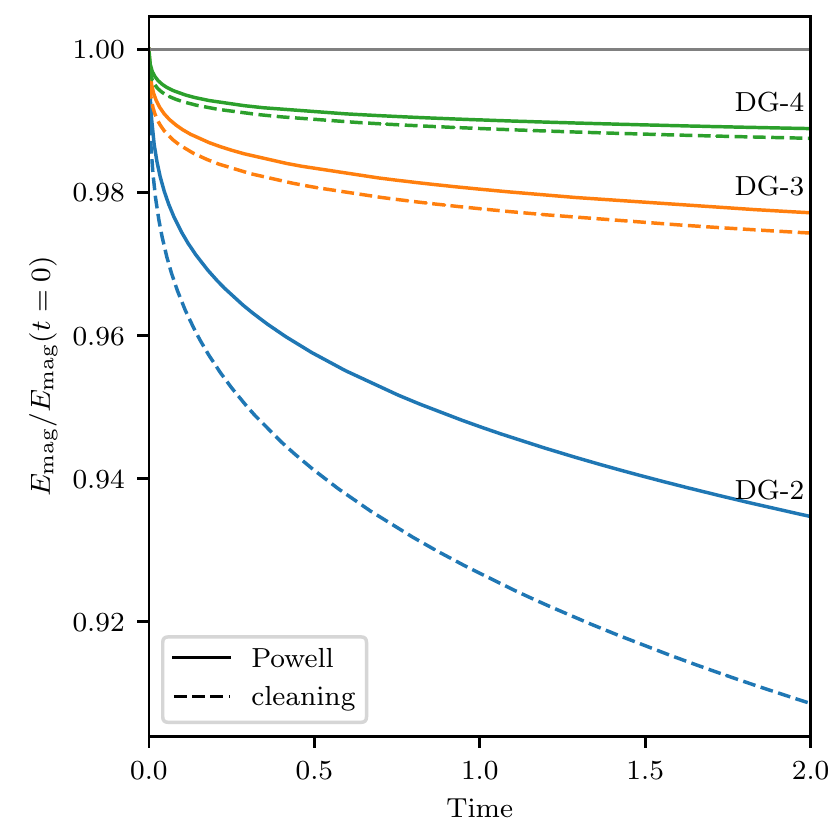}
  \caption{\emph{Time evolution of the magnetic energy in the field loop advection test},
    for the second, third, and fourth order schemes, at resolution $128^2$.
    The solid lines correspond to the Powell scheme,
    whereas the dashed lines use hyperbolic divergence cleaning.
  }
  \label{fig:test-loop-energies}
\end{figure}

The plot of Fig.~\ref{fig:test-loop-energies} follows the time evolution of the normalized total magnetic energy
during two horizontal domain crossings, at resolution $128^2$.
The results are shown for the Powell and hyperbolic cleaning schemes.
The hyperbolic cleaning scheme seems slightly more diffusive, especially at lower orders.
The corresponding decay time-scales may be fitted with the power law functional form proposed in \cite{gardiner_unsplit_2005}:
$E_\textrm{mag}(t) = A(1 - (t/\tau)^\alpha)$.
For the Powell scheme, we measure
$\tau \approx 1.8\times10^4$ for DG-2,
$\tau \approx 6.1\times10^7$ for DG-3, and
$\tau \approx 6.0\times10^{11}$ for DG-4,
which quantifies the strong reduction of the decay rate $\tau^{-1}$ with increasing scheme order.

\subsubsection{Current sheet problem}
\label{sec:test-current-sheet}

A very stringent test of the capability of the code to handle numerical divergence and maintain positivity is the so-called current sheet problem \citep{hawley_mocct:_1995,gardiner_unsplit_2005},
which follows the evolution of sharp interfaces between regions with magnetic fields of opposing directions.
The chosen fluid velocities will push and pull the interfaces along their normal direction,
letting the magnetic tension act as a restoring force which results in Alfv\'en oscillations.
At the interfaces---where the tangential component of the magnetic field flips sign---%
numerical reconnection will occur,
forming growing isolated islands of reconnected magnetic regions.
Similar configurations have also been used in the context of non-ideal MHD to measure reconnection rates
\citep[see e.g.][and references therein]{marinacci_nonidealmagnetohydrodynamicsmoving_2018}.

In our implementation of this test,
the computational domain is $[0, 1]^2$ with periodic boundary conditions.
The density and pressure are uniform with $\rho=1$, $\press=\frac{1}{2} B_0^2 \beta$.
We pick $B_0 = 1/\sqrt{4\pi}$ for our units, and we set the plasma-$\beta$ to $\beta=0.1$,
so that the configuration is strongly magnetized.
The domain is split into three regions forming vertical bands,
separated by two vertical interfaces located at $x=0.25$ and $x=0.75$.
In the left ($x < 0.25$) and right ($x > 0.75$) bands, we initially set $\Bfield = (0, B_0, 0)$.
In between the left and right interfaces, we initialize $\Bfield = (0, -B_0, 0)$,
so that the vertical (tangential) component of the magnetic field flips across each interface.
The initial velocity field is chosen as $\vel = (A \sin(2\pi y), 0, 0)$ with $A=0.1$.
The adiabatic constant is set to $\gamma = 5/3$,
and the evolution of the interfaces is followed until $t_f=10$.

Note that this setup corresponds to the results presented in the {\athena} code tests web page{\footnotemark},
\footnotetext{\url{https://www.astro.princeton.edu/~jstone/Athena/tests/current-sheet/current-sheet.html}.}
and is more stringent than the version of \citet{gardiner_unsplit_2005}:
the plasma-$\beta$ is twice as low, and the oscillations are chosen with a larger amplitude.

\begin{figure*}
  \includegraphics[width=\textwidth,trim={0 0mm 0 0mm},clip]{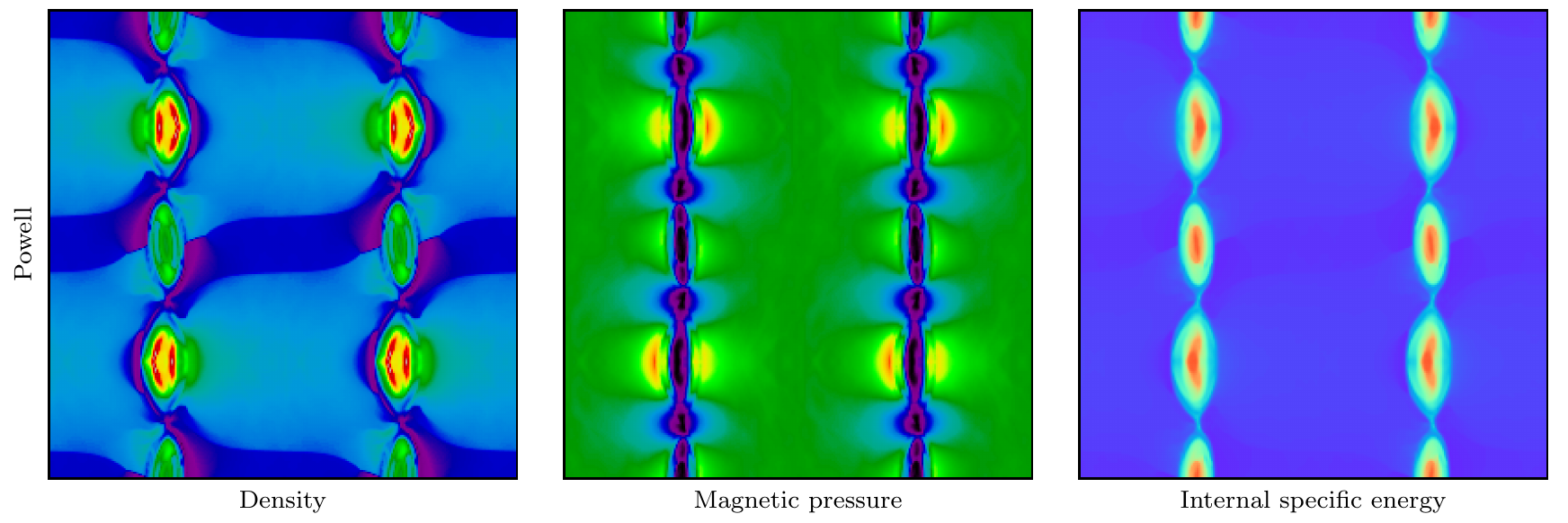}
  \includegraphics[width=\textwidth,trim={0 4mm 0 1mm},clip]{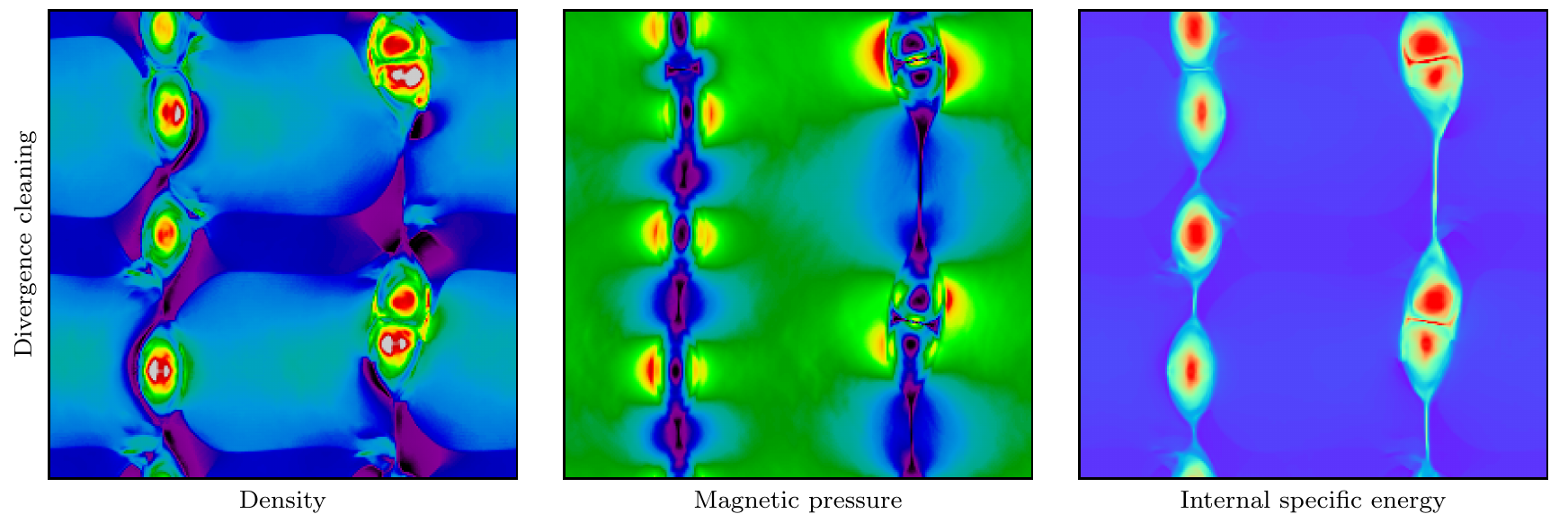}
  \caption{
    \emph{Current sheet test problem}.
    Maps of the density (left column),
    magnetic pressure (centre column) and internal specific energy (right column)
    are presented at final time $t=10$
    in the current sheet test problem, using the DG-3 scheme at $256^2$ resolution.
    The Powell scheme (top row) and divergence cleaning schemes (bottom row)
    can be compared directly.
    The divergence cleaning scheme produces a breaking of the problem symmetry
    resulting in reconnection islands coalescing.
  }
  \label{fig:test-current-sheet-maps}
\end{figure*}

The code is able to stably follow the evolution of the current sheet up to the final time $t=10$
with both the Powell and hyperbolic cleaning schemes.
The maps of the density, magnetic pressure, and internal specific energy
at final time $t=10$ are shown in Fig.~\ref{fig:test-current-sheet-maps}
for both divergence control methods.
Outside of the reconnection islands along the current sheets,
the two solutions are in very good agreement.
The main differences can be seen along the current sheets,
where numerical reconnection occurs at the visible ``nodes''.
The Powell scheme still maintains the symmetry of the problem perfectly at $t=10$,
unlike divergence cleaning for which reconnection islands have started merging along the two current sheets
due to numerical noise.
Note that the Powell scheme results in slightly lower temperatures inside the nodes.

\begin{figure*}
  \includegraphics[width=\textwidth]{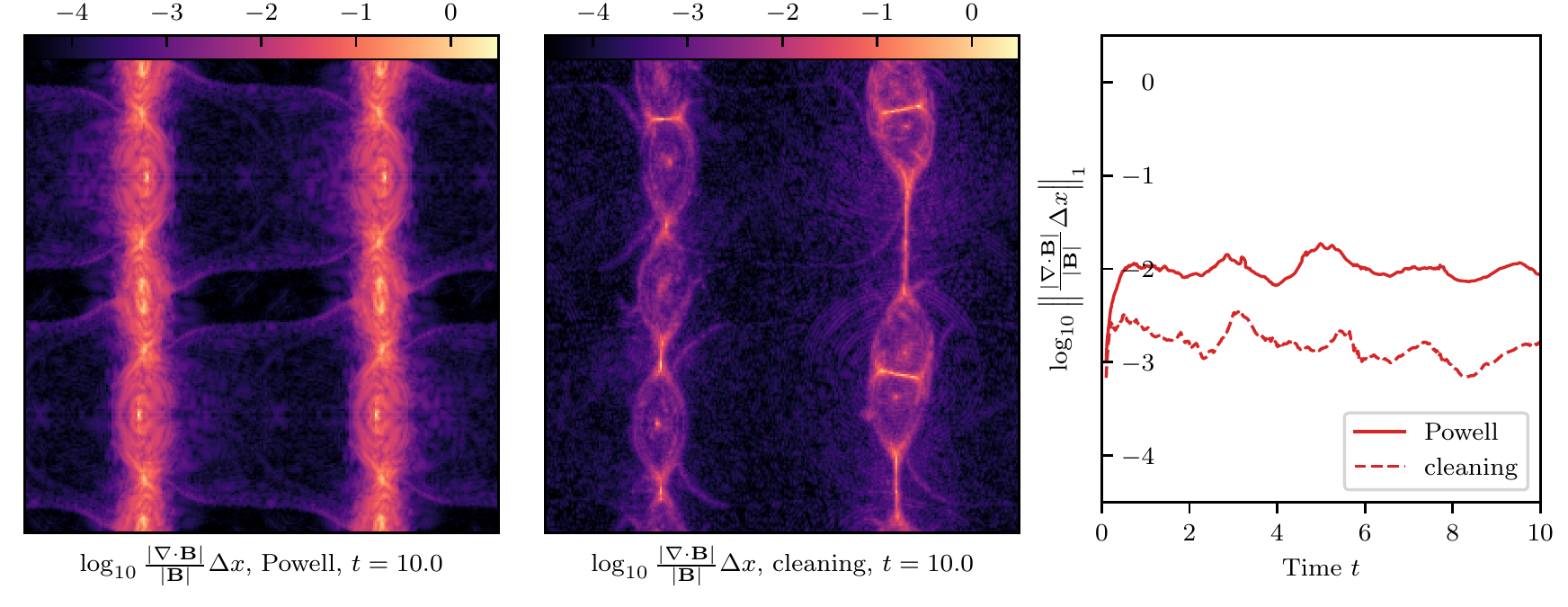}
  \caption{
    \emph{Magnetic field divergence in the current sheet test problem}.
    Maps of the normalized magnetic field divergence are shown for the DG-3 scheme at $256^2$ resolution
    using the Powell (left) and divergence cleaning (centre) schemes.
    The right panel shows the time evolution of the $L^1$ norm of the magnetic field divergence for both methods.
    The magnetic pressure maps of Fig.~\ref{fig:test-current-sheet-maps}
    show that $|\Bfield|$ vanishes at the nodes along the current sheet,
    which contributes to the large values of the normalized divergence
    at the heart of the reconnection islands.
}
  \label{fig:test-current-sheet-div}
\end{figure*}

Divergence maps, as well as the time evolution of the divergence,
are presented in Fig.~\ref{fig:test-current-sheet-div}.
The divergence is concentrated around the two current sheets,
at the sites of strong numerical reconnection.
%Interestingly, both methods produce noticeably different reconnection islands,
%due to a breaking of the solution symmetry occurring with the cleaning scheme.
The level of numerical divergence for both schemes is stable in time;
here again,
the divergence is about an order of magnitude higher with the Powell method compared to the cleaning scheme.

\begin{figure}
  \includegraphics[width=\columnwidth]{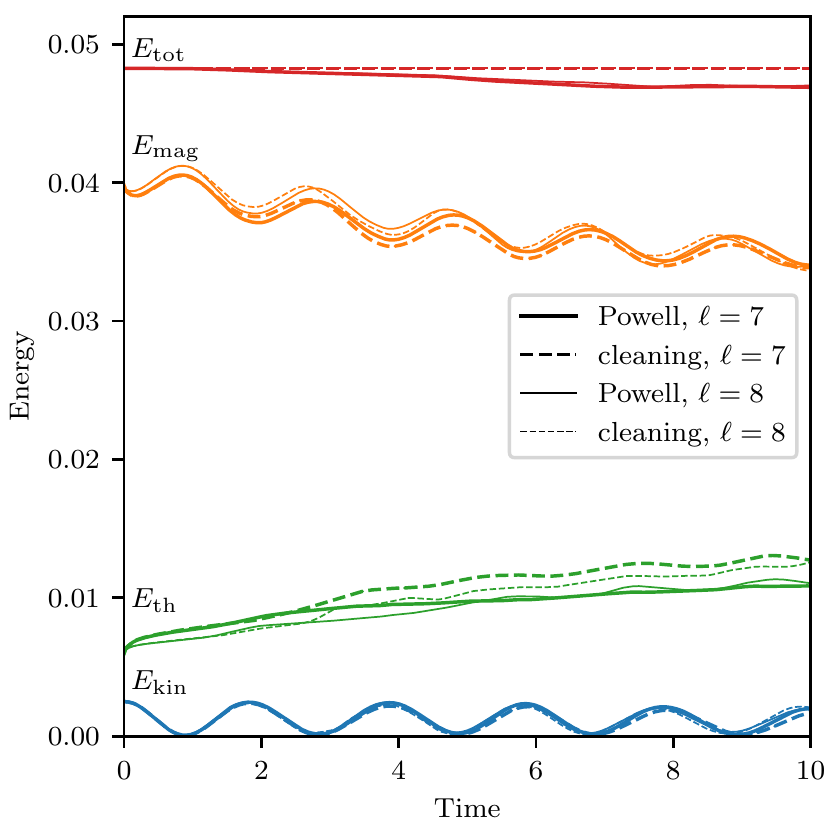}
  \caption{\emph{Time evolution of energies in the current sheet test}.
    The evolution of energies is shown for third-order DG scheme,
    computed with the Powell method (solid lines) and hyperbolic divergence cleaning (dashed lines),
    at resolution levels 7 ($128^2$ grid, thick lines) and 8 ($256^2$ grid, thin lines).
    The transfer of magnetic energy $E_\text{mag}$ into thermal energy $E_\text{th}$ is
    caused by numerical reconnection.
  }
  \label{fig:test-current-sheet}
\end{figure}

Fig.~\ref{fig:test-current-sheet} shows the evolution of the kinetic, thermal, magnetic and total energies
in the current sheet problem,
computed with both divergence control methods using the third-order DG-3 scheme,
at resolutions levels 7 and 8.
The evolution of the kinetic and magnetic energies are similar for both divergence control schemes:
while they exhibit some differences,
these are comparable in magnitude to the effect of resolution,
and it is therefore difficult to attribute them to the divergence control scheme.

However, the Powell scheme introduces noticeable non-conservation of the total energy (red curve),
of the order of 2--3\%,
which does not seem to disappear with increasing resolution.
This, in turn, translates into a deviation of the thermal energy,
whose relative magnitude can be severe for such strongly magnetized configurations.
In this case, the Powell scheme results in artificial cooling of the plasma,
which is consistent with the temperature maps of Fig.~\ref{fig:test-current-sheet-maps}.
%%%
This test is very challenging for the Powell method:
while initially the velocity and magnetic fields are orthogonal everywhere,
as soon as reconnection islands start to form, 
regions of collinear $\vel$ and $\Bfield$ will quickly develop.
In such regions, the energy injection component $\dotp{\vel}{\Bfield}$ of the Powell source term \eqref{eq:powell-q} will reach maximal magnitude,
and will be very sensitive to any local $\divB$ introduced in the numerical solution.

Finally, we note that the slope limiter has a marked overall impact on this problem.
First, the amount of energy injected by the Powell scheme seems to strongly depend on the choice of limiter threshold:
making the limiter more aggressive by decreasing $\tilde M=5$ down to $\tilde M=1$
reduces the deviation by more than one order of magnitude.
We advise that this complicates the comparison of the energy deviations at the two resolution levels,
as our choice of scaling of $M$ or $\tilde M$ with resolution may impact the results.
In addition, we see very little difference between the second and third-order DG-2 and DG-3 schemes for this problem.
A possible explanation is related to our second-order limiter,
which will trigger locally on the sharp discontinuities at the interfaces,
thereby degrading the scheme to at most second order precisely in places where numerical reconnection occurs.
It is therefore possible that
a more sophisticated high-order limiter could both improve energy conservation with the Powell scheme,
and provide clearer benefits of higher-order methods for this particular test configuration.
We discuss potential limiter improvements left for future work in Section~\ref{sec:discuss-limiters}.

\subsubsection{Rotated shock tube}
\label{sec:test-rotated-tube}

This problem, adapted from \citet{toth__2000}, is designed specifically to test the impact of Powell terms on jump conditions in the situation of a strong shock at an angle with the grid.
On the computational domain $[-1,1]^2$ with periodic boundary conditions,
we define a rotated frame at an angle $\theta = \arctan(2)$ with the lab frame $(x, y)$,
and whose origin lies at the box centre $(0, 0)$,
We label coordinates and vector components in the rotated frame with $(\parallel, \perp)$.
In this coordinate system, we define an interface at $x_\parallel = 0$,
where we initialize a shock.
For $x_\parallel < 0$, we set the velocity $(\vel_\parallel, \vel_\perp, \vel_z) = (10, 0, 0)$ and pressure $\press = 20$,
whereas for $x_\parallel \geq 0$ we initialize
$(\vel_\parallel, \vel_\perp, \vel_z) = (-10, 0, 0)$ and $\press = 1$.
This sets up two fast converging flows colliding at $x_\parallel = 0$.
The density is set to $\rho=1$ everywhere,
and the initial magnetic field is uniform with
$(\Bfield_\parallel, \Bfield_\perp, \Bfield_z) = (5, 5, 0)/\sqrt{4\pi}$.
We take $\gamma = 5/3$, and run the simulation until final time $t_f = 0.08 / \sqrt{5}$.

Note that in comparison to \citet{toth__2000},
we run the problem on a square domain with periodic boundary conditions for simplicity.
This will create shocks at domain boundaries,
so we use a bigger domain of size 2 to avoid contamination of the central region.
Correspondingly, we run at a higher spatial resolution of $512^2$,
focusing only on the central $256^2$ region of the domain which by time $t_f$ is not yet impacted by the boundary shocks.
As such, our results are directly comparable to \citet{toth__2000}.

In the configuration of this test problem, the $\divB = 0$ constraint imposes that
the component $\Bfield_\parallel$ of the magnetic field,
which is parallel to the shock propagation direction and normal to the interface,
should remain exactly at its initial constant value.
Achieving this is particularly challenging for Powell schemes,
because the angled shock will introduce jumps in both normal components of $\Bfield$,
and the resulting numerical divergence will act as sources and sinks
for conserved quantities through the Powell terms,
thereby impacting the jump conditions.
In addition, the initial shock at the convergence of the two regions is close to stationary with respect to the grid,
and as a result, the magnetic field divergence cannot be efficiently advected away by the Powell terms.

\begin{figure}
  \includegraphics[width=\columnwidth]{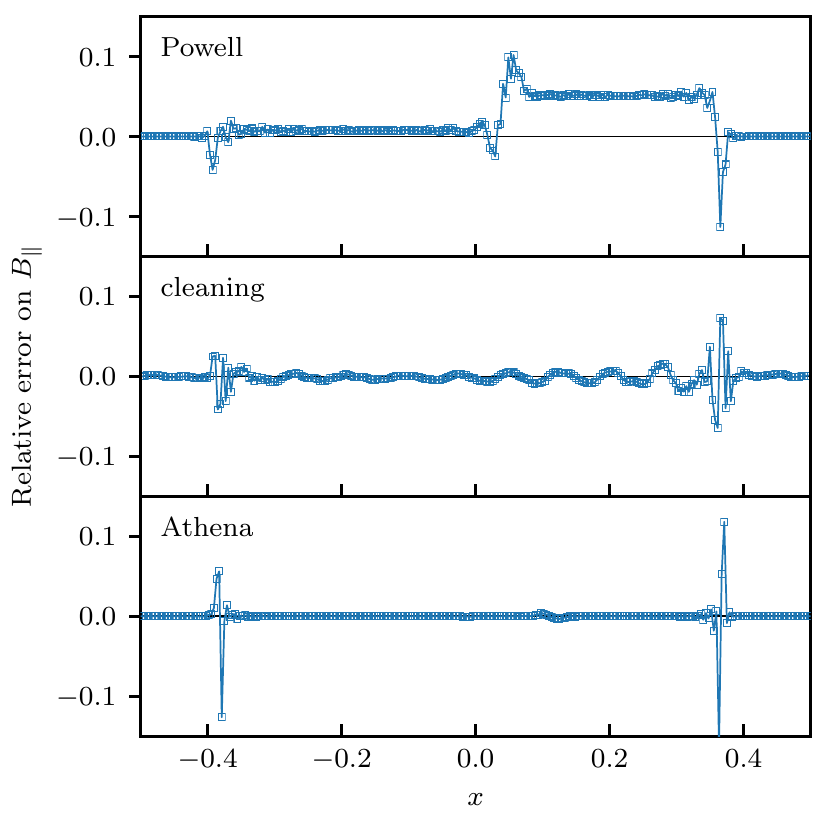}
  \caption{\emph{Rotated shock tube test}.
    The relative error on the parallel magnetic field $\Bfield_\parallel$ in the rotated shock tube test of \protect\cite{toth__2000}
    is shown for the DG-3 method with the Powell and hyperbolic cleaning schemes in the top and centre panel.
    The bottom panel shows the solution obtained with Athena, with the \nth{3}~order CTU (constrained transport) method.
    For the Powell method, between the left and right fast shocks ($x = \mp 0.4$), the value of $\Bfield_\parallel$ deviates with respect to the exact solution due to the nonconservative source terms.
    Apart from oscillations at the discontinuities, the Powell scheme results in systematic offsets of $\Bfield_\parallel$ of order $\lesssim 5$\%.
    The hyperbolic cleaning scheme features no such systematic offsets,
    but produces damped oscillations which overtake the fast shocks as the divergence gets advected away.
  }
  \label{fig:test-rotated-shocktube}
\end{figure}

Fig.~\ref{fig:test-rotated-shocktube} shows the errors on $\Bfield_\parallel$ across a slice along $x$ cutting though the shock region, for the DG-3 scheme.
Both the Powell and hyperbolic cleaning schemes are shown.

For the Powell method, we find systematic errors in the 5\% range in the inter-shock region,
consistent with the results of \cite{toth__2000}.
Our measured errors seem however lower than those reported by \cite{mignone_secondorderunsplitgodunov_2010} in their similar test,
where they measured relative errors greater than 10\%.
The Powell error may potentially be reduced by using a Riemann solver which propagates the divergence wave---for instance,
the 8-wave Rusanov flux produces much smaller errors, but it is also extremely diffusive.
We also noted in our experiments that the choice of face flux (see \ref{sec:normal-bx-choice})
plays a role in determining the location of the plateaus visible in Fig.~\ref{fig:test-rotated-shocktube},
and it is possible that $\Bnormal_F$ may be chosen optimally as to minimize the absolute deviations in this particular test.
Overall, this shows that DG cannot work around some of the pitfalls of nonconservative Powell schemes,
for the fundamental reason that shocks are essentially first-order features.

The solution obtained with hyperbolic cleaning is also represented in Fig.~\ref{fig:test-rotated-shocktube}.
Unlike the Powell solution, it does not feature systematic shifts in the value of $\Bfield_\parallel$,
however, large-scale oscillations due to the ongoing advection and damping of the divergence are visible.
More importantly, these oscillations also occur for $| x_\parallel | \gtrsim 0.4$,
outside of the region between the two shock fronts,
which is an unphysical effect,
and illustrates the fact that hyperbolic cleaning will propagate information faster than any physical MHD wave.

\section{Discussion}
\label{sec:discussion}

\subsection{High-order schemes for astrophysics}
\label{sec:discuss-hi-order}

Our tests have shown some promising results for higher-order schemes in astrophysical simulations,
and for MHD in particular.
% DG w/ AMR
We now discuss how the combination of DG with adaptive mesh refinement could allow very efficient computations in smooth regions of the flow with high order convergence,
while finely capturing the shocks and discontinuities with spatial refinement.

% order vs resolution convergence
We can interpret order/resolution convergence plots such as Fig.~\ref{fig:conv-smvrtx-2d} from two complementary points of view:
looking at a given spatial resolution on the $x$ axis, we can increase the order to reduce the solution error.
But at a fixed error on the $y$ axis, we may also increase the order and correspondingly reduce the spatial resolution.
We argue that this second vision is more relevant to many types of simulations in astrophysics,
as the spatial truncation errors need only be smaller or comparable to other types of errors,
stemming from uncertainties in the physical models, missing physics, subgrid recipes, etc.
We can therefore see order convergence as a way of getting away with fewer cells in smooth problems,
to the extent that we can efficiently ``patch'' smooth regions of the flow with coarser cells.

Cell-based adaptive mesh refinement provides a suitable framework to do this,
as it allows cell-by-cell resolution adaptivity to match the local feature size.
In a number of our test problems, we found higher-order to better resolve features close to the grid resolution.
The vortex problem of Fig.~\ref{fig:conv-smvrtx-2d} illustrates the advantage of higher orders
for capturing features which are barely resolved by the grid (shaded area):
at $32^2$ resolution, the \nth{2} order scheme is not yet resolving the vortex,
whereas the DG-4 method has already achieved its theoretical \nth{4} order convergence.
The same conclusion holds in the presence of shocks, as illustrated by the Shu-Osher MHD shock tube in Fig.~\ref{fig:test-shu-osher}
where we see a significant improvement from DG-2 to DG-3.
The loop advection problem further shows that sharp features (such as singular field derivatives appearing in the MHD current)
can also be captured within one cell by moderately increasing the spatial order:
going from \nth{2} to \nth{3} or \nth{4} order dramatically improves the loop sharpness,
while reducing the dissipation of magnetic energy.

% Problem with AMR: Galilean + CFL
Compared to Lagrangian methods,
a major issue with AMR Eulerian grid codes is that they require sufficient grid resolution
to avoid dissipation due to bulk flow velocities; i.e. they are only Galilean invariant for solutions sufficiently resolved to make advection errors negligible.
Because spatial resolution translates into tighter CFL constraints on the timestep, a compromise has to be reached between advection errors and compute time in practice.
The advected Orszag--Tang test of Section~\ref{sec:test-orszag-tang-adv} demonstrates that not only do higher order schemes help reduce advection errors and restore Galilean invariance,
but for smooth regions of the flow, it can actually be beneficial to increase the order while reducing spatial resolution.
Note that these test problems present MHD shocks, and it is encouraging to see that these positive features remain,
even though we find that they can be sensitive to the details of the limiter settings.

The combination of higher order methods with adaptive mesh refinement therefore seems particularly powerful.
We note that for most astrophysical situations, spatial refinement will likely be required,
because of the presence of shocks which are inherently first-order features,
but also whenever the fluid is self-gravitating.
% Diminishing returns
One may therefore ask what scheme order will turn out to be the optimal choice for a given problem.
While we discussed positive effects of higher orders, in practice we expect diminishing returns.
From the above discussion, it is clear the optimal global scheme order will depend on the volume filling fraction and geometry of shocks and other discontinuities,
as well as the acceptable truncation error,
both of which are very problem-dependent.
Higher orders will only be helpful to the extent that we can efficiently patch smooth regions of the flow with coarser and coarser cells.
In addition, the computational cost of DG becomes prohibitive for large orders,
in part because of the expensive quadrature operations,
but also because of the more restrictive CFL condition \eqref{eq:cfl}.
We note that this CFL constraint can be relaxed within the DG framework,
for example using so-called PNPM schemes \citep{dumbser_unified_2008},
where $N$ moments are evolved dynamically as in DG,
whereas high-order spatial reconstruction is used up to order $M \geq N$ to recover the remaining moments;
however this comes at the cost of a more extended pattern of ghost cells,
as with purely reconstruction-based schemes.

% Time integration
\label{sec:discuss-hi-order-time-integ}
Time integration for RKDG schemes is also both computationally and memory-expensive at high temporal orders,
as Runge-Kutta methods require multiple steps with intermediate storage.
These memory and computational requirements may be reduced using other time integration schemes such as ADER (see \ref{sec:local-timestepping-ader}).
Note that even though we match the RK time integration order to the spatial order of the scheme (up to RK4) in most of our runs,
this is only done to ensure that the time integration errors do not contaminate the convergence tests of Section~\ref{sec:conv-order}.
In practice, high-order RK time integration may be unnecessary,
for example when the CFL criterion enforces very small time steps due to high plasma temperatures,
or in presence of shorter time-scales in other physical processes.

For these reasons, based on the test problems presented in this work,
we expect third or fourth-order schemes (DG-3, DG-4) to achieve a good overall compromise when combined with adaptive mesh refinement.
\cite{anninos_cosmosdghpadaptivediscontinuous_2017} have reported similar practical limits for their astrophysical DG scheme,
based on compute time and floating point machine precision considerations.

\subsection{Discontinuous Galerkin for high-order}

Explicit DG schemes are often touted as a promising family of methods for high performance computing,
because they are both very local
(in the sense that information about all derivatives can be obtained locally within a cell,
instead of relying on multiple neighbouring cells as with reconstruction methods),
and compute-intensive.
Both properties are regarded as desirable in combination,
as they shift the load away from the communication and memory subsystems towards CPU cores,
which translates into greater parallel scalability.
In addition, most of the compute operations in DG involve dense linear algebra
and can be readily optimized to make use of the powerful floating point units of modern CPUs.

However, in practice, exploiting the full compute capabilities of modern CPUs,
such as vector units and fused multiply-add instructions,
requires exposing operations to the compiler at the right level of loop granularity,
which mandates careful organization of the code.
In some places, we resort to temporary data structures to pack and unpack data elements to benefit from vectorization.
We believe such optimization efforts are necessary if one intends to reap the benefits of DG,
because of the large additional computational costs compared to reconstruction methods.

In addition to being an asset for achieving high compute efficiency,
the local character of derivatives in DG could also have implications for the implementation of physical models.
We find that we can obtain a very sharp and clean signal for the derivatives within a cell,
as illustrated by the loop advection problem of Section~\ref{sec:test-loop}:
the MHD current is readily evaluated locally in each cell from the DG weights,
and higher orders significantly improve the capturing of singular features in the current (Fig.~\ref{fig:test-loop}).
We believe this property of DG schemes could have important applications for the modelling of physical processes
which require accurate derivatives of the fields, such as for example
shock detection \citep{schaal_shocks_2016},
or cosmic ray streaming in magnetic fields \citep[e.g.][]{wiener_cosmicraystreaming_2013,jiang_newnumericalscheme_2018}.

\subsection{Central divergence control methods}
\label{sec:discuss-div}
In this paper, we have explored the implementation of cell-centred divergence control techniques,
in the form of the 8-wave formalism with Powell terms, and hyperbolic divergence cleaning.
We find that although these techniques can be applied successfully to MHD test problems with DG,
they are no silver bullet against all divergence-related effects,
in particular if we insist on using the well-tested and low-diffusion HLLD fluxes.

\subsubsection{Hyperbolic divergence cleaning}

We find that the hyperbolic cleaning method of \cite{dedner_hyperbolic_2002} lends itself very well to a straightforward implementation with DG.
The scheme is very effective at eliminating divergence in test problems where a single spatial and temporal scale can be readily identified.
In addition, a major advantage is that we find the scheme to be equally stable with LDF but also Legendre bases for the magnetic field.
As a result, hyperbolic cleaning has been a popular implementation of cell-based divergence control in DG-MHD codes
\citep[e.g.][]{zanotti_solving_2015,dumbser_simple_2016,kidder_spectre:_2017}.
The scheme has a number of drawbacks however, which impact its suitability for large dynamic range simulations.

% Dimensional cleaning speed
First, the scheme requires at least one\footnotemark dimensional parameter to be set (the cleaning speed $c_h$),
which should be chosen faster than the fastest MHD wave at all points and all times in the simulation.
In the formulation of \citet{dedner_hyperbolic_2002}, $c_h$ is taken to be uniform in space and constant in time,
which is in practice intractable for simulations with large dynamic range because of the resulting global CFL constraint.
\cite{tricco_constrained_2016} showed that using the original divergence cleaning formulation with time-varying $c_h$ could result in locally creating divergence,
which we have observed in our test simulations in presence of adaptive mesh refinement.
To overcome this issue, they developed locally varying cleaning speeds in both space and time in the context of Lagrangian SPH methods.
Adapting this work to Eulerian grids could help alleviate the problem of choosing hyperbolic cleaning speeds for problems with large dynamic range.
\footnotetext{As discussed in Section~\ref{sec:dedner}, our current implementation requires three dimensional parameters,
  but they can be reduced to only one using the parametrization of \cite{mignone_secondorderunsplitgodunov_2010}.
}

% Non-physical effects
Beyond the question of parameter tuning, hyperbolic cleaning can also produce non-physical effects,
because it will propagate disturbances in the magnetic field at velocity $c_h$, i.e. faster than any physical wave of ideal MHD.
We have seen this effect with the rotated shock tube setup (Fig.~\ref{fig:test-rotated-shocktube}).
In our test problems, the effect is most impacting on the 3D blast wave problem,
where the issue is compounded by the very low ambient plasma $\beta$ of this particular setup:
hyperbolic cleaning will cause fluctuations in the magnetic energy to propagate outwards ahead of the shock into the background medium,
resulting in a loss of positivity of the thermal pressure.
In science applications, this has been found to produce spurious bending of field lines ahead of bow shocks by e.g. \cite{susanto_highorderfinitevolumeschemes_2014}.
Approaches have been developed to address positivity issues with divergence cleaning,
either by ensuring that the magnetic energy can only be decreased by the scheme \citep{tricco_constrainedhyperbolicdivergence_2012},
or through an entropy-stable formulation with entropy variables \citep{derigs_idealglmmhdentropy_2018}.
We have not investigated these directions in this work, but leave them as potential future improvements.

\subsubsection{Powell scheme}
In comparison to hyperbolic divergence cleaning, the Powell source term approach does not have any free parameter,
and is therefore intrinsically scale-free.
We find that for most test problems, the Powell approach produces results close to hyperbolic cleaning,
and in particular our implementation does not seem to suffer from some of the divergence-related problems
observed by other authors with pure source term approaches
(for example, comparing the MHD blast slices of Fig.~\ref{fig:test-mhd-blast-slices}
to Figure~18 of \citealt{hopkins_accurate_2016},
or the AMR rotor results of Fig.~\ref{fig:test-rotor-amr}
to Figure~7 of \citealt{derigs_idealglmmhdentropy_2018}).
The method handles strong shocks and low plasma-$\beta$ situations very robustly, as demonstrated in the 3D blast problem.
Interestingly, like \cite{balsara_staggered_1999}, we find that isolated moving shocks are not the strongest source of divergence with the Powell scheme.
Rather, interacting or standing shocks, and stagnation points in smooth flows, are most at risk of local corruption of the solution due to divergence, because the Powell scheme cannot advect it efficiently away.
Typical test setups exhibiting these issues are the rotated shock tube and the Orszag--Tang vortex,
the former featuring a standing shock, and the latter smooth stagnating flows, ``pinch'' points, and interacting shocks.

% Non conservation
An obvious drawback of the Powell method is its nonconservative character due to the source term which injects conserved quantities.
It is difficult to comprehensively assess the exact impact of this source term on the quality of the solution.
In terms of \emph{global} conservation, we found only a small impact of the source term in the Orszag--Tang vortex problem
(Fig.~\ref{fig:test-orszag-tang-energy}),
but we stress that the impact could get more severe over long time-scales.
The rotated shock tube provides an even more stringent test of \emph{local} conservation properties,
and we measure systematic deviations of about $5 \%$ in the magnetic field strength along the problematic direction.
We note that this error does not converge away with scheme order, since the shock is intrinsically a first-order feature.
In addition, we also observe that this error is constant with resolution as noted by \cite{toth__2000},
because it is caused by the effect of the \emph{integrated} Powell term which is singular at the jump,
and for which the cell size only acts as a regularization scale.
However, outside of shocks where the solution is smooth (possibly with steep gradients),
additional resolution produces smaller jumps at interfaces, thereby reducing the overall amount of divergence.
This is consistent with the findings of other studies \citep[e.g.][]{pakmor_simulations_2013,zhu_magnetized_2015}.
In addition to resolution, we showed that increasing the scheme order can very efficiently reduce the divergence in smooth regions of the flow,
as soon as the flow features are at least partially resolved (Fig.~\ref{fig:conv-smvrtx-2d-div}).

These non-conservative properties are a known weak point of Powell schemes,
and our study shows that DG and higher-order methods are not immune to them.
Our Powell scheme seems however very robust in high-Mach, low-$\beta$ environments, as illustrated by the 3D blast test.
We note that nonconservative source terms seem to be a recurring ingredient of cell-centred schemes with good positivity properties.
\cite{mignone_secondorderunsplitgodunov_2010} noticed that the behaviour of hyperbolic cleaning in the presence of shocks is improved by the addition of source terms \citep[the co-called EGLM approach of][]{dedner_hyperbolic_2002}.
Interestingly, entropy stability---which is related to positivity and robust behaviour at shocks---%
seems to call for non-conservative source terms,
which are themselves at odds with the underlying foundational formulation as conservation laws \citep{chandrashekar_entropy_2016,derigs_idealglmmhdentropy_2018}.

% Stability? Limiter interaction
\label{sec:discuss-limiter-powell-interaction}
The divergence control provided by the Powell scheme with our DG method is unfortunately not a turnkey solution.
While we did not experience divergence runaways with LDF bases and the Powell term prescriptions discussed in \ref{sec:powell-discretization} across all of our test runs,
the resulting solutions can, for some limiter settings, show signs of divergence damage when compared to runs with hyperbolic cleaning.
Generally speaking, if the characteristic slope limiter threshold $\tilde M$ is set too high (i.e. if the limiter is too lenient),
spurious extrema will contaminate all waves in the solution, a problem which is of course not specific to the magnetic field divergence.
The current sheet problem of Section~\ref{sec:test-current-sheet} illustrates a case where the Powell method suffers from too permissive limiter settings.
We also noticed, however, that $\tilde M$ may conversely be set \emph{too low}, so that the limiter becomes too aggressive and restricts the propagation of the \nth{8} divergence wave,
thereby reducing the efficiency of divergence control.
This can paradoxically result in divergence damage being greater with a more aggressive limiter setting.
This effect seems particularly noticeable on the Orszag--Tang vortex problem.
This does not happen with other variables or MHD waves, because they do not rely on one of the characteristics for stabilization against a runaway process.

% Need for LDF bases
Another important and likely related aspect of the Powell scheme in our DG implementation is that it seems to require LDF bases to achieve stability.
Expanding the magnetic field components on a Legendre basis, and accounting for a volume contribution of the Powell term to the DG integrals is not sufficient to ensure stability,
and can result in uncontrolled divergence blow-up.

% Better limiter?
These effects highlight the complex interactions between different ingredients of the Powell scheme,
which we attribute to the non-linear nature of the Powell source term.
We note that the slope limiter used in this work is rather primitive,
and other limiters could well help resolve this issue.
We review some possible limiter improvements in the dedicated discussion Section~\ref{sec:discuss-limiters}.

\subsubsection{Locally divergence-free bases}

As discussed above, we find that the LDF basis functions discussed in this paper are necessary for the stability of our Powell scheme.
They are, however, in no way sufficient as soon as a non-diffusive Riemann solver such as HLLD is employed.
More diffusive Riemann solvers, such as Rusanov fluxes, or entropy-stable fluxes of \cite{chandrashekar_entropy_2016},
can achieve stability with LDF bases and no other divergence treatment on simple problems:
these fluxes are generally more diffusive than HLLD,
and in particular include a normal component flux for the magnetic field which will dampen the normal jump of $\Bfield$ at the faces.
However, in the presence of strong shocks, we find that the HLLD solver provides sharper solutions,
and therefore constitutes a more sensible choice for high-order codes
attempting to resolve features close to the grid resolution.

LDF basis functions require significant additional complexity in the code and in the scheme itself,
from basis function generation (see Appendix~\ref{sec:apdx-div0})
to special limiting procedures (such as the one we introduced in \ref{sec:limiting-div0}).
Because LDF bases make use of the divergence-free condition, they require somewhat less storage than a component-wise Legendre basis for $\Bfield$.

\label{sec:perf-div0}
However, LDF bases turn out to be potentially more computationally expensive than Legendre bases for high orders:
the Legendre basis \eqref{eq:legendre-3d} possesses a tensor product structure (component $\times$ basis function)
which allows optimized computation of conserved states $\cns$ and Gaussian quadrature using sum factorization \cite[see][]{kronbichler_fast_2017}.
These optimizations cannot be readily used with LDF bases, because the coupling between components of $\Bfield$ breaks the tensor product structure.
In addition, precomputing and storing basis function values and gradients at quadrature points requires more memory for the LDF basis than the Legendre basis,
because all 3 components of $\Bfield$ have to be stored
(in addition to the Legendre basis, which is still required for hydro variables).
The memory footprint of the precomputed basis function data
can exceed the size of the L1 processor cache,
and for orders $\geq 4$ in 3D, the quadrature data for the LDF+Powell scheme will not even fit in a 256~kB L2 cache,
resulting in a strong performance degradation.
From a few practical tests,
we find that sum factorization optimizations and
careful memory footprint management can make a significant difference for orders $\geq 4$,
so users of very high order schemes may want to reconsider the use of divergence-free bases on computational efficiency grounds.

Finally, the convergence tests of Section~\ref{sec:conv-order} have shed light on a subtle instability with LDF bases
that reveals itself with HLLD fluxes in non-grid-aligned configurations.
We speculate that this may be related to the projection effects discussed in Appendix~\ref{sec:apdx-div0-proj}.
While we do not see this as a source of concern for actual simulations,
we note that the effective convergence order with LDF bases may in some cases be lower than expected.

\subsubsection{Conclusion on divergence control}

A natural way around the shortcomings of both schemes would be to rely on exactly globally divergence-free schemes.
Such methods have been proposed for DG,
for example based on vector potential techniques \cite[e.g.][]{rossmanith_high-order_2013}.
High-order versions of constrained transport have also been developed,
where the magnetic field is expanded on the interfaces,
reconstructed within the cell using a divergence-free reconstruction,
and evolved using a distinct evolution equation \citep[see e.g.][]{li_central_2011,balsara_von_2017}.
These methods involve complicated ingredients:
they require more complex mesh structures,
multidimensional Riemann solvers to obtain the electric fields,
and higher-order expansions for the magnetic field inside the cells.
In addition, because the dynamical degrees of freedom are split between cell- and face-centred discretizations,
it is less clear how to define a proper limiting procedure for high-order modes.

Although we find that both hyperbolic cleaning and the Powell methods have shortcomings
which are a source for concern for large dynamic range simulations,
we note that these methods have been successfully used in practice in production in a finite volume context.
The extensive code comparison study of \cite{kritsuk_comparing_2011}, focusing on decaying isothermal supersonic turbulence,
showed no significant systematic effect of the divergence control method on a number of physically relevant quantities;
the authors demonstrate that other aspects of codes are more important for accurate turbulence simulations.
In particular, they find that high-order codes achieve increased spectral bandwidth and effective Reynolds number.
In the context of the magneto-rotational instability (MRI),
\cite{flock_highordergodunovschemes_2010} see little difference between Powell and upwinded constrained transport schemes
on MRI growth rates and evolution of the magnetic energy.

In any case, we plan to apply both hyperbolic divergence cleaning and the Powell scheme to MHD turbulence simulations,
for which the impact of divergence control can be readily tested against published results.

\subsection{Future improvements}

Our current DG scheme relies on a number of simple ingredients which will likely require improvement for use in production astrophysical science applications.
The main current directions will be to replace the simple DG limiter described in \ref{sec:limiter} by more sophisticated alternatives,
add support for local timestepping,
and improve $\divB$ control, even within the framework of cell-centred divergence control schemes.
Fortunately, solutions to some of these issues have already been developed in the DG context,
and we plan to implement them as future improvements.

\subsubsection{Limiters}
\label{sec:discuss-limiters}

Our limiting procedure discussed in \ref{sec:limiter} relies on the choice of a dimensional parameter $\tilde M$ (or equivalently $M$),
which, although sufficient for smooth test problems where a single spatial scale dominates,
is not convenient for problems involving large dynamic range, or for which the spatial variation of the solution is not well controlled.
In addition, acting as a slope limiter, it can only ever retain second-order information in the cell whenever the limiter is triggered.

A number of more sophisticated limiters have been developed and successfully applied in the DG literature.
In particular, it is possible to limit the DG solution while preserving all its higher-order moments.
Such limiters were first introduced as a natural extension of the TVD limiter, for example by
\citet{biswas_paralleladaptivefinite_1994,krivodonova_limitershighorderdiscontinuous_2007}.
One advantage of these limiters is that they do not require a dimensional threshold parameter $M$.

Another interesting approach comes from the so-called weighted essentially non-oscillatory schemes (WENO),
whose weighted stencils may be used to reconstruct high-order solutions with reduced oscillations
\citep[see e.g.][]{jianxian_qiu_runge-kutta_2004,balsara_sub-cell_2007,zhong_simpleweightedessentially_2013}.
Divergence-free WENO reconstruction schemes have been developed for MHD \citep{balsara_divergence-free_2015,zhao_rungekutta_2017},
and could form a suitable basis for DG limiters with LDF bases.

More recently, so-called \emph{a posteriori} subcell limiters have been developed by \citet{clain_highorderfinitevolume_2011,sonntag_shockcapturingdiscontinuous_2014,dumbser_posteriori_2014,dumbser_simple_2016}.
They rely on after-the-fact detection of solution defects by first taking a timestep
and projecting the resulting tentative solution in each cell onto a finer subcell grid.
The subcell grid is then inspected for loss of positivity or spurious oscillations.
In case any subcell is marked as troubled, the solution in the parent cell is rolled back to the previous time point,
its subcell values are recomputed and evolved in time using a lower-order but more robust scheme,
and the updated subcell grid is used to reconstruct the DG moments of the parent cell at the final time.
This technique has been successfully used for DG implementations of MHD \citep{zanotti_solving_2015,fambri_aderdiscontinuousgalerkin_2018}.
In particular, simple criteria for solution admissibility have been developed,
and subcell oscillation detection can be performed without dimensional parameters \citep{dumbser_posteriori_2014}.
We intend to replace our current limiter with this promising technique in the near future.

\subsubsection{Local timestepping with ADER}
\label{sec:local-timestepping}
\label{sec:local-timestepping-ader}

While some astrophysical problems only span a limited range of spatial and temporal scales
(for example, simulations of idealized isotropic isothermal turbulence in absence of self-gravity),
many simulation setups require spatial resolution with AMR, or feature strongly locally-varying signal speeds.
In the latter case, the CFL constraint \eqref{eq:cfl} will impose a maximum $\Delta t$ which can vary significantly across cells,
and it becomes beneficial to update each cell at its own pace according to its maximal $\Delta t$ using local timestepping.

The RKDG method used in our scheme is not easily amenable to local timestepping,
because it treats the whole hyperbolic problem as a set of coupled ordinary differential equations through the semidiscrete scheme \eqref{eq:wdot},
in which the \emph{whole} vector of all cell weights is integrated forward in time using a Runge-Kutta method.

In the DG context, local timestepping may be achieved using the so-called ADER methods,
introduced by \citet{titarev_ader:_2002,dumbser_building_2006,dumbser_arbitrary_2007},
and more recently reformulated by \cite{dumbser_unified_2008}.
In its modern variant, ADER exploits a DG-like weak formulation on a high-order spacetime basis to first obtain a high-order predictor state local to each cell.
This state is then used within the DG framework to perform time integration as a predictor-corrector scheme.
This formulation has been used by e.g. \cite{zanotti_solving_2015,fambri_spacetimeadaptive_2017} to develop ADER-DG codes with adaptive mesh refinement and local timestepping.
Very recently, \citet{charrier_stoptalkingme_2018} have shown that this ADER-DG predictor-corrector time integration scheme can be made efficient on HPC architectures,
with a reduced memory footprint, increased compute intensity and optimized distributed memory data exchanges.
The above properties make ADER an attractive time integration algorithm for large dynamic range astrophysical applications,
as an improvement over our current RKDG scheme.

\subsubsection{Divergence control}

We have discussed divergence control in detail in \ref{sec:discuss-div}.
Short of implementing a full-fledged high-order version of constrained transport,
we would like to try to adapt improvements of the hyperbolic cleaning methodology \citep[following][]{tricco_constrainedhyperbolicdivergence_2012,mignone_secondorderunsplitgodunov_2010,tricco_constrained_2016} to a Eulerian DG framework,
to allow fully adaptive cleaning speeds,
and to reduce positivity issues with hyperbolic cleaning for very strong shocks.
For the Powell approach, a promising direction would be to experiment with an 8-wave version of the HLLD flux, such as proposed by \citet{fuchs_approximateriemannsolvers_2011}.

\section{Conclusions}

In this paper, we have described our implementation of a discontinuous Galerkin scheme with adaptive mesh refinement for magnetohydrodynamics,
implemented within the framework of the astrophysical code {\Arepo}.
Our scheme relies on the Runge-Kutta DG framework,
and can use two different types of cell-centred divergence control techniques:
locally divergence-free bases for the magnetic field together with Powell terms to control the global divergence,
and hyperbolic cleaning, which may be used with the same Legendre basis functions used for hydrodynamical variables.
We have introduced two main new numerical ingredients:
a non-linear limiting procedure for the magnetic field components,
and different discretizations of the Powell source term,
which we found to be a key ingredient for the stability and accuracy of the scheme.
We have also discussed some subtle numerical and performance-related properties of locally divergence-free bases,
which we encountered during our development of the scheme.

We have shown that the resulting method is accurate across a wide range of typical MHD test problems,
and can achieve the expected theoretical order of convergence.
Like previous authors, we found that higher-order schemes can efficiently reduce advection errors in Eulerian grid codes,
and we further showed that these results hold even with a simple limiting procedure and in presence of MHD shocks.

We showed that increasing the DG order of the scheme can benefit solutions which are barely resolved by the grid size,
by reducing not only numerical diffusion but also magnetic field divergence more efficiently than can be achieved by increasing grid resolution.
We argue that this makes higher order methods very attractive for resolution-limited astrophysical simulations.
In addition, DG provides instant access to local derivatives within the cell without smearing across neighbour stencils,
which can be important for astrophysical applications where a clean derivative signal is required for the physics,
such as shock detection or models of cosmic ray streaming.

We discussed the divergence control performance of the Powell scheme and hyperbolic cleaning,
and found that while both methods perform well across most test problems,
neither of them in our current implementation was fully satisfactory across all flow regimes.

Finally, we covered a few of the future improvements planned for our current DG implementation,
which we see as a base development platform for increasingly sophisticated high-order solvers in the {\Arepo} code,
in which to integrate more and more physical models while developing the solvers into highly scalable implementations.

We are now planning to further test and improve the scheme with high resolution simulations of MHD turbulence,
which will provide insights on the effectiveness of our numerical ingredients%%
---in particular the two divergence control techniques---%%
and more generally contribute to shedding light on the potential of DG methods for astrophysics.

\section*{Acknowledgements}

TG would like to thank Markus Zenk, Philipp Girichidis, Christoph Pfrommer, Matthew Bate and Pascal Tremblin
for very helpful discussions.
The authors would also like to thank the anonymous referee for their careful reading
and helpful comments which have contributed to significantly improving the manuscript.
The authors gratefully acknowledge the support of the Klaus Tschira Foundation.
We acknowledge financial support through subproject EXAMAG of the Priority Programme 1648 ``SPPEXA'' of the German Science Foundation,
and through the European Research Council through ERC-StG grant EXAGAL-308037.
This work was also partly supported by the ERC advanced grant No.~787361-COBOM.
PC acknowledges support from the Airbus Foundation Chair on Mathematics of Complex Systems at TIFR-CAM, Bangalore.

%%%%%%%%%%%%%%%%%%%%%%%%%%%%%%%%%%%%%%%%%%%%%%%%%%

%%%%%%%%%%%%%%%%%%%% REFERENCES %%%%%%%%%%%%%%%%%%

\bibliographystyle{mnras}
\bibliography{dgmhd}

\begin{thebibliography}{}
\makeatletter
\relax
\def\mn@urlcharsother{\let\do\@makeother \do\$\do\&\do\#\do\^\do\_\do\%\do\~}
\def\mn@doi{\begingroup\mn@urlcharsother \@ifnextchar [ {\mn@doi@}
  {\mn@doi@[]}}
\def\mn@doi@[#1]#2{\def\@tempa{#1}\ifx\@tempa\@empty \href
  {http://dx.doi.org/#2} {doi:#2}\else \href {http://dx.doi.org/#2} {#1}\fi
  \endgroup}
\def\mn@eprint#1#2{\mn@eprint@#1:#2::\@nil}
\def\mn@eprint@arXiv#1{\href {http://arxiv.org/abs/#1} {{\tt arXiv:#1}}}
\def\mn@eprint@dblp#1{\href {http://dblp.uni-trier.de/rec/bibtex/#1.xml}
  {dblp:#1}}
\def\mn@eprint@#1:#2:#3:#4\@nil{\def\@tempa {#1}\def\@tempb {#2}\def\@tempc
  {#3}\ifx \@tempc \@empty \let \@tempc \@tempb \let \@tempb \@tempa \fi \ifx
  \@tempb \@empty \def\@tempb {arXiv}\fi \@ifundefined
  {mn@eprint@\@tempb}{\@tempb:\@tempc}{\expandafter \expandafter \csname
  mn@eprint@\@tempb\endcsname \expandafter{\@tempc}}}

\bibitem[\protect\citeauthoryear{Anninos, Bryant, Fragile, Holgado, Lau  \&
  Nemergut}{Anninos
  et~al.}{2017}]{anninos_cosmosdghpadaptivediscontinuous_2017}
Anninos P.,  Bryant C.,  Fragile P.~C.,  Holgado A.~M.,  Lau C.,   Nemergut D.,
   2017, \mn@doi [Astrophys. J. Suppl. Ser.] {10.3847/1538-4365/aa7ff5}, 231,
  17

\bibitem[\protect\citeauthoryear{Balsara}{Balsara}{1998}]{balsara_total_1998}
Balsara D.~S.,  1998, \mn@doi [Astrophys. J. Suppl. Ser.] {10.1086/313093},
  116, 133

\bibitem[\protect\citeauthoryear{Balsara}{Balsara}{2004}]{balsara_second-order-accurate_2004}
Balsara D.~S.,  2004, Astrophys. J. Suppl. Ser., 151, 149

\bibitem[\protect\citeauthoryear{Balsara}{Balsara}{2009}]{balsara_divergence-free_2009}
Balsara D.~S.,  2009, \mn@doi [J. Comput. Phys.] {10.1016/j.jcp.2009.03.038},
  228, 5040

\bibitem[\protect\citeauthoryear{Balsara}{Balsara}{2017}]{balsara_higherorderaccuratespacetime_2017}
Balsara D.~S.,  2017, \mn@doi [Living Rev. Comput. Astrophys.]
  {10.1007/s41115-017-0002-8}, 3, 2

\bibitem[\protect\citeauthoryear{Balsara \& Dumbser}{Balsara \&
  Dumbser}{2015}]{balsara_divergence-free_2015}
Balsara D.~S.,  Dumbser M.,  2015, \mn@doi [J. Comput. Phys.]
  {10.1016/j.jcp.2015.07.012}, 299, 687

\bibitem[\protect\citeauthoryear{Balsara \& K\"appeli}{Balsara \&
  K\"appeli}{2017}]{balsara_von_2017}
Balsara D.~S.,  K\"appeli R.,  2017, \mn@doi [J. Comput. Phys.]
  {10.1016/j.jcp.2017.01.056}, 336, 104

\bibitem[\protect\citeauthoryear{Balsara \& Shu}{Balsara \&
  Shu}{2000}]{balsara_monotonicitypreservingweighted_2000}
Balsara D.~S.,  Shu C.-W.,  2000, \mn@doi [J. Comput. Phys.]
  {10.1006/jcph.2000.6443}, 160, 405

\bibitem[\protect\citeauthoryear{Balsara \& Spicer}{Balsara \&
  Spicer}{1999a}]{balsara_maintaining_1999}
Balsara D.~S.,  Spicer D.,  1999a, \mn@doi [J. Comput. Phys.]
  {10.1006/jcph.1998.6108}, 148, 133

\bibitem[\protect\citeauthoryear{Balsara \& Spicer}{Balsara \&
  Spicer}{1999b}]{balsara_staggered_1999}
Balsara D.~S.,  Spicer D.~S.,  1999b, \mn@doi [J. Comput. Phys.]
  {10.1006/jcph.1998.6153}, 149, 270

\bibitem[\protect\citeauthoryear{Balsara, Altmann, Munz  \& Dumbser}{Balsara
  et~al.}{2007}]{balsara_sub-cell_2007}
Balsara D.~S.,  Altmann C.,  Munz C.-D.,   Dumbser M.,  2007, \mn@doi [J.
  Comput. Phys.] {10.1016/j.jcp.2007.04.032}, 226, 586

\bibitem[\protect\citeauthoryear{Balsara, Rumpf, Dumbser  \& Munz}{Balsara
  et~al.}{2009}]{balsara_efficient_2009}
Balsara D.~S.,  Rumpf T.,  Dumbser M.,   Munz C.-D.,  2009, \mn@doi [J. Comput.
  Phys.] {10.1016/j.jcp.2008.12.003}, 228, 2480

\bibitem[\protect\citeauthoryear{Bauer \& Springel}{Bauer \&
  Springel}{2012}]{bauer_subsonic_2012}
Bauer A.,  Springel V.,  2012, \mn@doi [Mon. Not. R. Astron. Soc.]
  {10.1111/j.1365-2966.2012.21058.x}, 423, 2558

\bibitem[\protect\citeauthoryear{Biswas, Devine  \& Flaherty}{Biswas
  et~al.}{1994}]{biswas_paralleladaptivefinite_1994}
Biswas R.,  Devine K.~D.,   Flaherty J.~E.,  1994, \mn@doi [Appl. Numer. Math.]
  {10.1016/0168-9274(94)90029-9}, 14, 255

\bibitem[\protect\citeauthoryear{Boscheri \& Dumbser}{Boscheri \&
  Dumbser}{2017}]{boscheri_arbitrary-lagrangianeulerian_2017}
Boscheri W.,  Dumbser M.,  2017, \mn@doi [J. Comput. Phys.]
  {10.1016/j.jcp.2017.06.022}, 346, 449

\bibitem[\protect\citeauthoryear{Boscheri, Dumbser  \& Balsara}{Boscheri
  et~al.}{2014}]{boscheri_high_2014}
Boscheri W.,  Dumbser M.,   Balsara D.,  2014, \mn@doi [Int. J. Numer. Methods
  Fluids] {10.1002/fld.3947}, 76, 737

\bibitem[\protect\citeauthoryear{Brackbill \& Barnes}{Brackbill \&
  Barnes}{1980}]{brackbill_effect_1980}
Brackbill J.~U.,  Barnes D.~C.,  1980, \mn@doi [J. Comput. Phys.]
  {10.1016/0021-9991(80)90079-0}, 35, 426

\bibitem[\protect\citeauthoryear{Brio \& Wu}{Brio \&
  Wu}{1988}]{brio_upwind_1988}
Brio M.,  Wu C.~C.,  1988, \mn@doi [J. Comput. Phys.]
  {10.1016/0021-9991(88)90120-9}, 75, 400

\bibitem[\protect\citeauthoryear{Chakravarthy, Arora  \&
  Chakraborty}{Chakravarthy
  et~al.}{2015}]{chakravarthy_simplehybridfinite_2015}
Chakravarthy V.~K.,  Arora K.,   Chakraborty D.,  2015, \mn@doi [Int. J. Numer.
  Methods Fluids] {10.1002/fld.4000}, 77, 707

\bibitem[\protect\citeauthoryear{Chandrashekar \& Klingenberg}{Chandrashekar \&
  Klingenberg}{2016}]{chandrashekar_entropy_2016}
Chandrashekar P.,  Klingenberg C.,  2016, \mn@doi [SIAM J. Numer. Anal.]
  {10.1137/15M1013626}, 54, 1313

\bibitem[\protect\citeauthoryear{Charrier \& Weinzierl}{Charrier \&
  Weinzierl}{2018}]{charrier_stoptalkingme_2018}
Charrier D.~E.,  Weinzierl T.,  2018, ArXiv180108682 Cs

\bibitem[\protect\citeauthoryear{Clain, Diot  \& Loub\`ere}{Clain
  et~al.}{2011}]{clain_highorderfinitevolume_2011}
Clain S.,  Diot S.,   Loub\`ere R.,  2011, \mn@doi [J. Comput. Phys.]
  {10.1016/j.jcp.2011.02.026}, 230, 4028

\bibitem[\protect\citeauthoryear{Cockburn \& Shu}{Cockburn \&
  Shu}{1989}]{cockburn_tvb_1989-1}
Cockburn B.,  Shu C.-W.,  1989, \mn@doi [Math. Comput.] {10.2307/2008474}, 52,
  411

\bibitem[\protect\citeauthoryear{Cockburn \& Shu}{Cockburn \&
  Shu}{1998}]{cockburn_localdiscontinuousgalerkin_1998}
Cockburn B.,  Shu C.,  1998, \mn@doi [SIAM J. Numer. Anal.]
  {10.1137/S0036142997316712}, 35, 2440

\bibitem[\protect\citeauthoryear{Cockburn, Lin  \& Shu}{Cockburn
  et~al.}{1989}]{cockburn_tvb_1989}
Cockburn B.,  Lin S.-Y.,   Shu C.-W.,  1989, \mn@doi [J. Comput. Phys.]
  {10.1016/0021-9991(89)90183-6}, 84, 90

\bibitem[\protect\citeauthoryear{Cockburn, Li  \& Shu}{Cockburn
  et~al.}{2004}]{cockburn_locally_2004}
Cockburn B.,  Li F.,   Shu C.-W.,  2004, \mn@doi [J. Comput. Phys.]
  {10.1016/j.jcp.2003.09.007}, 194, 588

\bibitem[\protect\citeauthoryear{Collins, Xu, Norman, Li  \& Li}{Collins
  et~al.}{2010}]{collins_cosmological_2010}
Collins D.~C.,  Xu H.,  Norman M.~L.,  Li H.,   Li S.,  2010, \mn@doi
  [Astrophys. J. Suppl. Ser.] {10.1088/0067-0049/186/2/308}, 186, 308

\bibitem[\protect\citeauthoryear{Dai \& Woodward}{Dai \&
  Woodward}{1998}]{dai_divergence-free_1998}
Dai W.,  Woodward P.~R.,  1998, \mn@doi [Astrophys. J.] {10.1086/305176}, 494,
  317

\bibitem[\protect\citeauthoryear{Dedner, Kemm, Kr\"oner, Munz, Schnitzer  \&
  Wesenberg}{Dedner et~al.}{2002}]{dedner_hyperbolic_2002}
Dedner A.,  Kemm F.,  Kr\"oner D.,  Munz C.-D.,  Schnitzer T.,   Wesenberg M.,
  2002, \mn@doi [J. Comput. Phys.] {10.1006/jcph.2001.6961}, 175, 645

\bibitem[\protect\citeauthoryear{Derigs, Winters, Gassner  \& Walch}{Derigs
  et~al.}{2016}]{derigs_novel_2016}
Derigs D.,  Winters A.~R.,  Gassner G.~J.,   Walch S.,  2016, \mn@doi [J.
  Comput. Phys.] {10.1016/j.jcp.2016.04.048}, 317, 223

\bibitem[\protect\citeauthoryear{Derigs, Winters, Gassner, Walch  \&
  Bohm}{Derigs et~al.}{2018}]{derigs_idealglmmhdentropy_2018}
Derigs D.,  Winters A.~R.,  Gassner G.~J.,  Walch S.,   Bohm M.,  2018, \mn@doi
  [J. Comput. Phys.] {10.1016/j.jcp.2018.03.002}, 364, 420

\bibitem[\protect\citeauthoryear{Dubois \& Teyssier}{Dubois \&
  Teyssier}{2010}]{dubois_magnetisedwindsdwarf_2010}
Dubois Y.,  Teyssier R.,  2010, \mn@doi [Astron. Astrophys.]
  {10.1051/0004-6361/200913014}, 523, A72

\bibitem[\protect\citeauthoryear{Dumbser \& Loub\`ere}{Dumbser \&
  Loub\`ere}{2016}]{dumbser_simple_2016}
Dumbser M.,  Loub\`ere R.,  2016, \mn@doi [J. Comput. Phys.]
  {10.1016/j.jcp.2016.05.002}, 319, 163

\bibitem[\protect\citeauthoryear{Dumbser \& Munz}{Dumbser \&
  Munz}{2006}]{dumbser_building_2006}
Dumbser M.,  Munz C.-D.,  2006, \mn@doi [J. Sci. Comput.]
  {10.1007/s10915-005-9025-0}, 27, 215

\bibitem[\protect\citeauthoryear{Dumbser, K\"aser  \& Toro}{Dumbser
  et~al.}{2007}]{dumbser_arbitrary_2007}
Dumbser M.,  K\"aser M.,   Toro E.~F.,  2007, \mn@doi [Geophys. J. Int.]
  {10.1111/j.1365-246X.2007.03427.x}, 171, 695

\bibitem[\protect\citeauthoryear{Dumbser, Balsara, Toro  \& Munz}{Dumbser
  et~al.}{2008}]{dumbser_unified_2008}
Dumbser M.,  Balsara D.~S.,  Toro E.~F.,   Munz C.-D.,  2008, \mn@doi [J.
  Comput. Phys.] {10.1016/j.jcp.2008.05.025}, 227, 8209

\bibitem[\protect\citeauthoryear{Dumbser, Zanotti, Loub\`ere  \& Diot}{Dumbser
  et~al.}{2014}]{dumbser_posteriori_2014}
Dumbser M.,  Zanotti O.,  Loub\`ere R.,   Diot S.,  2014, \mn@doi [J. Comput.
  Phys.] {10.1016/j.jcp.2014.08.009}, 278, 47

\bibitem[\protect\citeauthoryear{Evans \& Hawley}{Evans \&
  Hawley}{1988}]{evans_simulation_1988}
Evans C.~R.,  Hawley J.~F.,  1988, \mn@doi [Astrophys. J.] {10.1086/166684},
  332, 659

\bibitem[\protect\citeauthoryear{Fambri, Dumbser  \& Zanotti}{Fambri
  et~al.}{2017}]{fambri_spacetimeadaptive_2017}
Fambri F.,  Dumbser M.,   Zanotti O.,  2017, \mn@doi [Comput. Phys. Commun.]
  {10.1016/j.cpc.2017.08.001}, 220, 297

\bibitem[\protect\citeauthoryear{Fambri, Dumbser, K\"oppel, Rezzolla  \&
  Zanotti}{Fambri et~al.}{2018}]{fambri_aderdiscontinuousgalerkin_2018}
Fambri F.,  Dumbser M.,  K\"oppel S.,  Rezzolla L.,   Zanotti O.,  2018,
  \mn@doi [Mon. Not. R. Astron. Soc.] {10.1093/mnras/sty734}, 477, 4543

\bibitem[\protect\citeauthoryear{Federrath, {Roman-Duval}, Klessen, Schmidt  \&
  Mac~Low}{Federrath
  et~al.}{2010}]{federrath_comparingstatisticsinterstellar_2010}
Federrath C.,  {Roman-Duval} J.,  Klessen R.~S.,  Schmidt W.,   Mac~Low M.-M.,
  2010, \mn@doi [Astron. Astrophys.] {10.1051/0004-6361/200912437}, 512, A81

\bibitem[\protect\citeauthoryear{Flock, Dzyurkevich, Klahr  \& Mignone}{Flock
  et~al.}{2010}]{flock_highordergodunovschemes_2010}
Flock M.,  Dzyurkevich N.,  Klahr H.,   Mignone A.,  2010, \mn@doi [Astron.
  Astrophys.] {10.1051/0004-6361/200912443}, 516, A26

\bibitem[\protect\citeauthoryear{Fromang, Hennebelle  \& Teyssier}{Fromang
  et~al.}{2006}]{fromang_high_2006}
Fromang S.,  Hennebelle P.,   Teyssier R.,  2006, \mn@doi [Astron. Astrophys.]
  {10.1051/0004-6361:20065371}, 457, 371

\bibitem[\protect\citeauthoryear{Fryxell et~al.,}{Fryxell
  et~al.}{2000}]{fryxell_flashadaptivemesh_2000}
Fryxell B.,  et~al., 2000, \mn@doi [Astrophys. J. Suppl. Ser.]
  {10.1086/317361}, 131, 273

\bibitem[\protect\citeauthoryear{Fuchs, McMurry, Mishra, Risebro  \&
  Waagan}{Fuchs et~al.}{2011}]{fuchs_approximateriemannsolvers_2011}
Fuchs F.~G.,  McMurry A.~D.,  Mishra S.,  Risebro N.~H.,   Waagan K.,  2011,
  \mn@doi [Commun. Comput. Phys.] {10.4208/cicp.171109.070510a}, 9, 324

\bibitem[\protect\citeauthoryear{Gaburov \& Nitadori}{Gaburov \&
  Nitadori}{2011}]{gaburov_astrophysical_2011}
Gaburov E.,  Nitadori K.,  2011, \mn@doi [Mon. Not. R. Astron. Soc.]
  {10.1111/j.1365-2966.2011.18313.x}, 414, 129

\bibitem[\protect\citeauthoryear{Gardiner \& Stone}{Gardiner \&
  Stone}{2005}]{gardiner_unsplit_2005}
Gardiner T.~A.,  Stone J.~M.,  2005, \mn@doi [J. Comput. Phys.]
  {10.1016/j.jcp.2004.11.016}, 205, 509

\bibitem[\protect\citeauthoryear{Godunov}{Godunov}{1972}]{godunov_symmetric_1972}
Godunov S.~K.,  1972, Numer. Methods Mech. Contin. Medium, 1, 26

\bibitem[\protect\citeauthoryear{Gottlieb}{Gottlieb}{2005}]{gottlieb_high_2005}
Gottlieb S.,  2005, \mn@doi [J. Sci. Comput.] {10.1007/s10915-004-4635-5}, 25,
  105

\bibitem[\protect\citeauthoryear{Hawley \& Stone}{Hawley \&
  Stone}{1995}]{hawley_mocct:_1995}
Hawley J.~F.,  Stone J.~M.,  1995, \mn@doi [Comput. Phys. Commun.]
  {10.1016/0010-4655(95)00190-Q}, 89, 127

\bibitem[\protect\citeauthoryear{Helzel, Rossmanith  \& Taetz}{Helzel
  et~al.}{2011}]{helzel_unstaggeredconstrainedtransport_2011}
Helzel C.,  Rossmanith J.~A.,   Taetz B.,  2011, \mn@doi [J. Comput. Phys.]
  {10.1016/j.jcp.2011.02.009}, 230, 3803

\bibitem[\protect\citeauthoryear{Hesthaven \& Warburton}{Hesthaven \&
  Warburton}{2008}]{hesthaven_nodaldiscontinuousgalerkin_2008}
Hesthaven J.~S.,  Warburton T.,  2008, Nodal Discontinuous {{Galerkin}}
  Methods: Algorithms, Analysis, and Applications.
No.~54 in Texts in Applied Mathematics, {Springer}, New York

\bibitem[\protect\citeauthoryear{Hopkins}{Hopkins}{2016}]{hopkins_constrained-gradient_2016}
Hopkins P.~F.,  2016, \mn@doi [Mon. Not. R. Astron. Soc.]
  {10.1093/mnras/stw1578}, 462, 576

\bibitem[\protect\citeauthoryear{Hopkins \& Raives}{Hopkins \&
  Raives}{2016}]{hopkins_accurate_2016}
Hopkins P.~F.,  Raives M.~J.,  2016, \mn@doi [Mon. Not. R. Astron. Soc.]
  {10.1093/mnras/stv2180}, 455, 51

\bibitem[\protect\citeauthoryear{Janhunen}{Janhunen}{2000}]{janhunen_positive_2000}
Janhunen P.,  2000, \mn@doi [J. Comput. Phys.] {10.1006/jcph.2000.6479}, 160,
  649

\bibitem[\protect\citeauthoryear{Jiang \& Oh}{Jiang \&
  Oh}{2018}]{jiang_newnumericalscheme_2018}
Jiang Y.-F.,  Oh S.~P.,  2018, \mn@doi [Astrophys. J.]
  {10.3847/1538-4357/aaa6ce}, 854, 5

\bibitem[\protect\citeauthoryear{Jiang \& Shu}{Jiang \&
  Shu}{1996}]{jiang_efficient_1996}
Jiang G.-S.,  Shu C.-W.,  1996, \mn@doi [J. Comput. Phys.]
  {10.1006/jcph.1996.0130}, 126, 202

\bibitem[\protect\citeauthoryear{Jiang \& Wu}{Jiang \&
  Wu}{1999}]{jiang_high-order_1999}
Jiang G.-S.,  Wu C.-c.,  1999, \mn@doi [J. Comput. Phys.]
  {10.1006/jcph.1999.6207}, 150, 561

\bibitem[\protect\citeauthoryear{Karami~Halashi \& Luo}{Karami~Halashi \&
  Luo}{2016}]{karami_halashi_reconstructed_2016}
Karami~Halashi B.,  Luo H.,  2016, \mn@doi [J. Comput. Phys.]
  {10.1016/j.jcp.2016.08.055}, 326, 258

\bibitem[\protect\citeauthoryear{Kemm}{Kemm}{2013}]{kemm_origin_2013}
Kemm F.,  2013, \mn@doi [Commun. Appl. Math. Comput. Sci.]
  {10.2140/camcos.2013.8.1}, 8, 1

\bibitem[\protect\citeauthoryear{Kidder et~al.,}{Kidder
  et~al.}{2017}]{kidder_spectre:_2017}
Kidder L.~E.,  et~al., 2017, \mn@doi [J. Comput. Phys.]
  {10.1016/j.jcp.2016.12.059}, 335, 84

\bibitem[\protect\citeauthoryear{Kritsuk et~al.,}{Kritsuk
  et~al.}{2011}]{kritsuk_comparing_2011}
Kritsuk A.~G.,  et~al., 2011, \mn@doi [Astrophys. J.]
  {10.1088/0004-637X/737/1/13}, 737, 13

\bibitem[\protect\citeauthoryear{Krivodonova}{Krivodonova}{2007}]{krivodonova_limitershighorderdiscontinuous_2007}
Krivodonova L.,  2007, \mn@doi [J. Comput. Phys.] {10.1016/j.jcp.2007.05.011},
  226, 879

\bibitem[\protect\citeauthoryear{Kronbichler, Kormann, Pasichnyk  \&
  Allalen}{Kronbichler et~al.}{2017}]{kronbichler_fast_2017}
Kronbichler M.,  Kormann K.,  Pasichnyk I.,   Allalen M.,  2017, in High
  {{Performance Computing}}. Lecture {{Notes}} in {{Computer Science}}.
{Springer}, pp 237--255, \mn@doi{10.1007/978-3-319-58667-0_13}

\bibitem[\protect\citeauthoryear{LeVeque}{LeVeque}{2002}]{leveque_finitevolumemethods_2002}
LeVeque R.~J.,  2002, Finite {{Volume Methods}} for {{Hyperbolic Problems}}.
{Cambridge University Press}

\bibitem[\protect\citeauthoryear{Li \& Shu}{Li \& Shu}{2005}]{li_locally_2005}
Li F.,  Shu C.-W.,  2005, \mn@doi [J. Sci. Comput.]
  {10.1007/s10915-004-4146-4}, 22-23, 413

\bibitem[\protect\citeauthoryear{Li \& Xu}{Li \& Xu}{2012}]{li_arbitrary_2012}
Li F.,  Xu L.,  2012, \mn@doi [J. Comput. Phys.] {10.1016/j.jcp.2011.12.016},
  231, 2655

\bibitem[\protect\citeauthoryear{Li, Xu  \& Yakovlev}{Li
  et~al.}{2011}]{li_central_2011}
Li F.,  Xu L.,   Yakovlev S.,  2011, \mn@doi [J. Comput. Phys.]
  {10.1016/j.jcp.2011.03.006}, 230, 4828

\bibitem[\protect\citeauthoryear{Liu, Shu  \& Zhang}{Liu
  et~al.}{2018}]{liu_entropystablehigh_2018}
Liu Y.,  Shu C.-W.,   Zhang M.,  2018, \mn@doi [J. Comput. Phys.]
  {10.1016/j.jcp.2017.10.043}, 354, 163

\bibitem[\protect\citeauthoryear{Mac~Low}{Mac~Low}{1999}]{maclow_energydissipationrate_1999}
Mac~Low M.-M.,  1999, \mn@doi [Astrophys. J.] {10.1086/307784}, 524, 169

\bibitem[\protect\citeauthoryear{Marinacci, Vogelsberger, Kannan, Mocz, Pakmor
  \& Springel}{Marinacci
  et~al.}{2018}]{marinacci_nonidealmagnetohydrodynamicsmoving_2018}
Marinacci F.,  Vogelsberger M.,  Kannan R.,  Mocz P.,  Pakmor R.,   Springel
  V.,  2018, \mn@doi [Mon. Not. R. Astron. Soc.] {10.1093/mnras/sty397}, 476,
  2476

\bibitem[\protect\citeauthoryear{Meurer et~al.,}{Meurer
  et~al.}{2017}]{meurer_sympysymboliccomputing_2017}
Meurer A.,  et~al., 2017, \mn@doi [PeerJ Comput. Sci.] {10.7717/peerj-cs.103},
  3, e103

\bibitem[\protect\citeauthoryear{Mignone \& Tzeferacos}{Mignone \&
  Tzeferacos}{2010}]{mignone_secondorderunsplitgodunov_2010}
Mignone A.,  Tzeferacos P.,  2010, \mn@doi [J. Comput. Phys.]
  {10.1016/j.jcp.2009.11.026}, 229, 2117

\bibitem[\protect\citeauthoryear{Mignone, Bodo, Massaglia, Matsakos, Tesileanu,
  Zanni  \& Ferrari}{Mignone et~al.}{2007}]{mignone_plutonumericalcode_2007}
Mignone A.,  Bodo G.,  Massaglia S.,  Matsakos T.,  Tesileanu O.,  Zanni C.,
  Ferrari A.,  2007, \mn@doi [Astrophys. J. Suppl. Ser.] {10.1086/513316}, 170,
  228

\bibitem[\protect\citeauthoryear{Mignone, Zanni, Tzeferacos, van Straalen,
  Colella  \& Bodo}{Mignone et~al.}{2012}]{mignone_plutocodeadaptive_2012}
Mignone A.,  Zanni C.,  Tzeferacos P.,  van Straalen B.,  Colella P.,   Bodo
  G.,  2012, \mn@doi [Astrophys. J. Suppl. Ser.] {10.1088/0067-0049/198/1/7},
  198, 7

\bibitem[\protect\citeauthoryear{Miyoshi \& Kusano}{Miyoshi \&
  Kusano}{2005}]{miyoshi_multistatehllapproximate_2005}
Miyoshi T.,  Kusano K.,  2005, \mn@doi [J. Comput. Phys.]
  {10.1016/j.jcp.2005.02.017}, 208, 315

\bibitem[\protect\citeauthoryear{Mocz, Vogelsberger, Sijacki, Pakmor  \&
  Hernquist}{Mocz et~al.}{2014}]{mocz_discontinuous_2014}
Mocz P.,  Vogelsberger M.,  Sijacki D.,  Pakmor R.,   Hernquist L.,  2014,
  \mn@doi [Mon. Not. R. Astron. Soc.] {10.1093/mnras/stt1890}, 437, 397

\bibitem[\protect\citeauthoryear{Mocz, Pakmor, Springel, Vogelsberger,
  Marinacci  \& Hernquist}{Mocz et~al.}{2016}]{mocz_moving_2016}
Mocz P.,  Pakmor R.,  Springel V.,  Vogelsberger M.,  Marinacci F.,   Hernquist
  L.,  2016, \mn@doi [Mon. Not. R. Astron. Soc.] {10.1093/mnras/stw2004}, 463,
  477

\bibitem[\protect\citeauthoryear{Nelson, Vogelsberger, Genel, Sijacki, Kere{\v
  s}, Springel  \& Hernquist}{Nelson et~al.}{2013}]{nelson_moving_2013}
Nelson D.,  Vogelsberger M.,  Genel S.,  Sijacki D.,  Kere{\v s} D.,  Springel
  V.,   Hernquist L.,  2013, \mn@doi [Mon. Not. R. Astron. Soc.]
  {10.1093/mnras/sts595}, 429, 3353

\bibitem[\protect\citeauthoryear{Pakmor \& Springel}{Pakmor \&
  Springel}{2013}]{pakmor_simulations_2013}
Pakmor R.,  Springel V.,  2013, \mn@doi [Mon. Not. R. Astron. Soc.]
  {10.1093/mnras/stt428}, 432, 176

\bibitem[\protect\citeauthoryear{Pakmor, Bauer  \& Springel}{Pakmor
  et~al.}{2011}]{pakmor_magnetohydrodynamics_2011}
Pakmor R.,  Bauer A.,   Springel V.,  2011, \mn@doi [Mon. Not. R. Astron. Soc.]
  {10.1111/j.1365-2966.2011.19591.x}, 418, 1392

\bibitem[\protect\citeauthoryear{Pakmor, Marinacci  \& Springel}{Pakmor
  et~al.}{2014}]{pakmor_magneticfieldscosmological_2014}
Pakmor R.,  Marinacci F.,   Springel V.,  2014, \mn@doi [Astrophys. J. Lett.]
  {10.1088/2041-8205/783/1/L20}, 783, L20

\bibitem[\protect\citeauthoryear{Pakmor et~al.,}{Pakmor
  et~al.}{2017}]{pakmor_magneticfieldformation_2017}
Pakmor R.,  et~al., 2017, \mn@doi [Mon. Not. R. Astron. Soc.]
  {10.1093/mnras/stx1074}, 469, 3185

\bibitem[\protect\citeauthoryear{Powell}{Powell}{1994}]{powell_approximate_1994}
Powell K.~G.,  1994, Technical report, Approximate {{Riemann}} Solver for
  Magnetohydrodynamics (That Works in More than One Dimension).
{NASA ICASE}

\bibitem[\protect\citeauthoryear{Powell, Roe, Linde, Gombosi  \&
  De~Zeeuw}{Powell et~al.}{1999}]{powell_solution-adaptive_1999}
Powell K.~G.,  Roe P.~L.,  Linde T.~J.,  Gombosi T.~I.,   De~Zeeuw D.~L.,
  1999, \mn@doi [J. Comput. Phys.] {10.1006/jcph.1999.6299}, 154, 284

\bibitem[\protect\citeauthoryear{Qiu \& Shu}{Qiu \&
  Shu}{2004}]{jianxian_qiu_runge-kutta_2004}
Qiu J.,  Shu C.-w.,  2004, \mn@doi [SIAM J. Sci. Comput.]
  {10.1137/S1064827503425298}, 26, 907

\bibitem[\protect\citeauthoryear{Rieder \& Teyssier}{Rieder \&
  Teyssier}{2016}]{rieder_smallscaledynamofeedbackdominated_2016}
Rieder M.,  Teyssier R.,  2016, \mn@doi [Mon. Not. R. Astron. Soc.]
  {10.1093/mnras/stv2985}, 457, 1722

\bibitem[\protect\citeauthoryear{Rieder \& Teyssier}{Rieder \&
  Teyssier}{2017}]{rieder_smallscaledynamofeedbackdominated_2017b}
Rieder M.,  Teyssier R.,  2017, \mn@doi [Mon. Not. R. Astron. Soc.]
  {10.1093/mnras/stx2276}, 472, 4368

\bibitem[\protect\citeauthoryear{Robertson, Kravtsov, Gnedin, Abel  \&
  Rudd}{Robertson
  et~al.}{2010}]{robertson_computationaleulerianhydrodynamics_2010}
Robertson B.~E.,  Kravtsov A.~V.,  Gnedin N.~Y.,  Abel T.,   Rudd D.~H.,  2010,
  \mn@doi [Mon. Not. R. Astron. Soc.] {10.1111/j.1365-2966.2009.15823.x}, 401,
  2463

\bibitem[\protect\citeauthoryear{Robinet, Gressier, Casalis  \&
  Moschetta}{Robinet et~al.}{2000}]{robinet_shockwaveinstability_2000}
Robinet J.-C.,  Gressier J.,  Casalis G.,   Moschetta J.-M.,  2000, \mn@doi [J.
  Fluid Mech.] {10.1017/S0022112000001129}, 417, 237

\bibitem[\protect\citeauthoryear{Rossmanith}{Rossmanith}{2006}]{rossmanith_unstaggeredhighresolution_2006}
Rossmanith J.,  2006, \mn@doi [SIAM J. Sci. Comput.] {10.1137/050627022}, 28,
  1766

\bibitem[\protect\citeauthoryear{Rossmanith}{Rossmanith}{2013}]{rossmanith_high-order_2013}
Rossmanith J.~A.,  2013, ArXiv13104251 MathNA

\bibitem[\protect\citeauthoryear{Ryu, Ostriker, Kang  \& Cen}{Ryu
  et~al.}{1993}]{ryu_cosmological_1993}
Ryu D.,  Ostriker J.~P.,  Kang H.,   Cen R.,  1993, \mn@doi [Astrophys. J.]
  {10.1086/173051}, 414, 1

\bibitem[\protect\citeauthoryear{Ryu, Miniati, Jones  \& Frank}{Ryu
  et~al.}{1998}]{ryu_divergence-free_1998}
Ryu D.,  Miniati F.,  Jones T.~W.,   Frank A.,  1998, \mn@doi [Astrophys. J.]
  {10.1086/306481}, 509, 244

\bibitem[\protect\citeauthoryear{Schaal}{Schaal}{2016}]{schaal_shocks_2016}
Schaal K.,  2016, PhD thesis, Universit\"at Heidelberg

\bibitem[\protect\citeauthoryear{Schaal, Bauer, Chandrashekar, Pakmor,
  Klingenberg  \& Springel}{Schaal et~al.}{2015}]{schaal_astrophysical_2015}
Schaal K.,  Bauer A.,  Chandrashekar P.,  Pakmor R.,  Klingenberg C.,
  Springel V.,  2015, \mn@doi [Mon. Not. R. Astron. Soc.]
  {10.1093/mnras/stv1859}, 453, 4279

\bibitem[\protect\citeauthoryear{Schekochihin, Cowley, Taylor, Maron  \&
  McWilliams}{Schekochihin
  et~al.}{2004}]{schekochihin_simulationssmallscaleturbulent_2004}
Schekochihin A.~A.,  Cowley S.~C.,  Taylor S.~F.,  Maron J.~L.,   McWilliams
  J.~C.,  2004, \mn@doi [Astrophys. J.] {10.1086/422547}, 612, 276

\bibitem[\protect\citeauthoryear{Shu}{Shu}{1987}]{shu_tvbuniformlyhighorder_1987}
Shu C.-W.,  1987, \mn@doi [Math. Comput.] {10.1090/S0025-5718-1987-0890256-5},
  49, 105

\bibitem[\protect\citeauthoryear{Shu}{Shu}{1998}]{shu_essentiallynonoscillatoryweighted_1998}
Shu C.-W.,  1998, in Quarteroni A.,  ed., , Vol.~1697, Advanced {{Numerical
  Approximation}} of {{Nonlinear Hyperbolic Equations}}.
{Springer Berlin Heidelberg}, Berlin, Heidelberg, pp 325--432,
  \mn@doi{10.1007/BFb0096355}

\bibitem[\protect\citeauthoryear{Shu}{Shu}{2013}]{shu_brief_2013}
Shu C.-W.,  2013, Adv. Mech., 43, 541

\bibitem[\protect\citeauthoryear{Shu \& Osher}{Shu \&
  Osher}{1989}]{shu_efficientimplementationessentially_1989}
Shu C.-W.,  Osher S.,  1989, in , Upwind and {{High}}-{{Resolution Schemes}}.
{Springer, Berlin, Heidelberg}, pp 328--374,
  \mn@doi{10.1007/978-3-642-60543-7_14}

\bibitem[\protect\citeauthoryear{Sonntag \& Munz}{Sonntag \&
  Munz}{2014}]{sonntag_shockcapturingdiscontinuous_2014}
Sonntag M.,  Munz C.-D.,  2014, in Springer {{Proceedings}} in {{Mathematics}}
  \& {{Statistics}}, Finite {{Volumes}} for {{Complex Applications
  VII}}-{{Elliptic}}, {{Parabolic}} and {{Hyperbolic Problems}}.
{Springer, Cham}, pp 945--953, \mn@doi{10.1007/978-3-319-05591-6_96}

\bibitem[\protect\citeauthoryear{Spiegel, Huynh  \& DeBonis}{Spiegel
  et~al.}{2015}]{spiegel_surveyisentropiceuler_2015}
Spiegel S.~C.,  Huynh H.~T.,   DeBonis J.~R.,  2015, in , 22nd {{AIAA
  Computational Fluid Dynamics Conference}}.
{American Institute of Aeronautics and Astronautics},
  \mn@doi{10.2514/6.2015-2444}

\bibitem[\protect\citeauthoryear{Springel}{Springel}{2010}]{springel_pursimuove_2010}
Springel V.,  2010, \mn@doi [Mon. Not. R. Astron. Soc.]
  {10.1111/j.1365-2966.2009.15715.x}, 401, 791

\bibitem[\protect\citeauthoryear{Stone \& Norman}{Stone \&
  Norman}{1992}]{stone_zeus-2d:_1992}
Stone J.~M.,  Norman M.~L.,  1992, \mn@doi [Astrophys. J. Suppl. Ser.]
  {10.1086/191681}, 80, 791

\bibitem[\protect\citeauthoryear{Stone, Ostriker  \& Gammie}{Stone
  et~al.}{1998}]{stone_dissipationcompressiblemagnetohydrodynamic_1998}
Stone J.~M.,  Ostriker E.~C.,   Gammie C.~F.,  1998, \mn@doi [Astrophys. J.
  Lett.] {10.1086/311718}, 508, L99

\bibitem[\protect\citeauthoryear{Stone, Gardiner, Teuben, Hawley  \&
  Simon}{Stone et~al.}{2008}]{stone_athena:_2008}
Stone J.~M.,  Gardiner T.~A.,  Teuben P.,  Hawley J.~F.,   Simon J.~B.,  2008,
  \mn@doi [Astrophys. J. Suppl. Ser.] {10.1086/588755}, 178, 137

\bibitem[\protect\citeauthoryear{Susanto}{Susanto}{2014}]{susanto_highorderfinitevolumeschemes_2014}
Susanto A.,  2014, PhD thesis, University of Waterloo

\bibitem[\protect\citeauthoryear{Teyssier}{Teyssier}{2002}]{teyssier_cosmologicalhydrodynamicsadaptive_2002}
Teyssier R.,  2002, \mn@doi [Astron. Astrophys.] {10.1051/0004-6361:20011817},
  385, 337

\bibitem[\protect\citeauthoryear{Teyssier, Fromang  \& Dormy}{Teyssier
  et~al.}{2006}]{teyssier_kinematicdynamosusing_2006}
Teyssier R.,  Fromang S.,   Dormy E.,  2006, \mn@doi [J. Comput. Phys.]
  {10.1016/j.jcp.2006.01.042}, 218, 44

\bibitem[\protect\citeauthoryear{Titarev \& Toro}{Titarev \&
  Toro}{2002}]{titarev_ader:_2002}
Titarev V.~A.,  Toro E.~F.,  2002, \mn@doi [J. Sci. Comput.]
  {10.1023/A:1015126814947}, 17, 609

\bibitem[\protect\citeauthoryear{T\'oth}{T\'oth}{2000}]{toth__2000}
T\'oth G.,  2000, J. Comput. Phys., 161, 605

\bibitem[\protect\citeauthoryear{Tricco \& Price}{Tricco \&
  Price}{2012}]{tricco_constrainedhyperbolicdivergence_2012}
Tricco T.~S.,  Price D.~J.,  2012, \mn@doi [J. Comput. Phys.]
  {10.1016/j.jcp.2012.06.039}, 231, 7214

\bibitem[\protect\citeauthoryear{Tricco, Price  \& Bate}{Tricco
  et~al.}{2016}]{tricco_constrained_2016}
Tricco T.~S.,  Price D.~J.,   Bate M.~R.,  2016, \mn@doi [J. Comput. Phys.]
  {10.1016/j.jcp.2016.06.053}, 322, 326

\bibitem[\protect\citeauthoryear{Velasco~Romero, Han~Veiga, Teyssier  \&
  Masset}{Velasco~Romero
  et~al.}{2018}]{velascoromero_planetdiscinteractions_2018}
Velasco~Romero D.~A.,  Han~Veiga M.,  Teyssier R.,   Masset F.~S.,  2018,
  \mn@doi [Mon. Not. R. Astron. Soc.] {10.1093/mnras/sty1192}, 478, 1855

\bibitem[\protect\citeauthoryear{Waagan}{Waagan}{2009}]{waagan_positivemusclhancockscheme_2009}
Waagan K.,  2009, \mn@doi [J. Comput. Phys.] {10.1016/j.jcp.2009.08.020}, 228,
  8609

\bibitem[\protect\citeauthoryear{Waagan, Federrath  \& Klingenberg}{Waagan
  et~al.}{2011}]{waagan_robust_2011}
Waagan K.,  Federrath C.,   Klingenberg C.,  2011, \mn@doi [J. Comput. Phys.]
  {10.1016/j.jcp.2011.01.026}, 230, 3331

\bibitem[\protect\citeauthoryear{Wang \& Abel}{Wang \&
  Abel}{2009}]{wang_magnetohydrodynamicsimulationsdisk_2009}
Wang P.,  Abel T.,  2009, \mn@doi [Astrophys. J.] {10.1088/0004-637X/696/1/96},
  696, 96

\bibitem[\protect\citeauthoryear{Wang, Zhang, Shu  \& Ning}{Wang
  et~al.}{2012}]{wang_robust_2012}
Wang C.,  Zhang X.,  Shu C.-W.,   Ning J.,  2012, \mn@doi [J. Comput. Phys.]
  {10.1016/j.jcp.2011.10.002}, 231, 653

\bibitem[\protect\citeauthoryear{Warburton \& Karniadakis}{Warburton \&
  Karniadakis}{1999}]{warburton_discontinuousgalerkinmethod_1999}
Warburton T.~C.,  Karniadakis G.~E.,  1999, \mn@doi [J. Comput. Phys.]
  {10.1006/jcph.1999.6248}, 152, 608

\bibitem[\protect\citeauthoryear{Wiener, Oh  \& Guo}{Wiener
  et~al.}{2013}]{wiener_cosmicraystreaming_2013}
Wiener J.,  Oh S.~P.,   Guo F.,  2013, \mn@doi [Mon. Not. R. Astron. Soc.]
  {10.1093/mnras/stt1163}, 434, 2209

\bibitem[\protect\citeauthoryear{Winters \& Gassner}{Winters \&
  Gassner}{2016}]{winters_affordable_2016}
Winters A.~R.,  Gassner G.~J.,  2016, \mn@doi [J. Comput. Phys.]
  {10.1016/j.jcp.2015.09.055}, 304, 72

\bibitem[\protect\citeauthoryear{Xu \& Liu}{Xu \& Liu}{2016}]{xu_new_2016}
Xu Z.,  Liu Y.,  2016, \mn@doi [J. Comput. Phys.] {10.1016/j.jcp.2016.09.044},
  327, 203

\bibitem[\protect\citeauthoryear{Yakovlev, Xu  \& Li}{Yakovlev
  et~al.}{2013}]{yakovlev_locally_2013}
Yakovlev S.,  Xu L.,   Li F.,  2013, \mn@doi [J. Comput. Sci.]
  {10.1016/j.jocs.2012.05.002}, 4, 80

\bibitem[\protect\citeauthoryear{Zanotti, Fambri  \& Dumbser}{Zanotti
  et~al.}{2015}]{zanotti_solving_2015}
Zanotti O.,  Fambri F.,   Dumbser M.,  2015, \mn@doi [Mon. Not. R. Astron.
  Soc.] {10.1093/mnras/stv1510}, 452, 3010

\bibitem[\protect\citeauthoryear{Zhang \& Shu}{Zhang \&
  Shu}{2010}]{zhang_positivity-preserving_2010}
Zhang X.,  Shu C.-W.,  2010, J. Comput. Phys., 229, 8918

\bibitem[\protect\citeauthoryear{Zhao \& Tang}{Zhao \&
  Tang}{2017}]{zhao_rungekutta_2017}
Zhao J.,  Tang H.,  2017, \mn@doi [J. Comput. Phys.]
  {10.1016/j.jcp.2017.04.027}, 343, 33

\bibitem[\protect\citeauthoryear{Zhao, Yang  \& Seyler}{Zhao
  et~al.}{2014}]{zhao_positivity-preserving_2014}
Zhao X.,  Yang Y.,   Seyler C.~E.,  2014, \mn@doi [J. Comput. Phys.]
  {10.1016/j.jcp.2014.08.044}, 278, 400

\bibitem[\protect\citeauthoryear{Zhong \& Shu}{Zhong \&
  Shu}{2013}]{zhong_simpleweightedessentially_2013}
Zhong X.,  Shu C.-W.,  2013, \mn@doi [J. Comput. Phys.]
  {10.1016/j.jcp.2012.08.028}, 232, 397

\bibitem[\protect\citeauthoryear{Zhu, Feng, Xia, Shu, Gu  \& Fang}{Zhu
  et~al.}{2013}]{zhu_turbulence_2013}
Zhu W.,  Feng L.-l.,  Xia Y.,  Shu C.-W.,  Gu Q.,   Fang L.-Z.,  2013, \mn@doi
  [Astrophys. J.] {10.1088/0004-637X/777/1/48}, 777, 48

\bibitem[\protect\citeauthoryear{Zhu, Pakmor, van Kerkwijk  \& Chang}{Zhu
  et~al.}{2015}]{zhu_magnetized_2015}
Zhu C.,  Pakmor R.,  van Kerkwijk M.~H.,   Chang P.,  2015, \mn@doi [Astrophys.
  J. Lett.] {10.1088/2041-8205/806/1/L1}, 806, L1

\makeatother
\end{thebibliography}

%%%%%%%%%%%%%%%%%%%%%%%%%%%%%%%%%%%%%%%%%%%%%%%%%%

%%%%%%%%%%%%%%%%% APPENDICES %%%%%%%%%%%%%%%%%%%%%

\appendix

\section{Locally divergence-free basis}
\label{sec:apdx-div0}

\subsection{Divergence-free vector space}

In this section, we take $D$ to be the number of spatial dimensions,
and work in spatial coordinates of the reference cell, $(\xi_1, \ldots, \xi_D) \in [-1,1]^D$,
which are the indeterminates of our polynomials.
Let $\mathcal{P}_d$ be the vector space of $D$-variate polynomials of degree $\leq d$:
\begin{equation}
  \mathcal{P}_d = \left\{ \sum_{k_1+\ldots+k_D \leq d} c_{k_1 \cdots k_D} \xi_1^{k_1} \cdots \xi_D^{k_D},
    \quad c_{k_1 \cdots k_D} \in \mathbb{R}
  \right\}.
\end{equation}
This vector space has dimension:
\begin{align}
  n(d) \coloneqq \dim \mathcal{P}_d & = \sum_{k_1+\ldots+k_D \leq d} 1 \\
                                    & = \frac{1}{2}(d+1)(d+2) \quad \text{ for }D=2 \\
                                    & = \frac{1}{6}(d+1)(d+2)(d+3) \quad \text{ for }D=3.
\end{align}
Let $\mathcal{V}_d$ be the space of polynomial vector fields of degree $\leq d$:
\begin{equation}
  \mathcal{V}_d = \left\{
    \begin{bmatrix} Q_1 \\ \vdots \\ Q_D \end{bmatrix},
      \quad (Q_1, \ldots, Q_D) \in \mathcal{P}_d^D
  \right\},
\end{equation}
and it follows that
\begin{equation}
  \dim \mathcal{V}_d = D \dim \mathcal{P}_d = Dn(d).
\end{equation}

We are interested in $\mathcal{V}_d^\text{div}$, the vector space of divergence-free polynomial vectors:
\begin{equation}
  \mathcal{V}_d^\text{div} = \left\{
    \bm V \in \mathcal{V}_d,
    \sum_{\alpha=1}^D \frac{\partial V_\alpha}{\partial \xi_\alpha} = 0
  \right\}.
\end{equation}
We can obtain the dimension of $\mathcal{V}_d^\text{div}$ from the rank-nullity theorem.
Consider the divergence as a linear application $\delta$:
\begin{align}
  \delta: \mathcal{V}_d & \rightarrow \mathcal{P}_{d-1} \\
  \bm V & \mapsto \dotp{\bm \nabla}{\bm V}.
\end{align}
By definition of $\mathcal{V}_d^\text{div}$, $\mathcal{V}_d^\text{div} = \text{Ker}\,\delta$.
Furthermore, for any $\bm V \in \mathcal{V}_d$, $\delta(\bm V)$ is a scalar polynomial of degree $d-1$,
so $\text{Im}\,\delta \subset \mathcal{P}_{d-1}$.
But the reverse inclusion also holds, since for any scalar polynomial $Q \in \mathcal{P}_{d-1}$, we can construct $\bm V \in \mathcal{V}_d$ such that $\delta(\bm V) = Q$, for example:
\begin{equation}
\bm V = \begin{bmatrix} \int Q \,\mathrm{d}\xi_1 \\ 0 \\ \vdots \\ 0 \end{bmatrix} \in \mathcal{V}_d,
\end{equation}
from which we conclude that $\text{Im}\,\delta = \mathcal{P}_{d-1}$.
Then, the rank-nullity theorem applied to $\delta$ states:
\begin{align}
  \dim \mathcal{V}_d &= \dim \text{Im}\,\delta + \dim \text{Ker}\,\delta \\
                     &= \dim \mathcal{P}_{d-1} + \dim \mathcal{V}_d^\text{div},
\end{align}
from which we find:
\begin{equation}
  N_\text{div} \coloneqq \dim \mathcal{V}_d^\text{div} = D n(d) - n(d-1).
  \label{eq:dimVdiv}
\end{equation}

\begin{table}
  \centering
  \begin{tabular}{lrrrrr}
    \toprule
    {} & \multicolumn{5}{l}{$N_\text{div}$} \\
    $d$ &              0 &   1 &   2 &   3 &   4 \\
    $D$ &                &     &     &     &     \\
    \midrule
    2   &              2 &   5 &   9 &  14 &  20 \\
    3   &              3 &  11 &  26 &  50 &  85 \\
    \bottomrule
  \end{tabular}
  \caption{Number $N_\text{div}$ of divergence-free basis functions,
    corresponding to the number of degrees of freedom for the magnetic field,
    as a function of the number of spatial dimensions $D$
  and polynomial degree $d$.}
  \label{tab:ndiv}
\end{table}

\subsection{Generation of basis functions}
Table~\ref{tab:ndiv} shows the number of divergence-free basis functions for $D\in\{2,3\}$ and some values of the degree $d$.
The number of basis functions grows rapidly with $d$, and determining them manually quickly becomes error prone, if not intractable.
In addition, we not only need the functions $\bfunc_l$ but also their $D$ spatial derivatives and the diagonal of the mass matrix.
We therefore resort to symbolic computation using the Python package {\SymPy} \citep{meurer_sympysymboliccomputing_2017}
to compute all the required symbolic expressions and generate corresponding C code.

We adopt a simple, brute-force approach to basis function generation, which we now illustrate in 3D for clarity.
We start by choosing a sequence $\{ Q_i \}$ of $n(d+1)$ polynomials which form a basis of $\mathcal{P}_{d+1}$.
At this stage, we can choose the Legendre polynomials, or simply the canonical basis of monomials.
We then generate a pool of vectors of $\mathcal{V}_{d+1}$ of the form:
\begin{equation}
\left\{ \begin{bmatrix} Q \\ 0 \\ 0 \end{bmatrix}, Q \in \{ Q_i \} \right\} \cup
\left\{ \begin{bmatrix} 0 \\ Q \\ 0 \end{bmatrix}, Q \in \{ Q_i \} \right\} \cup
\left\{ \begin{bmatrix} 0 \\ 0 \\ Q \end{bmatrix}, Q \in \{ Q_i \} \right\}.
\end{equation}
For each vector $\bm V$ of this pool, $\bm \nabla \times \bm V$ is a vector of $\mathcal{V}_d^\text{div}$ since the curl is divergence-free, and at least one degree less than $\bm V$.
We discard the $\bm V$ whose curl vanishes, and
sort the remaining $\bm \nabla \bm \times \bm V$ by increasing total degree, and optionally some other arbitrary choices such as ordering of degree on vector components.

Finally, we start from the first (lowest-degree) vector polynomial in the resulting sequence,
and perform a Gram-Schmidt orthogonalization of the sequence vectors with respect to the dot product:
\begin{equation}
  \innerp{\bm V}{\bm W} = \int_{[-1,1]^3} (V_1 W_1 + V_2 W_2 + V_3 W_3) \, \diffxcell,
\end{equation}
which is exactly the dot product defined at \eqref{eq:innerp} restricted to the components of the magnetic field.
Once we have collected $N_\text{div}$ mutually orthogonalized non-null vectors, the process is complete and we have obtained an orthogonal basis of $\mathcal{V}_d^\text{div}$,
sorted by increasing total degree.

\section{Projection effects with LDF bases}
\label{sec:apdx-div0-proj}

In this section, we illustrate that care should be taken when $L^2$-projecting non-divergence-free functions onto LDF basis vectors.
Such a situation could arise for example with our Powell scheme,
when projecting limited solutions back onto DG weights, as discussed in Section~\ref{sec:limiting-div0}.

\begin{table}
\centering
\begin{tabular}{ccr}
\toprule
Degree & LDF basis index $l$ & Projection           \\
\midrule
0 & 0  &                                        $0$ \\
  & 1  &                                        $0$ \\
\midrule
1 & 2  &                                        $0$ \\
  & 3  &                                        $0$ \\
  & 4  &           $\frac{4}{3} \left(a - b\right)$ \\
\midrule
2 & 5  &                                        $0$ \\
  & 6  &                                        $0$ \\
  & 7  &                                        $0$ \\
  & 8  &                                        $0$ \\
\midrule
3 & 9  &                                        $0$ \\
  & 10 &                                        $0$ \\
  & 11 &       $\frac{152}{249} \left(a + b\right)$ \\
  & 12 &         $- \frac{2}{3} \left(a + b\right)$ \\
  & 13 &                                        $0$ \\
\midrule
4 & 14 &                                        $0$ \\
  & 15 &                                        $0$ \\
  & 16 &                                        $0$ \\
  & 17 &                                        $0$ \\
  & 18 &                                        $0$ \\
  & 19 &                                        $0$ \\
\midrule
5 & 20 &                                        $0$ \\
  & 21 &                                        $0$ \\
  & 22 &    $\frac{1268}{28629} \left(a + b\right)$ \\
  & 23 &        $- \frac{2}{45} \left(a + b\right)$ \\
  & 24 &  $- \frac{220}{121263} \left(a + b\right)$ \\
  & 25 &                                        $0$ \\
  & 26 &                                        $0$ \\
\bottomrule
\end{tabular}
\caption{Contamination of higher-order coefficients
  resulting from the $L^2$ projection of $\bm u = (ax, by)$
  onto a locally divergence-free basis.
 }
\label{tab:ldf-proj-contamination}
\end{table}

The main issue with these projections is that lower-degree contributions can contaminate higher-degree terms.
Table~\ref{tab:ldf-proj-contamination} illustrates this effect
with a simple first-degree 2D vector field $\bm u = (ax, by)$
by showing the projections $\int_K \bm u \cdot \bfunc_l$ for each 2D LDF basis function $\bfunc_l$.
Note that $\bm \nabla \bm \cdot \bm u = a+b$,
and even though $\bm u$ has only degree 1,
the projection can impact arbitrarily high degrees of the LDF basis whenever $a+b \neq 0$.
This effect is due to the LDF basis not spanning the whole space of polynomial vector fields,
and can be seen as a special form of polynomial aliasing with LDF bases.

\section{Divergence}
\label{sec:apdx-divergence}

\subsection{A prescription for the amount of divergence}

Different prescriptions to estimate the global divergence of the magnetic field have been proposed in the context of MHD or DG schemes \citep[see e.g.][]{cockburn_locally_2004,winters_affordable_2016}.
In this work, we adopt a definition suitable to DG methods
which is similar to that given in Section~2.1 of \citet{cockburn_locally_2004}.
We define the divergence of the magnetic field in a cell $K$ of volume $V_K$ as:
\begin{align}
  \left| \divB \right|_K \coloneqq
     \frac{1}{V_K} \int_{K} \left| \divB \right| \, \diffxlab
     + \frac{1}{V_K} \int_{\partial K} \left| \Bnormal_F - \Bnormal \right| \, \diffS,
  \label{eq:divb-cell}
\end{align}
where $\Bnormal = \dotp{\Bfield}{\facenormal}$ with $\facenormal$ the face normal vector,
and $\Bnormal_F$ is the normal component of the magnetic field at the face (see \ref{sec:normal-bx-choice}).
The first integral is the $L^1$ norm of the divergence inside the cell volume,
while the second is the $L^1$ norm of the jump in the normal magnetic field component
over all faces of the cell.
The face value $\Bnormal_F$ is used here as a way to distribute the total jump $\Bnormal_R - \Bnormal_L$
between neighbouring cells $L$ and $R$
in a way that is consistent with the face value seen by the Riemann solver
(see Fig.~\ref{fig:powell}).
The face and volume integrals are computed with the same Gauss-Legendre quadrature
used in the DG scheme \citep[see][]{schaal_astrophysical_2015}.

We define the global divergence norm of the solution over all cells in the simulation volume as:
\begin{align}
  \left\| \divB \right\|_1 \coloneqq \frac{1}{V} \sum_K V_K \left| \divB \right|_K.
  \label{eq:divb-total}
\end{align}
This definition corresponds to the norm of the divergence given in Section~2.1 of \cite{cockburn_locally_2004}
whenever $\Bnormal_F \in \left[ \Bnormal_L , \Bnormal_R \right]$,
which is the case for both definitions of $\Bnormal_F$ proposed in \ref{sec:normal-bx-choice}.
As noted by \cite{cockburn_locally_2004},
$\left\| \divB \right\|_1 = 0$ implies that the field is exactly globally divergence free everywhere,
and as such it is a stringent measure of the amount of divergence present in the solution.

\subsection{Normalized divergence}
\label{sec:apdx-divergence-normalized}

Following other authors
\citep[e.g.][among others]{pakmor_magnetohydrodynamics_2011,tricco_constrainedhyperbolicdivergence_2012,hopkins_constrained-gradient_2016},
we define the normalized divergence in a cell $K$ of size $\Delta x$ as
\begin{align}
  \frac{\left| \divB \right|_K}{\left| \Bfield \right|_K} \Delta x,
  \label{eq:divb-normalized}
\end{align}
where $\left| \Bfield \right|_K$ is the norm of the average magnetic field in the cell.
This quantity is dimensionless, and measures the relative error on the magnetic field
that can be attributed to the violation of the divergence constraint in the cell.
In that sense, it measures the relative severity of the magnetic field divergence.

\subsection{Signed divergence}

Note that by definition, $\left| \divB \right|_K$ of Eq.~\eqref{eq:divb-cell} is always positive.
In some instances, it is useful to give a signed definition of the divergence,
for example to assess the sign of the contribution of the Powell term \eqref{eq:powell-src} to various conserved quantities.
In such cases, the local divergence can be estimated as the cell average:
\begin{align}
  \left( \divB \right)_K \coloneqq \frac{1}{V} \int_{\partial K} \Bnormal_F \, \diffS,
  \label{eq:divb-gauss}
\end{align}
where the integral runs over all faces.
Here, the normal component $\Bnormal_F$ at the face may be chosen as in \ref{sec:normal-bx-choice} with the HLLD solver
(or returned by the Riemann solver, if using an 8-wave solver which provides an interface state).
This definition arises from averaging $\divB$ over the cell volume, applying the divergence theorem, and setting $\Bfield \coloneqq \Bfield_F$ at the face.
It adopts the finite volume formulation of DG,
while being sensitive to the choice of normal component of $\Bfield$ which is seen by the Riemann solver.
However, it is less stringent than the definition \eqref{eq:divb-cell}
because cancellations can occur in the integral.

%%%%%%%%%%%%%%%%%%%%%%%%%%%%%%%%%%%%%%%%%%%%%%%%%%

% Don't change these lines
\bsp	% typesetting comment
\label{lastpage}
\end{document}